\documentclass[traditabstract]{aa}  
\usepackage{natbib}
\bibpunct{(}{)}{;}{a}{}{,}
\usepackage{amssymb}
\usepackage{longtable}
\usepackage{txfonts}
\usepackage{graphicx}
\usepackage{epstopdf}
\usepackage{pdflscape}
\usepackage{color}
\newcommand{\teff}{T$_{\rm eff}$~} 
\newcommand{\logg}{\ensuremath{\log g}}
\newcommand{\vmic}{v$_\mathrm{turb}$}
\newcommand{\msun}{M$_\odot$~}
\newcommand{\lsun}{L$_\odot$~}

\newcommand{\kms}{\,km\,s$^{-1}$}    %kms -1
\newcommand {\bv} {{\it B--V\/}}
\newcommand {\vi} {{\it V--I\/}}
\newcommand {\vh} {{\it V--H\/}}
\newcommand {\vj} {{\it V--J\/}}
\newcommand {\vk} {{\it V--K\/}}
\newcommand {\mb} {{\it B\/}}
\newcommand {\mv} {{\it V\/}}
\newcommand {\mi} {{\it I\/}}
\begin{document} 

   \title{What is the Milky Way outer halo made of? \thanks{Based on ESO program 093.B-0615(A)},\thanks{Based on observations obtained with
   the Hobby-Eberly Telescope, which is a joint project of the University of Texas at Austin,
   the Pennsylvania State University, Stanford University, Ludwig-Maximilians-Universit\"at
   M\"unchen, and Georg-August-Universit\"at G\"ottingen.},\thanks{This paper presents data gathered with the
   {\it Magellan} Telescopes at Las Campanas Observatory, Chile.}}

   \subtitle{High resolution spectroscopy of distant red giants}

   \author{G. Battaglia\inst{1,2}
          \and
          P. North\inst{3}
          \and
    P. Jablonka\inst{3,4}
          \and
    M. Shetrone\inst{5}
          \and
    D. Minniti\inst{6,7,8}
          \and
          M. D\'{i}az\inst{9}
          \and
	  E. Starkenburg\inst{10}
    \and
    M. Savoy\inst{3}
     }

   \institute{Instituto de Astrofisica de Canarias, calle Via Lactea s/n, E-38205 La Laguna, Tenerife, Spain\\
     \email{gbattaglia@iac.es}
           \and
     Universidad de La Laguna, Dpto. Astrofisica, E-38206 La Laguna, Tenerife, Spain
           \and
     Institute of Physics, Laboratory of Astrophysics, 
           \'Ecole Polytechnique F\'ed\'erale de Lausanne (EPFL),
           Observatoire de Sauverny,
           CH-1290 Versoix, Switzerland
           \and
            GEPI, Observatoire de Paris, CNRS, Universit\'e de Paris Diderot, F-92195
           Meudon, Cedex, France
           \and
     University of Texas at Austin, McDonald Observatory, USA
           \and
     Instituto Milenio de Astrofisica, Santiago, Chile
           \and
     Departamento de Fisica, Facultad de Ciencias Exactas, Universidad Andres Bello,
     Av. Fernandez Concha 700, Las Condes, Santiago, Chile
           \and
     Vatican Observatory, V00120 Vatican City State, Italy
           \and
     Departamento de Astronomia, Universidad de Chile, Camino el Observatorio 1515,
     Las Condes, Santiago, Chile, Casilla 36-D
           \and
     Leibniz-Institut f\"ur Astrophysik Potsdam (AIP)
     An der Sternwarte 16, 14482 Potsdam, Germany
      }

   \date{Received: 01/09/2017; accepted: 02/10/2017}

   \abstract{In a framework where galaxies form hierarchically,
     extended stellar haloes are predicted 
     to be an ubiquitous feature around Milky Way-like galaxies and to consist mainly of the shredded stellar component
     of smaller galactic systems.
     The type of accreted stellar systems are expected to vary according to the specific accretion and merging history 
of a given galaxy, and so is the fraction of stars formed in-situ versus accreted. Analysis of the  
chemical properties of Milky Way halo stars out to large Galactocentric 
radii can provide important insights into the properties of the environment in which the stars that contributed to the 
build-up of different regions of the Milky Way stellar halo formed. 
In this work we focus on the outer regions of the Milky Way stellar halo, by determining 
chemical abundances of halo stars with large present-day  
Galactocentric distances, $>$15 kpc. The data-set we acquired consists of high resolution
HET/HRS, Magellan/MIKE and VLT/UVES spectra for 28
red giant branch stars covering a wide metallicity range, $-3.1 \lesssim$[Fe/H]$\lesssim -0.6$. 
We show that the ratio of $\alpha$-elements over Fe as a function of [Fe/H] for our 
sample of outer halo stars is not dissimilar from the pattern shown by MW halo stars from 
solar neighborhood samples. On the other hand, significant differences appear at [Fe/H]$\gtrsim -1.5$
when considering chemical abundance ratios such as [Ba/Fe], [Na/Fe], [Ni/Fe], [Eu/Fe], [Ba/Y].
Qualitatively, this type of chemical abundance trends are observed in massive dwarf galaxies, such
as Sagittarius and the Large Magellanic Cloud.
This appears to suggest a larger contribution
in the outer halo of stars formed in an environment with high initial star formation rate and already
polluted by asymptotic giant branch stars with respect to inner halo samples.}


   \maketitle
%
%________________________________________________________________
   
   \section{Introduction}

Extended stellar haloes containing a significant amount of streams and substructures 
are observed around the Milky Way (MW), M31, and there are indications that they may be a ubiquitous 
component of galaxies down to the scale of Large Magellanic Cloud (LMC)-like objects \citep[e.g., for recent works and reviews, 
see][]{Helmi2008, Mcconnachie2009, 
Martinez-Delgado2009, Mouhcine2011, Rich2012}. 

Within the $\Lambda$ Cold Dark Matter ($\Lambda$CDM) framework, stellar haloes are a natural outcome of the hierarchical 
build-up of structures \citep[e.g.,][]{Bullock2005, Cooper2010}.
High resolution N-body and hydro-dynamical simulations \citep[e.g.,][]{Zolotov2009, Tissera2012, Tissera2013, Tissera2014, Pillepich2014}
show that stellar haloes are expected to consist both of stars formed within the virial radius of the main progenitor, for instance 
in a disk structure from the dissipative collapse of smoothly accreted cold gas at high-redshift and
later on put on halo orbits by some violent event, and by the shredded stellar component
of smaller galactic systems accreted onto the main progenitor\footnote{Some of these systems will still be
  gas-rich when accreted and will continue forming stars, formally
  within the virial halo of the main progenitor; even though \citet{Tissera2013} treat them as
  a separate component, here we consider them as ``accreted stars'', since their
  properties should reflect the chemical enrichment and star formation history of the accreted
  sub-galactic system.}. The relative dominance of these mechanisms depends upon the specific
build-up history of a given galaxy, but there is general consensus that the outer parts of
stellar haloes almost exclusively host accreted stars, whilst the inner regions
might also contain an important component of stars originated within the main progenitor. Another expectation
from simulations is that stars from early accretion events are to be preferentially
found in the inner regions of haloes, while the outer parts are in general dominated by late accretion events.

Observational evidence for spatial variations in the properties of the MW stellar halo, likely related to
different dominant formation mechanisms at play, was 
first put forward by \citet{Searle1978} on the basis of the metallicity and horizontal branch colors
of MW globular clusters. The wealth of information brought about in the last years by very large area photometric and spectroscopic surveys, such as
those within SDSS, has painted a complex observational picture, which confirms the existence of
spatial variations in the kinematic, metallicity, abundance and age properties of the stellar halo, as well as
the existence of a variety of substructures. The stellar halo of our Galaxy can be broadly described
as consisting of at least two partially overlapping components:
a flatter inner-halo population, with a small net prograde rotation and 
a metallicity distribution function peaking at [Fe/H]$\sim -1.6$, and a more extended and approximately
spherical outer-halo population, showing no
or little retrograde rotation \citep[but see the recent work by][]{Deason2017}
and a metallicity distribution function peaking at [Fe/H]$\sim -2.3$
\citep[e.g.,][]{Carollo2007, Carollo2010, deJong2010, Beers2012, AllendePrieto2014, Fernandez-Alvar2015, Das2016}.

The shift in dominance between the inner- and outer- halo component occurs at Galactocentric distances
of $\sim$15kpc, which is also the distance range in which such transition is seen in the simulations.
The age distribution of field blue horizontal branch (BHB) stars
(e.g., Santucci et al. \citeyear{Santucci2015}: 4700 stars with spectroscopic from SDSS;
Carollo et al. \citeyear{Carollo2016}: 130,000 color-selected BHB stars from SDSS
photometry),
shows a decrease of 1-1.5 Gyr in its mean value and a larger age spread
from the inner 15kpc to the outer
regions probed by these studies (45-50kpc); an age difference between stars likely associated to
the inner- and outer-halo population,
with the former being older than the latter, was also pointed out by
\citet{schuster12} with a much smaller sample of solar neighborhood stars with halo kinematics
for which exquisite high resolution spectroscopic data were obtained. 
From the analysis of 100,000 main-sequence
turn-off stars with spectroscopy from SDSS out to 15kpc from the Sun, \citet{Lee2017} show that the median [C/Fe]
increases as a function of height from the Galactic disk mid-plane, as so does
the fraction of CEMP-no stars (those with no over-abundance of heavy neutron-capture elements) versus
CEMP-s stars (those with over-abundance of heavy neutron-capture elements associated to the s-process);
the authors argue this effect might be related to the mass function of the sub-haloes in which the
stars in the inner- and outer-halo region formed. From an analysis of $\sim$4500 K-giants likely
belonging to the halo with spectroscopy from SEGUE and SEGUE-2, \citet{Janesh2016} conclude that
a larger amount of substructure is seen at [Fe/H] $>-1.2$ with respect to the stars in the sample with lower metallicities,
and for those located beyond
30kpc from the MW center with respect to those found at smaller Galactocentric distances
\citep[see also e.g.,][]{Xue2011}; the Sagittarius (Sag) stream appears
responsible for the increase in substructure seen in metallicity and distance, in particular at [Fe/H] $> -1.9$.

In general, the above observational picture appears consistent with the inner regions of the halo having
been assembled earlier and in an environment experiencing a faster initial chemical enrichment than the outer halo; the
outer halo containing the remnants of later accretion events, some of which
have been identified as substructures even when
lacking the full 6D phase space information, due to the larger dynamical mixing times at those distances. 

It is clear then that even though stellar haloes typically account for only a few percent of a galaxy stellar mass, 
studying their spatially varying properties allows us to retrieve crucial information on the galaxy build-up history,
going back to its earliest phases.

Satellite galaxies are
arguably the sub-haloes that escaped tidal disruption during the halo assembly
and survived until present day. The comparison between the chemical abundance
patterns of their stars to those of halo stars offers then 
a particularly illuminating and direct way of identifying in which type of environment halo stars formed, 
as well as for constraining the timescales of accretion events. 

Until recently, this type of analysis, which requires high resolution spectroscopy,
was by necessity restricted to samples of solar neighborhood stars,
providing however already a wealth of information.
The ratio of $\alpha$- elements over Fe in solar neighborhood sample
was found to be super-solar for the overwhelming majority of stars with halo kinematics
\citep[e.g.,][]{venn04}. A dichotomy is present at intermediate metallicities
(especially visible in the range $-1.5 \lesssim$ [Fe/H] $\lesssim -0.8$), with a
sequence of ``high-$\alpha$'' stars, showing an almost constant value at all metallicities
(e.g., [Mg/Fe], [Si/Fe] $\sim$+0.3,+0.4; NLTE [O/Fe] $\sim$+0.5) and a sequence of ``low-$\alpha$'' stars,
at 0.1,0.2dex lower values and with a slightly declining trend for increasing [Fe/H] 
\citep[e.g.,][]{Nissen1997, nissenschuster10, Ramirez2012, Hawkins2015};
the ``high-$\alpha$'' and ``low-$\alpha$''
sequences can be also traced in a number of other elements, as [Cu/Fe], [Zn/Fe], [Ba/Y], [Na/Fe], [Al/Fe], [Ni/Fe],
[(C+N)/Fe] \citep[e.g.,][]{nissenschuster11, Hawkins2015}.
Such chemical patterns have been interpreted as ``high-$\alpha$'' stars likely forming in regions with
a star formation rate high enough that only massive stars and type~II supernovae contributed to the chemical enrichment;
on the other hand, ``low-$\alpha$'' stars most likely originated in environments with a slower chemical evolution,
experiencing enrichment
also from type Ia supernovae and low-mass asymptotic giant branch (AGB) stars. Typically
the ``low-alpha'' sequence is attributed to accreted systems.

Interestingly, the analysis of space velocities of a few dozen halo stars of the solar neighborhood shows that
``low-$\alpha$'' stars have on average more eccentric orbits than ``high-$\alpha$'' stars, allowing them
to reach larger apocenter distances and larger heights above the Galactic plane. This essentially
indicates that ``low-$\alpha$'' stars, possibly born in accreted systems, are likely to belong to the outer-halo population \citep[e.g.,][]{Nissen1997, Fulbright2002, Roederer2009, nissenschuster10}.
On the other hand, APOGEE spectra of 3200 giants have shown that
the ``high-alpha'' sequence appears chemically indistinguishable from the canonical thick disk, with both components
exhibiting a high degree of chemical homogeneity 
\citep{Hawkins2015}; an interpretation offered by Hawkins et al. 
for this finding is that the gas from which the inner regions of the MW halo
formed was also the precursor of the thick disk.
However, the aforementioned work does not
carry out an analysis of how the chemical abundance properties of the various Galactic component might vary
spatially, hence it is not possible to establish where the transition between ``canonical thick disk \& halo'' to
accreted halo occurs.

\citet{Fernandez-Alvar2015} analyze the trends in Ca, Mg, and Fe abundances as a function of Galactocentric distance
out to $\sim$80 kpc 
for almost 4000 stars with low resolution spectroscopy from surveys within SDSS and, in the
range $-1.6 <$[Fe/H]~$<-0.4$, find a decreasing trend for
[Ca/Fe] but an increasing trend for [Mg/Fe]. \citet{Fernandez-Alvar2017} extend their previous analysis
to several other chemical elements, but over a smaller range in Galactocentric distances ($5 < r_g \mathrm{[kpc]} < 30$),
using infrared spectroscopy of giants from APOGEE; the median [X/Fe] for $\alpha$ elements is lower by 
0.1dex (or more for O, Mg, S) for stars at $r_g >$15kpc at [M/H]$> -1.1$ with respect to closer by stars, and differences
are also detected in other elements (Ni, K, Na and Al). This confirms that ``low-$\alpha$'' stars are found at large
Galactocentric distances. 

It should be pointed out that even the so-called ``low-$\alpha$'' stars have chemical abundance patterns that
do not match those of stars in the early-type dwarf galaxy satellites of the MW when compared at the same metallicity;
in other words, in
MW early type satellites the decline (``knee'') in [$\alpha$/Fe] (or individual $\alpha$-elements over Fe) is detected
at much lower metallicities than in the halo 
(e.g., in [Mg/Fe]: at [Fe/H]$\sim -2$ in the Fornax dwarf spheroidal galaxy [hereafter, dSph], Hendricks et al. \citeyear{hendricks14},
Lemasle et al. \citeyear{lemasle14};
[Fe/H]$\sim -1.6$ in the Sculptor dSph, Tolstoy, Hill \& Tosi \citeyear{Tolstoy2009};
[Fe/H]$\sim -2.8$ in the Draco dSph, Cohen \& Huang \citeyear{Cohen2009}). The surviving MW
satellites are on orbits that make them unlikely to contribute halo stars passing through the Solar
Neighborhood \citep{Roederer2009}.
The low metallicity part of the halo might then be compatible 
with having been assembled by the shredded ancient (prior to the time
where SNIa started having a significant contribution to the Fe production) stellar component of systems resembling
the MW early-type satellite progenitors;
however, 
the super-solar values seen in the halo out to much larger metallicity most likely
require the supposedly accreted component to have been formed
in environments experiencing a higher initial SFR anyway \citep[see e.g.,][]{Fulbright2002, venn04}.
Recently, low resolution spectroscopic data from SDSS/SEGUE allowed to place the
$\alpha$-knee of likely Sag stream member stars at [Fe/H]$\sim -1.3$, only slightly
 more metal-poor than in the MW halo \citep{deBoer2014}: this might indicate that a small number of
 large systems
 experiencing an initial chemical enrichment like Sag stream stars might have contributed
 substantially to the build up of the MW stellar halo at old times.

 Detailed chemical abundance properties of distant halo stars
 for a variety of elements with different nucleosynthetic origins allows to make a more refined comparison
 to the properties of MW satellites and of the inner versus outer halo.
 This requires time-consuming high resolution spectroscopy of typically faint stars. 
 The only study in which detailed chemical abundance properties of the stellar halo have been analyzed as a function of
distance from the Galactic center for a large number of elements derived from high resolution spectroscopy is the
work by \citet{Fernandez-Alvar2017} based on APOGEE infrared spectra. However, the authors do not carry out a
comparison between the halo chemical abundance properties and those of MW satellites and do not take explicitly into
account the possible presence of Sag stream stars and how this might affect the interpretation. 

 In this work, we derive chemical abundances from high resolution optical spectroscopy for a sample
 of 28 halo stars which can be considered as highly likely outer-halo objects, due to their
 large present-day Galactocentric distances $>15$ kpc and
 heights from the MW mid-plane |z|$>$9kpc; we can in this way by-pass the typically large uncertainties in the
 measurements of space kinematics and related orbital properties in assigning
 stars presently found in the solar neighborhood to the inner- or outer-halo population.
 We then interpret the chemical abundance
 trends for this sample of outer halo stars in the light of what has been observed for solar neighborhood samples,
 MW satellites, including the LMC, and factoring in the possible presence of Sag
 stream stars in our sample. Our sample is genuinely new, with only one star overlapping with APOGEE \citep{Ahn2014}
 measurements, is competitive in size (e.g., there are $\sim$50 stars beyond a Galactocentric distance of 15kpc in Fernandez-Alvar et l. 2017) and provides abundances for several elements which are 
 not measured from APOGEE spectra (Sc, Cr, Co, Cu, Zn, Sc, Y, Zr, Ba, La, Ce, Pr, Nd, Sm, Eu) as well as 
 a set of elements in common (C, Mg, Ti, Si, Ca, O, Al, Na, Ni, Mn, V), handy for comparisons of trends.

 The article is organized as follows: in Sect.~\ref{sec:data} we describe the sample selection,
 observing facilities used and data reduction procedure; Sects.~\ref{sec:ews} summarizes the
 method for equivalent widths and radial velocity determination, and in Sect.~\ref{sec:analysis}
 we explain the analysis performed for deriving the elemental abundances. We proceed to 
 comparing our radial velocities, metallicities and distances with results in the literature
 for the same stars in Sect.~\ref{sec:comp}. In Sect.~\ref{sec:trends} we comment on the
 chemical abundance trends found for our sample of halo stars, and compare them with
 those of solar neighborhood samples and MW satellites, including Sag and the LMC.
 Sect.~\ref{sec:substructures} is devoted to exploring whether the stars in our sample might belong to
 known substructures. We conclude with a discussion and summary in Sect.~\ref{sec:summary}.

\section{Sample, observations and data reduction} \label{sec:data}
\begin{figure*}
\centering
\includegraphics[width=\hsize]{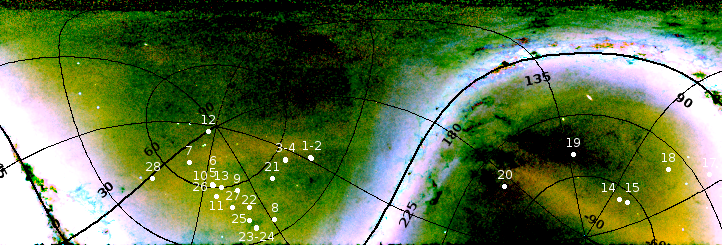}
\caption{Location of our targets overlaid on an RGB rendering of the distribution of Milky Way halo stars in an Equatorial- Carr\'{e} view.
  The latter has been produced by combining the 8 (blue channel), 15 (green channel), 25 (red channel) kpc slices from the Pan-STARRS1 3$\pi$ Survey data-set
  from https:\/\/zenodo.org\/record\/60518\#.WHTxMHqvelM, obtained by applying to the Pan-STARRS1 3$\pi$ Survey data-set 
  matched-filter technique using an [Fe/H]=-1.5 and 12 Gyr old model in the g,r bands 
  (for details, we refer the reader to Bernard et al. \citeyear{Bernard2016}). The grid is in Galactic coordinates.}
\label{fig:location}
\end{figure*}

Our sample consists of 28 individual red giant branch (RGB) stars observed at high spectral resolution
with HET/HRS, Magellan/MIKE and VLT/UVES.
Their location on the sky is shown in Fig.~\ref{fig:location}. Three stars were
observed both with HET/HRS and VLT/UVES to test
the consistency of results from the different facilities. 
The details of the target selection, instrument set-ups and data-reduction will be given below.
Table~\ref{table:atmo} lists the Julian dates of the observations, total exposure times per target and
combined signal to noise per pixel, together with the adopted stellar parameters and radial velocities.
Table~\ref{table:Gaia} lists the equatorial and galactic coordinates provided by the Gaia satellite
in the first data release, as well as the Gaia DR1 ID and the m$_\mathrm{G}$ magnitude
\citep{BVP16,PBB16,LLB16,vLED17,ALB17,ERD17,CEM16}.

\subsection{The sample}
Most of our target RGB stars were selected from the Spaghetti 
survey sample \citep[e.g.,][]{Morrison2000, Morrison2001, Dohm-Palmer2001, Morrison2003}
as published in Starkenburg et al. \citeyear{Starkenburg2010, Starkenburg2011} (21 stars); of the remaining 7 stars, 
 six targets were drawn from the catalog of distance determinations
 for the SEGUE K Giants by \citet{Xue2014} and one from the APOGEE sample.
 The main criterion for
selection was for the giants to be placed at Galactocentric distances\footnote{We assumed that the Sun is found at 8 kpc
  from the Galactic center.} $r_g>$15\,kpc. Additionally, we preferred targets 
bright enough to be within the reach of high resolution 
spectroscopic observations on the chosen facilities in a reasonable exposure time ($V \lesssim$17.5) and 
made sure that the selected targets would cover a large
metallicity range, in particular the [Fe/H] region where the abundances of $\alpha$-elements in solar neighborhood halo samples
differs the most from those of stars in classical dSphs, that is beyond [Fe/H]$=-1.6$.

\subsection{Observations}
The observations were carried out with three high resolution spectrographs:
\begin{itemize}
\item MIKE (Magellan Inamori Kyocera Echelle) attached to the $6.5$m Magellan telescope
at Las Campanas Observatory, Chile \citep{BSG03}. Its blue and red arms cover simultaneously the
wavelength ranges $\sim3350-5060$\AA\ and $4860-9400$\AA\,; however, we only use the portion of the spectra
redder than 3700\AA\, due to the low signal to noise ratio (S/N) of the spectra 
of our cool target stars in the bluest regions of the wavelength range.
With the chosen $1$\arcsec\ slit,
the resolving powers are $\sim 28\,000$ and $\sim 22\,000$ in the blue and red ranges respectively.
A $2\times 2$ pixels binning was adopted on the detectors. The total exposure times varied
between $4400$ and $5400$ seconds, distributed into two or three successive exposures. Observations
were carried out in visitor mode in 2 nights (Mar 8 and June 19, 2014) under program CN2014A-20 (PI: Minniti). 
\item UVES (UV-Visual Echelle Spectrograph) attached to the $8.2$m Kueyen UT2 unit of the
VLT telescope at Paranal Observatory operated by ESO, Chile \citep{DDK00}.
It was used in the Dichroic \#1 mode, providing a spectral coverage
$\sim3500-4525$\AA\ in the blue, and $\sim 4784-5759$\AA\ and $\sim 5838-6805$\AA\ in
the red, with a resolving power $R\sim 45\,000$ for a $1$\arcsec\ slit. A $2\times 2$ pixels
binning was used, as for MIKE. The exposure time was either $5\times 3000$ seconds ($5$ exposures)
or $3\times 2400$ seconds.
Observations were carried out in service mode under program 093.B-0615 (PI: Battaglia, 45h), and the different exposures on a given star
are often separated in time by as much as a whole year.
\item HRS (High Resolution Spectrograph) attached to the $10$m Hobby-Eberly Telescope (HET)
at McDonald Observatory, Texas, USA \citep{T98,RAB98}. The instrument was configured to
HRS\_15k\_central\_600g5822\_2as\_2sky\_IS0\_GC0\_2x5 to
achieve R=18,000 spectra covering 4825\AA\ to 6750\AA\,.
The data were acquired as part of normal queue scheduled observing \citep{Shetrone2007}
under program UT13-2-007 (PI: Shetrone).  
The targets were observed between 
February and June of 2013 for a total of 33.7 hours of shutter open time.
Directly after each target a Th-Ar exposure was
taken and on nearly every night (weather permitting) a radial velocity 
standard was observed during twilight.
The total exposure times varied between $6000$ and $12000$ seconds,
distributed into two to five exposures of  $1800$, $2400$, or $3000$ seconds
spread over a few weeks. 
\end{itemize}

\subsection{Data reduction}

\subsubsection{MIKE}
  The spectra were reduced with the pipeline written by
Dan Kelson\footnote{http://code.obs.carnegiescience.edu/mike} and based
on the Carnegie Python Distribution
(CarPy)\footnote{http://code.obs.carnegiescience.edu/carnegie-python-distribution}.
The pipeline provides flat-fielded, optimally extracted and wavelength calibrated 1-D spectra
for the successive spectral orders. We normalized them to the continuum in a preliminary
way and merged them following the method proposed by \citet{A08}, using the IRAF
package\footnote{IRAF is distributed by the National Optical
  Astronomy Observatory (NOAO), which is operated by the Association of
  Universities for Research in Astronomy (AURA), Inc., under cooperative agreement
  with the U.S. National Science Foundation}.
The 1-dimensional spectra resulting from order merging were then
visually examined and the remaining obvious cosmics removed by hand, using the
IRAF \verb+splot+ subroutine.

\subsubsection{UVES}
The reduction was done with the ESO UVES pipeline (release 5.09) with optimal
extraction. Cosmics and other anomalies (e.g., poorly subtracted telluric emission
lines) were suppressed by hand, as for the MIKE spectra.
  
\subsubsection{HRS}
The spectra were reduced with IRAF ECHELLE scripts.  The standard IRAF scripts for
overscan removal, bias subtraction, flat fielding and scattered light
removal were employed.  For the HRS flat field we masked out
the Li I, H I and Na D regions because the HET HRS flat field lamp 
suffered from faint emission lines. The sky fibers were extracted 
in the same manner as the star fibers.   We also extracted the spectra
of a sky flat and used that to determine the throughput differences
between the sky fibers and the object fibers.   The sky fibers were
then scaled by this value and subtracted from the star flux.
The spectra were combined into a
single long spectrum for the blue and red chips.

\section{Equivalent widths and radial velocity measurements} \label{sec:ews}
\subsection{MIKE}\label{section:MIKE}
To normalize the spectra to the continuum and determine line equivalent widths (EW), we
have applied the 4DAO avatar \citep{M13} of the DAOSPEC code\footnote{DAOSPEC has been written
by P. B. Stetson for the Dominion Astrophysical Observatory of the Herzberg Institute
of Astrophysics, National Research Council, Canada.} \citep{SP08,SP10} to each of the blue and
red spectral ranges, after masking the numerous telluric lines that plague the red
range. In some cases we have divided the blue range in two portions, typically below
and above $4200$~\AA, because of the very poor S/N of the bluer side, whose contribution
is penalized with a low weight. The code also computes the radial velocity (RV) for each
identified line and gives the mean RV and its rms scatter for each
spectral range. The RVs given in Table~\ref{table:atmo}
are weighted means of the RVs determined for each spectral range treated by the 4DAO code.
The weights are the reciprocal of the squared respective errors, the latter being
defined as the rms scatter given by DAOSPEC divided by the square root of the number
of identified lines. The errors on the mean RV values correspond to their rms scatter
divided by the square root of the number of spectral ranges (only 2 or
3 in practice).

\begin{figure}
\centering
\includegraphics[width=\hsize]{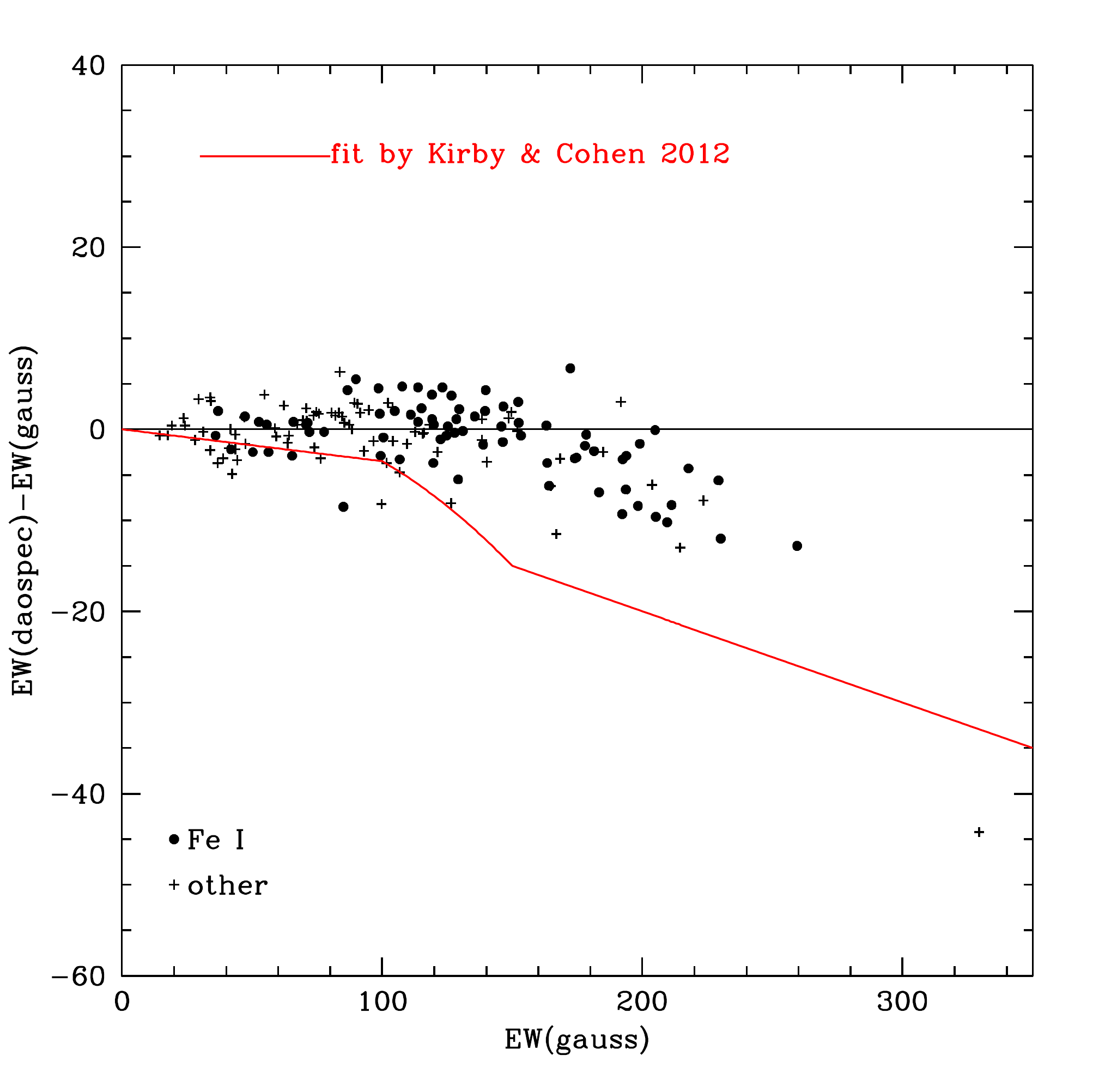}
\caption{Comparison of the EWs determined by DAOSPEC with those obtained by ``manual'' 
  Gaussian fits for clean lines, for star \#13, of the MIKE sample. The red curve
shows the bias found by Kirby \& Cohen
(2012) for their spectra with a resolution $R\gtrsim 60\,000$.}
\label{fig:mike_ew_bias}
\end{figure}

The DAOSPEC code fits saturated Gaussians to the line profiles and succeeds well for
faint lines, but its EW estimate is biased for strong lines \citep{KC12}. The limiting
EW above which the DAOSPEC determination becomes significantly
biased depends on spectral resolution: the higher the resolution, the sooner a bias
appears with increasing EW. To evaluate the bias, we have determined manually the EWs
of a set of well isolated lines in the star \#13, using both direct integration and
Gaussian fit. Both methods give similar results, and due to the modest spectral
resolution used here, the bias appears only for $EW > 160$\,m\AA, as shown in
Fig.~\ref{fig:mike_ew_bias}. Therefore, we decided to keep only lines with $EW\leq 160$\,m\AA\
for our analysis.

The DAOSPEC code provides formal errors on EWs, that do not include systematic
effects like uncertain continuum level. To make them more realistic, we arbitrarily
add a $5$\% error in a quadratic way to the error computed by DAOSPEC. This avoids
assigning too contrasted weights to some lines when computing mean abundances. 

\subsection{UVES}
\begin{figure}
\centering
\includegraphics[width=\hsize]{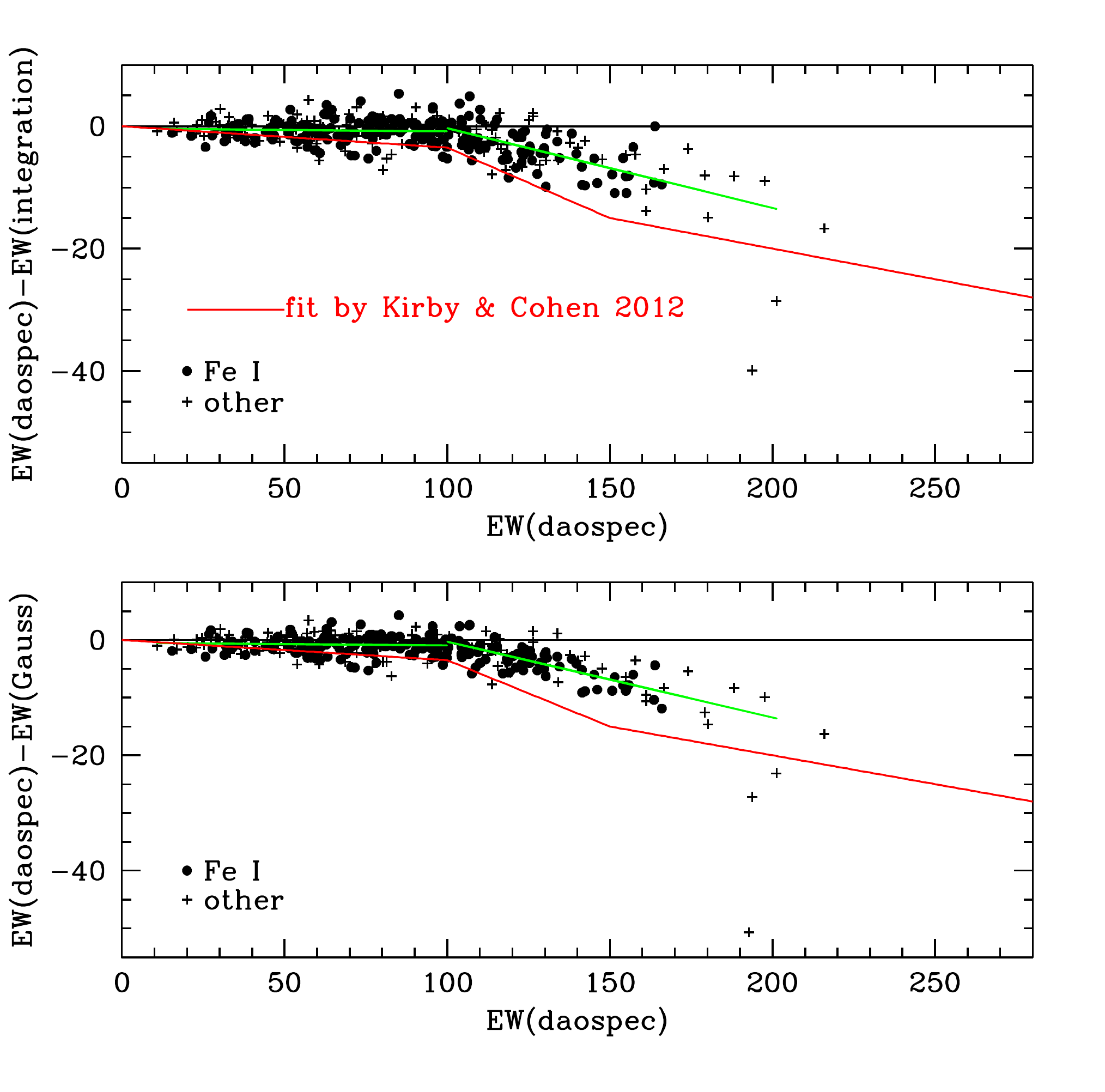}
\caption{Comparison of the EWs determined by DAOSPEC with those obtained by manual
  direct integration (top panel) and Gaussian fits (bottom panel) for clean lines,
for all ten stars of the UVES sample.
The red curve shows the bias found by Kirby \& Cohen (2012) for their own spectra
($R\gtrsim 60\,000$). The green lines show the linear regressions below and above
100\,m\AA.}
\label{fig:uves_ew_bias}
\end{figure}

We applied the same method as for MIKE to determine the continuum and EWs.
However, because of the higher resolving power of UVES, we had to correct all EWs for
the bias that we determined in the same way as above, but here for all stars of
our UVES sample. The difference between EWs
determined with DAOSPEC and manually is shown on Fig.~\ref{fig:uves_ew_bias}
We still exclude very strong lines ($EW > 210$\,m\AA) and adopt the following
correction to the EWs given by 4DAO:
\begin{equation}
EW = 1.0044\cdot EW(4DAO)+0.5 ~~\mathrm{for}~~ EW < 100\,\mathrm{m\AA,}  
\end{equation}
\begin{equation}
  EW = 1.11\cdot EW(4DAO)-10.06 \mathrm{for}~ 100 < EW < 210\,\mathrm{m\AA.}
\end{equation}  
%\begin{eqnarray}
%EW &=& 1.0044\cdot EW(4DAO)+0.5 ~~\mathrm{for}~~ EW < 100\,\mathrm{m\AA,}\\
%EW &=& 1.11\cdot EW(4DAO)-10.06 \\
%   & & \mathrm{for}~ 100 < EW < 210\,\mathrm{m\AA.}
%\label{eq:ew_corr}
%\end{eqnarray}
This function is continuous at $EW=100$\,m\AA. The correction being small, we
keep the same error on EW as provided by DAOSPEC, after adding quadratically a
$5$\% error to them, as for the MIKE sample.

The RVs were determined in the same way as for MIKE, and the values
given in Table~\ref{table:atmo} are averaged over three determinations
corresponding to the three UVES spectral ranges. The table lists also the velocities
from the individual exposures, since some were observed as much as a whole year apart.
However, we did not find any sign of evident RV variability in any of the ten UVES stars.

\subsection{HRS}
We have used the same method as for MIKE and UVES to determine the continuum and EWs.
The resolution is lower than that of MIKE spectra, and we verified that the EWs given
by the 4DAO code do not need any correction.
The S/N is lower on average for this sample than for the MIKE and
UVES ones, so we used the UVES data rather than the HRS ones for the three stars
that were observed with both instruments. For the sake of completeness and for
comparison purposes, we have nevertheless determined independently the stellar parameters
and abundances of these three objects (designated \#21 UVES, \#26 UVES, and \#28 UVES in
the tables) on the basis of HRS spectra alone. We note that the spectrum of \#28 UVES
has a very poor S/N, so the corresponding results must be taken with caution.

The EWs are provided in Tables \ref{table:lines_a}, \ref{table:lines_b}, \ref{table:lines_c},
\ref{table:lines_d}, \ref{table:lines_e}, \ref{table:lines_f}, and \ref{table:lines_g}.

Radial velocities were determined from cross-correlation using the IRAF
task fxcor using the Arcturus spectra \citep{Hinkle2000}.  
The heliocentric correction was made using the IRAF task rvcorrect.
A correction for zero points was made based on the radial velocity standard 
taken in twilight.   The final heliocentric velocities are given in Table~\ref{table:atmo};
also in this case we list radial velocities from the individual exposures, due to the
different dates of observation in some cases. In this sample, star \#03 is found to be a RV variable.
The spectra were then shifted and combined using the IRAF tasks dopcor and scombine.

\section{Analysis} \label{sec:analysis}

\begin{figure}
\centering
\includegraphics[width=0.5\textwidth]{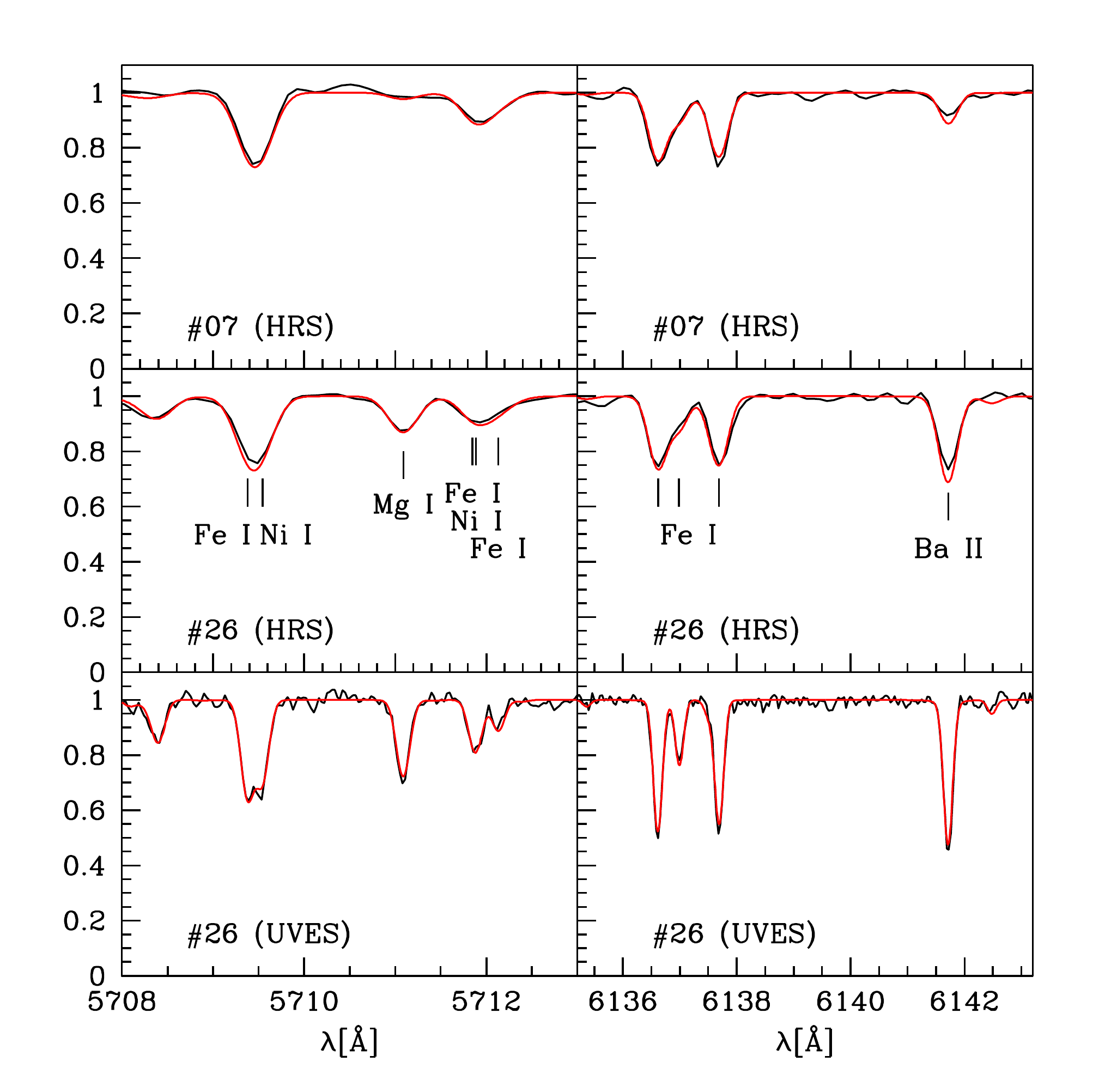}
\caption{Examples of observed spectra and best-fitting models over a wavelength region encompassing
  absorption lines from different elements (see labels). The first row refers to star \#7, observed with
HET/HRS. The second and last rows show star~\#26, observed with HET/HRS and VLT/UVES, respectively.}
\label{fig:spectra}
\end{figure}

\subsection{Models}
We used the MARCS 1-D spherical atmosphere models with standard abundances,
downloaded from the MARCS web site\footnote{marcs.astro.uu.se}
\citep{GEE08}, and interpolated using Thomas Masseron's \verb+interpol_modeles+
code available on the same site. ``Standard composition'' means that the
abundances are scaled solar, except for the $\alpha$ elements that are 
overabundant relative to solar by $0.4$~dex for [Fe/H]$\leq -1.0$, by $0.3$~dex
for [Fe/H]$=-0.75$, by $0.2$~dex for [Fe/H]$=-0.5$ and by $0.1$~dex for [Fe/H]$=-0.25$.
We adopted those computed for a microturbulence velocity $v_\mathrm{turb}= 2$~\kms\ and for
$1$ solar mass.

\subsection{Codes}
All abundances were computed with the \verb+turbospectrum+ code \citep{AP98,P12}.
It assumes local thermodynamic equilibrium (LTE) but is able to compute the line transfer
in spherical geometry, and  includes continuum scattering in the source function.
We used this code in the ABFIND mode to get abundances from the EWs.

In a few cases
we also used \verb+turbospectrum+ to compute synthetic spectra and determine abundances
through a fit to the observed spectrum, especially for blended zirconium lines.
For this we developed a python routine \verb+fitspec.py+ which generates synthetic
spectra with \verb+turbospectrum+ for various abundances of the element of interest,
and convolves them with Gaussians with various FWHMs, and finds
the optimal abundance, FWHM and $\lambda$ shift in the relevant short spectral range.
The convolution is done with the \verb+faltbo3+ utility code provided with
\verb+turbospectrum+. This code was also used to determine the carbon abundance, based on part of the
CH molecular G band, in the $4322-4325$\,\AA\ spectral range. As in \citet{JNM15},
we assumed a solar [N/Fe] ratio, a carbon isotopic ratio $^{12}$C/$^{12}$C$=6$
typical of the tip of the RGB, and we adopted the [Mg/Fe] ratio as a proxy for
[O/Fe] when the latter was not measured. To compute hyperfine structure (HFS) corrections of some lines, we used Chris Sneden's
MOOG code\footnote{http://www.as.utexas.edu/$\sim$chris/moog.html} (2010 version) with the
\verb+blend+ driver. More details are given below.
Examples of observed spectra, with overlaid the best-fitting
models, are shown in Fig.~\ref{fig:spectra}.

\subsection{Line list and solar abundances}
We adopted the line list of \citet{RFS08,RST10}, and complemented it with
data from the VALD database \citep{PKR95,RPK97,KPR99,KRP00,RPK15}. The line wavelengths
and oscillator strengths are given in Table~\ref{table:lines_a}, \ref{table:lines_b},
\ref{table:lines_c}, \ref{table:lines_d}, \ref{table:lines_e}, \ref{table:lines_f}, and
\ref{table:lines_g} .
The adopted solar abundances are taken from \citet{AG89} and \citet{GS98} and displayed
in Table~\ref{tab:abundances}.

\subsection{Preliminary stellar parameters}\label{sec:prel}
In order to derive the stars' atmospheric parameters, we feed the spectroscopic analysis with
first guesses for the effective temperature \teff, gravity \logg, microturbulent 
velocity $v_\mathrm{turb}$ and [Fe/H].
As initial [Fe/H] values we adopt those given in the original works from which the targets were selected
(that have low
spectral resolution, except for the star drawn from APOGEE);
\teff and \logg\, are instead derived photometrically, and $v_\mathrm{turb}$ follows from
the empirical relation $v_\mathrm{turb} = 2.0 - 0.2 \times log g$ by \citet{Anthony-Twarog2013}. 

The majority of the stars have SDSS ugriz and JHK photometry from the UKIDSS large 
area spectroscopic survey\footnote{The only exception is star \#21 that lacks  
  a measurement of the J-magnitude in the UKIDSS survey and for which we used the 2MASS value.}, which allows
us to constrain the star' photometric effective temperature $T_\mathrm{eff, p}$ by using 
several of the \citet{Ramirez2005} effective temperature-color-[Fe/H] 
calibrations $T_\mathrm{eff, p, col}$ 
(specifically T$_\mathrm{VI}$, T$_\mathrm{VJ}$, T$_\mathrm{VH}$ and T$_\mathrm{VK}$); on the other hand,
for stars \#20, 22, 23, 24, 25, 27, we have used only $T_\mathrm{eff, p, VI}$, due to the availability of only Washington photometry,
transformed into \mv \& \vi\ using the relation in Morrison et al. \citeyear{Morrison2003}.  

We transformed the SDSS photometry into Johnson-Cousins \mb, \mv\ and \mi\ 
using the transformations involving $gri$ in Lupton (2005)\footnote{\tiny{http://www.sdss3.org/dr8/algorithms/sdssUBVRITransform.php\#Lupton2005}}. The  
\bv, \vi, \vj, \vh, \vk\ colors are dereddened using the E(B-V) reddening from the \citet{Schlegel1998} maps,
with the \citet{Schlafly2011} recalibration. 

$T_\mathrm{eff, p}$ is determined as the weighted average of the individual $T_\mathrm{eff, p, col}$ relations with the error given by 
the scatter between $T_\mathrm{eff, p, col}$  
to which we add in quadrature the average error in T$_\mathrm{eff, p, col}$.\footnote{The error in each of the 
T$_\mathrm{eff, p, col}$ is derived by propagating the error in [Fe/H] (assumed to be a conservative 0.2dex) 
and in color, where the latter includes the scatter in the transformations between 
SDSS and Johnson-Cousin bands. To this we add in quadrature the scatter found by Ramirez \& Melendez 
(2005) around their \teff-[Fe/H]-color relations (40, 38, 32, 28K for T$_\mathrm{VI}$, T$_\mathrm{VJ}$, T$_\mathrm{VH}$, T$_\mathrm{VK}$ respectively). 
Finally, other 50K are added in quadrature to the average T$_\mathrm{eff,p, col}$ to account for the 
scatter between direct temperatures and those from the infrared flux method.}
The error in $T_\mathrm{eff, p}$ typically ranges between 60 \& 80K. We exclude T$_\mathrm{BV}$ as it is the most sensitive to uncertainties in [Fe/H]. 

The first guess logg is obtained by finding the point along the RGB locus (logg $\le$ 3.5) with the closest matching \teff 
and [Fe/H] in a set of Dartmouth isochrones \citep{Dotter2008} of different ages, 
[Fe/H] and [$\alpha$/Fe] (age= 4, 8, 12 Gyr; -2.4 $<$ [Fe/H] $<$ -0.60, with a spacing of 0.2dex; 
[$\alpha$/Fe]=0, 0.2, 0.4); we explore different values for the ages and [$\alpha$/Fe] because 
these quantities are in practice unknown (at this stage) and we wish not to fix them a priori.
We derive the error in logg$_{\rm p}$ 
by repeating N=100 times the search for the best-matching isochrone, where each time the effective temperature 
and [Fe/H] are randomly drawn from a Gaussian distribution centered on the input
$T_\mathrm{eff, p}$  and [Fe/H] and with $\sigma$ given by the corresponding errors.

\subsection{Adopted spectroscopic stellar parameters}
We started from the photometrically estimated effective temperatures, surface
gravities and microturbulent velocities as described in the previous section,
and applied the usual spectroscopic diagnostics
by plotting the \ion{Fe}{i} abundance as a function of excitation potential to
constrain \teff\ and of EW to constrain \vmic. Following \cite{tafel2010} and
\cite{JNM15}, we discarded the Fe lines with $\chi_\mathrm{exc}< 1.4$~eV in order
to minimize NLTE effects as much as possible, as well as lines fainter than
$\sim20$\,m\AA\ and stronger than $\sim 200$\,m\AA\ ($160$\,m\AA\ for the MIKE
sample). In the \ion{Fe}{i} abundance vs. $EW$ diagram, we used the {\sl predicted}
rather than observed EWs to avoid any bias on the \vmic\ determination, following
\cite{M84}.

The surface gravity was determined as usual from the ionization balance, by
requiring the equality of the \ion{Fe}{i} and \ion{Fe}{ii} abundances.

When a change in \teff\ and \logg\ was required by the relevant spectroscopic diagnostic,
we tried to remain within $2\sigma$ of the photometric estimate, allowing
the slopes of the diagnostic plots to differ from  zero by no more than $2\sigma$
either. In only three cases we had to lower \teff\ by as much as $\sim2.5\,\sigma$
in order to fulfill the spectroscopic diagnostic (\#7, 9, 19).
Regarding \logg, it was possible to maintain it within $2\sigma$
of the photometric estimate in all case but one (\#6).

This trade-off between photometric and spectroscopic criteria often implies a spectroscopic
\teff\ slightly lower than the photometric one (and/or a slightly negative slope
in the \ion{Fe}{i} versus excitation potential plot) and a larger \ion{Fe}{ii}
abundance relative to the \ion{Fe}{i} one. The latter difference $\Delta_{II-I}$
does not exceed
$2\sigma$ in most cases ($\sigma$ being defined as the rms dispersion of the
individual \ion{Fe}{ii} abundances divided by the square root of the number of lines),
especially in the MIKE sample (where the only exception is star \#9: $3\sigma$)
and in the HRS one (except for star \#7: $2.3\sigma$). In the UVES sample,
which benefits from the best resolution
and S/N, half the stars show differences larger than $2\sigma$ (between $2.2$ and
$4.3\sigma$). This indicates a higher visibility of systematics like NLTE effects
in this high quality data.

We chose to be more tolerant regarding ionization
equilibrium in the case of very metal poor stars ([Fe/H]$\lesssim -2$), because
there are indications in the literature that NLTE effects are more pronounced
for them than at solar metallicity. For instance, \citet{MFT12} compute the NLTE
correction for both \ion{Fe}{i} and \ion{Fe}{ii} abundances for two models with
\teff$=4500$\,K, \logg$=1.0$, [Fe/H]$=-1.50$ and [Fe/H]$=-2.00$; they find
$\Delta_{II-I}=0.13$ for [Fe/H]$=-1.50$ and $\Delta_{II-I}=0.15$ for [Fe/H]$=-2.00$,
when the stars are analyzed assuming LTE. This is consistent with our empirical
$-0.02 < \Delta_{II-I} < 0.20$ range for UVES. $\Delta_{II-I}=0.24$ in two HRS
stars, but the S/N of their spectra is poor and this represents no more than $2\sigma$.

Disregarding the photometric \logg\ estimate and strictly forcing ionization
equilibrium would imply decreasing \logg\, but also \teff\ because these quantities
are correlated. We preferred to remain not too far from the photometric estimates,
which are physically sound.

The final stellar parameters are given in
Table~\ref{table:atmo}.  The typical error on \teff\ is about $100$~K, that on
\logg\ about $0.2$~dex and that on \vmic\ about $0.2$~\kms.

\subsection{Hyperfine structure}
The hyperfine structure (HFS) mainly affects elements with an odd atomic mass,
namely Sc, Mn, Co, Cu, and Eu, but also odd isotopes of Ba. It broadens the line,
thereby alleviating its saturation, so that estimating the abundance directly from
the EW results in an overestimate, if one does not take into account the HFS structure.
For a given EW, the HFS correction increases in absolute value as \vmic\ decreases,
because both tend to desaturate the line. Faint lines remain unaffected, being far
from saturation, but for strong lines on the plateau of the curve of growth, the
HFS correction may exceed $1.5$~dex in absolute value (the HFS correction is always
negative).

As mentioned before, we estimated HFS corrections to the raw EW abundances using the
MOOG code, running it with both the \verb+blend+ driver and the \verb+abfind+ driver
and computing  the abundance difference, as described in \citet{NCJ12} and \cite{JNM15}.
We used the HFS components with their oscillator strengths given in the
Kurucz web site\footnote{http://kurucz.harvard.edu/linelists.html} for Co, Cu, most
Ba lines and Eu, and by \citet{PMW00} for Sc and Mn.

For \ion{Ba}{ii} (and especially for the $\lambda 4934$ line which has a
wide HFS), which has both odd and even isotopes, we assumed a solar mix with an
$18$\% fraction of odd isotopes \citep{LPG09}. For Eu, we assumed equal abundances
of the $A=151$ and $153$ isotopes \citep{ZCZ10}.

\subsection{Abundances}

The final abundances given in Table~\ref{tab:abundances} are the weighted mean
of the individual line abundances, the weights being the inverse variances of the
single line abundances.  These variances were propagated by \verb+turbospectrum+
from the errors on the corresponding EWs. We also give in the same table some upper
limits evaluated from the small EW of marginally identified lines, for stars \#10
(O), \#12 (Pr), and \#23 (Mn). Such upper limits concern only abundances given
by one or two lines and are indicated in the relevant tables by a \verb+<+ sign before their value.

The carbon abundances are given in Table~\ref{tab:carbon_ab}. They are given only
for the stars measured with MIKE and with UVES, because the spectral range of HRS
does not include the CH G band. The spectrum of star \#12 is too noisy in the wavelength
range of interest to provide a reliable estimate of the C abundance.

For species represented by only one line, the
errors indicated have to be considered as lower limits, because they include only
the formal error given by the 4DAO code (plus 5\%), but not the uncertainty on the
continuum level, oscillator strength, possible blend with a small telluric line
(especially in the case of oxygen), or faint unrecognized cosmic hit.
For the stars measured with HET/HRS, we did not discard very strong lines, and the
abundance of some elements like Na, Mg, and Ba sometimes rely on lines with $EW > 200$\,m\AA\
and must be taken with caution: the Gaussian fit may not be quite appropriate for such
lines, and abundances depend more critically on broadening parameters that may be uncertain.

\subsubsection{Errors}

As mentioned in Sect.~\ref{section:MIKE}, we added quadratically a 5\%
error to the EW error estimated by the DAOSPEC code to make the errors on EW more realistic;
thus, no EW has an error smaller than 5\%. These errors are generally larger
than those obtained from the \citet{C88} formula revised by \citet{BIT08}. They
are listed in Tables \ref{table:lines_a} to \ref{table:lines_g}.

The errors listed in Table~\ref{tab:abundances} are defined
in the same way as in \citet{JNM15}.

\section{Comparison to previous works} \label{sec:comp}

Both the Spaghetti survey \citep[e.g.,][see also Starkenburg et al. 2010 for additional constraints onto the luminosity classification]{Morrison2000, Dohm-Palmer2001} and the SDSS-SEGUE
survey are carried out at relatively low spectral resolution ($\sim$2.5-3.5\AA\,). Here we compare the
heliocentric velocity and [Fe/H] estimates from these original sources with our determinations at ~$\sim$10x larger
spectral resolving power. Similarly, we revise the stars's distance determinations on the basis of the 
stellar gravity, temperature, [Fe/H] and [$\alpha$/Fe] from our analysis.

\subsection{Heliocentric velocity and [Fe/H]}

\begin{figure*}
\centering
\includegraphics[width=0.8\textwidth]{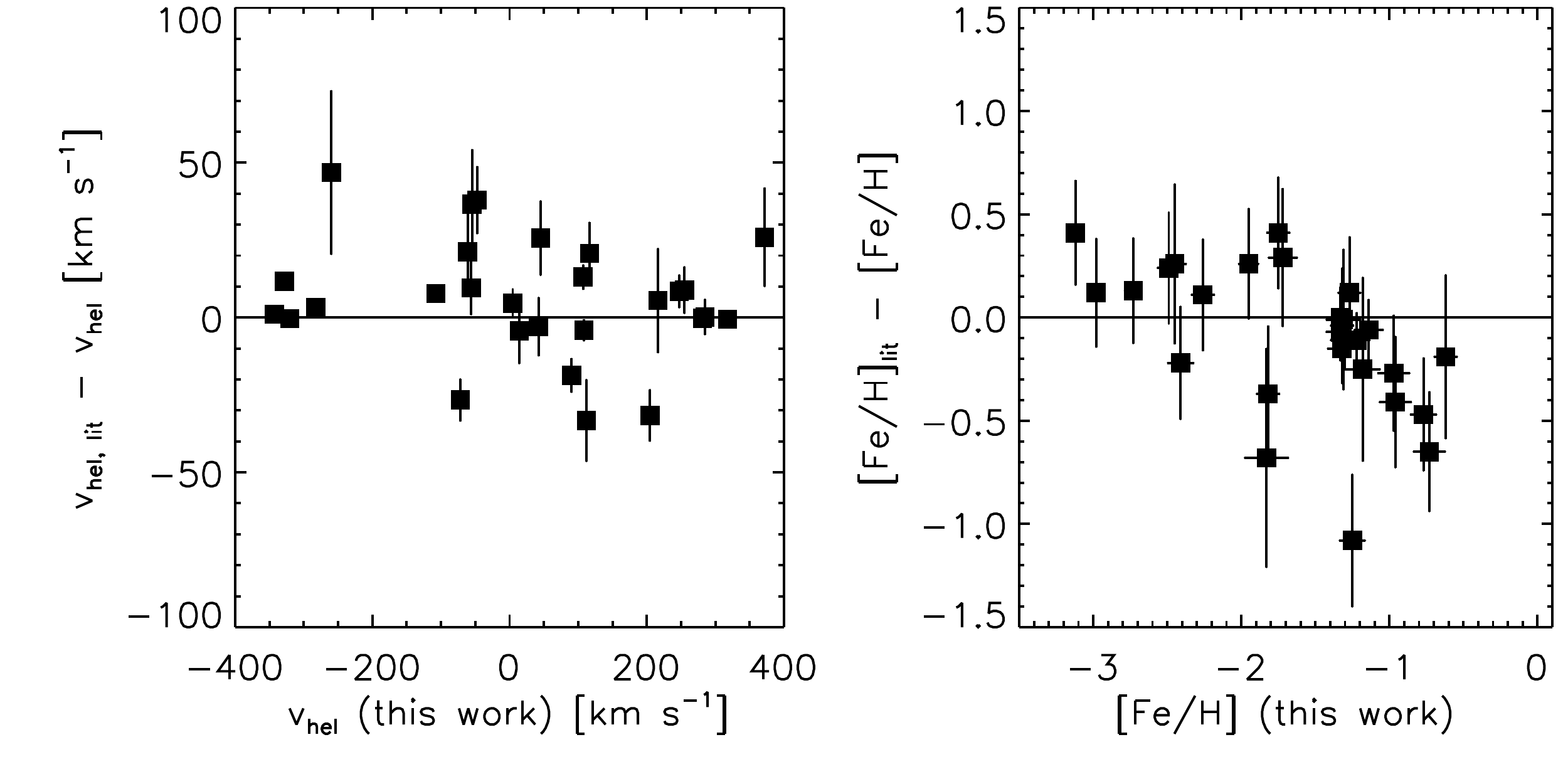}
\caption{Comparison between the heliocentric radial velocities (left) and metallicities
[Fe/H] (right) obtained in this work to those obtained in the original sources.}
\label{fig:comparisonvel}
\end{figure*}

Figure~\ref{fig:comparisonvel} shows the comparison of the heliocentric radial velocities (left) and metallicities
[Fe/H] (right) obtained in this work to those obtained in the original sources. As mentioned before, the sample contains two stars detected as
radial velocity binaries (star \#3 as determined from our HET observations and \#11, also known as 2M12490495-0743456,
with the binarity detected by APOGEE multiple visits), however their velocities happen to agree well with the measurements in the literature;
there are only a few cases for which the
velocities disagree beyond the 3$\sigma$ level, and we cannot exclude that these velocity differences are due to
the stars being unidentified radial velocity binaries. As for the metallicity, the comparison can be deemed satisfactory.
As expected, the high 
resolution observations have more precise [Fe/H] determinations with respect to the original measurements at low spectral resolution. 
There is only one star with clearly discrepant [Fe/H] (\#28): its Spaghetti survey metallicity was derived from the Mg triplet index,
but was found to be very (and unusually) discrepant from the metallicity derived from the Ca~K line index, which returned [Fe/H] $= -1.30$,
much closer to the high-resolution spectrum value (Starkenburg, priv. communication).

Star 2M12490495-0743456 was also observed by APOGEE and its SDSS DR13 abundances are in excellent agreement with our determinations,
except for [Co/Fe] and [Ni/Fe], which anyway agree within 2$\sigma$; the abundance of [Al/Fe] and [V/Fe] is undetermined in our data 
(SDSS DR13 : [Fe/H]$=-0.97 \pm 0.04$;   [O/Fe]$= 0.21\pm      0.06$; [Mg/Fe]$=       0.18\pm      0.05$; [Al/Fe]$= -0.05\pm       0.14$; [Si/Fe]$=       0.17\pm        0.05$;  [Ca/Fe]$=       0.27\pm      0.07$; [V/Fe]$= 0.57 \pm      0.24$; [Mn/Fe]$=      -0.39 \pm     0.05$; [Co/Fe]$= 0.32 \pm       0.23$; [Ni/Fe]$=     -0.01 \pm     0.04$. This work: [Fe/H]$=-0.96 \pm 0.11$;   [O/Fe]$= 0.25 \pm      0.04$; [Mg/Fe]$=       0.24 \pm      0.11$; [Al/Fe]$=$ ND; [Si/Fe]$=       0.12 \pm        0.07$;  [Ca/Fe]$=       0.30 \pm      0.13$; [V/Fe]$=$ ND; [Mn/Fe]$=      -0.49 \pm     0.12$; [Co/Fe]$= -0.23 \pm       0.19$; [Ni/Fe]$=     -0.17 \pm     0.09$).

\subsection{Distances}
In both the Spaghetti survey and \citet{Xue2014} the distance modulus
(and hence, the heliocentric distance $d_h$)
to a star is found by comparing the star's apparent magnitude
with the absolute magnitude of globular clusters giant-branch color-luminosity fiducials,
interpolated at the observed star color and spectroscopic [Fe/H].

In the Spaghetti survey the distance errors are obtained from 
Montecarlo simulations in which the effect of the color and metallicity error are factored in; to the latter 
is added in quadrature 0.25dex to account for possible systematic calibration errors.
On the other hand, \citet{Xue2014} adopt a probabilistic approach to propagate the errors in metallicities, magnitudes,
and colors into distance uncertainties, and be able to fold in also prior information
about the giant-branch luminosity function and
the different metallicity distributions of the SEGUE K-giant targeting sub-categories. 

This type of determination implicitly assumes that the age and [$\alpha$/Fe] of halo stars 
compare well to those of the globular clusters fiducials,
which might not hold for the whole sample: for instance, a few halo stars are
known to have sub-solar [$\alpha$/Fe] and we also want to account for the possibility of the late accretion of stars
originated in satellites with prolonged star formation histories (e.g., Sag).

\begin{figure}
\centering
\includegraphics[width=0.4\textwidth]{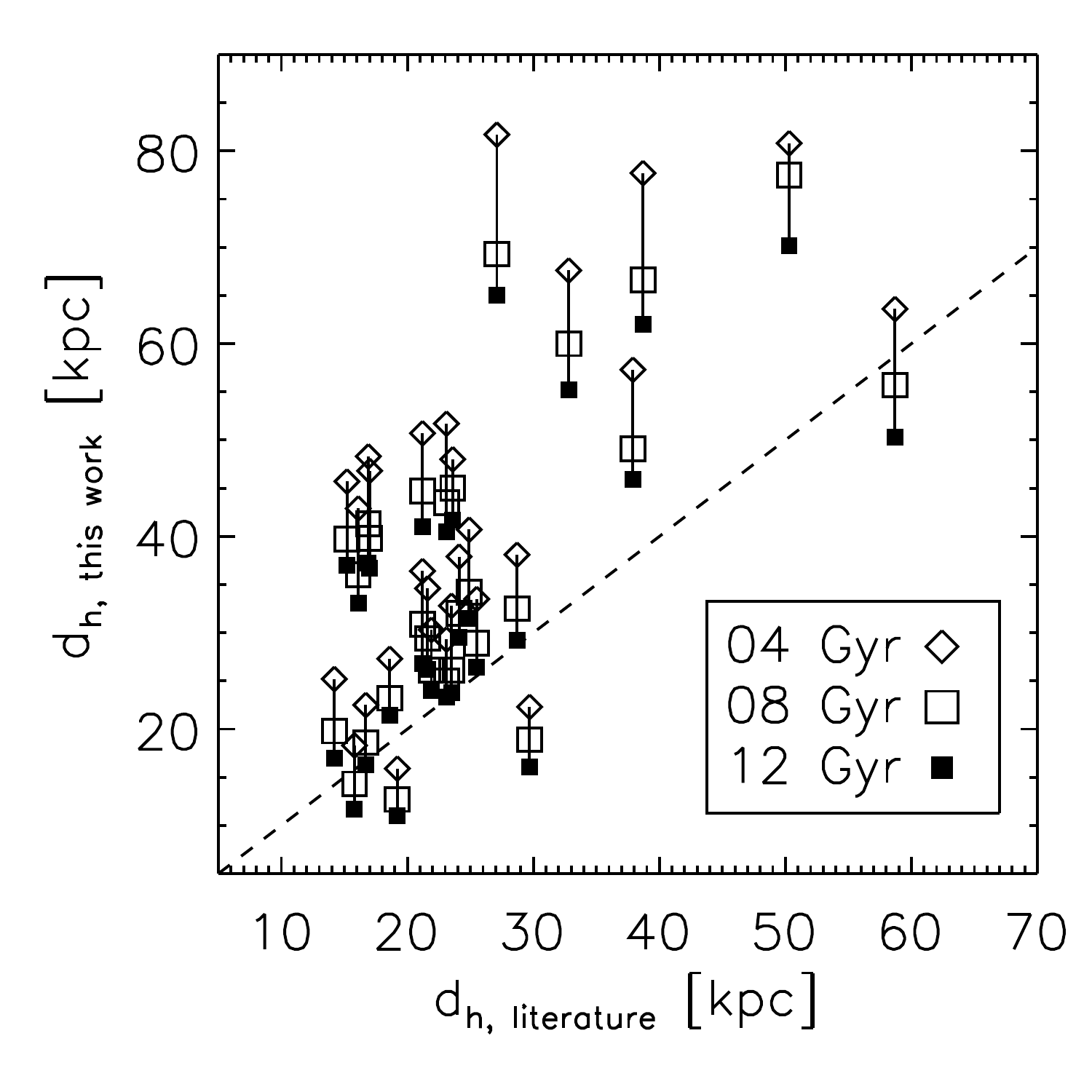}
\caption{Comparison between the heliocentric distances derived in the literature and in this work; the different
symbols indicate the different ages assumed in the isochrone fitting (see legend).}
\label{fig:comparisondh}
\end{figure}

Hence we relax the above assumption and revise the stars' distance determinations by applying the same method as in
Sect.~\ref{sec:prel} but
using the stars's \logg, \teff, [Fe/H] and [$\alpha$/Fe] (= ([Ca/Fe] + [Mg/Fe] + [Ti/Fe])/3)
derived in our spectroscopic analysis and repeating the fit for different ages (4, 8 and 12 Gyr); the 
range of values for the isochrones [$\alpha$/Fe] grid goes now from -0.2 to +0.6, with steps of 0.2dex. 
Figure~\ref{fig:comparisondh} shows that
our revised distances are typically larger than those in the literature, placing  
the great majority of the objects in the sample beyond the
nominal separation between inner and outer halo.
Assuming an age of 4 Gyr rather than 12 Gyr results in a 20-30\% distance increase for most of the stars.

The determination of the distance errors is non-trivial: at face value, the 68\% confidence level obtained by
repeating the fit with hundreds 
MonteCarlo realizations of the stars's \logg, \teff, [Fe/H] and [$\alpha$/Fe]
drawn from Gaussian distributions centered around the spectroscopically determined values would yield relative errors of
30-40\%. However, this is likely to be an overestimate, because in practice several combinations of the 
randomly drawn \logg, \teff, [Fe/H] and [$\alpha$/Fe] would be rejected by the spectroscopic analysis because not
compatible with the spectroscopic diagnostics.
Since in the following the distance estimates are only used as additional, possible indicators of the stars's belonging
to known substructures, we will consider as error the range of values due to the different ages assumed.

\begin{figure}
\centering
\includegraphics[width=\hsize]{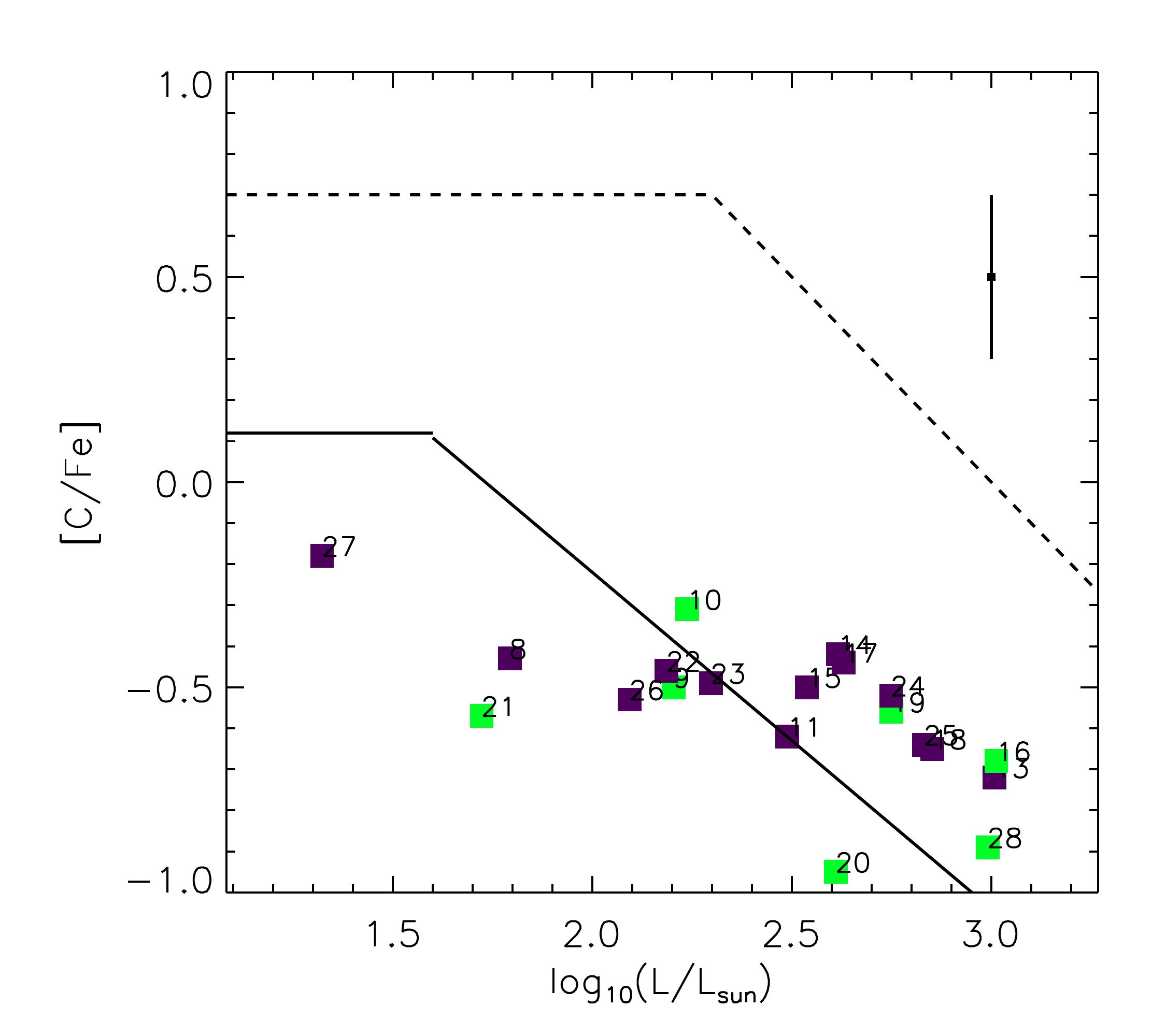}
\caption{[C/Fe] as a function of luminosity for our sample (squares), split in stars
  that have spatial and kinematic properties compatible with membership to the Sag stream (green) and not
  compatible (purple), as described in Sect.~\ref{sec:substructures} (see also Fig.~\ref{fig:sag}). The dashed line
  indicates the \citet{Aoki2007} criterion for identifying carbon-enhanced metal-poor stars when taking into account
  evolutionary effects after the first dredge up; the solid line shows the trend of declining [C/Fe] with increasing stellar luminosity
  in three MW globular clusters as in \citep[see][]{Kirby2015}, which also agrees with the trends observed in that same work for
  the MW classical dSphs Ursa Minor, Draco, Sculptor and Fornax. In the corner we give a representative error-bar.}
\label{fig:cfe}
\end{figure}

\begin{figure*}
\centering
\includegraphics[width=\hsize]{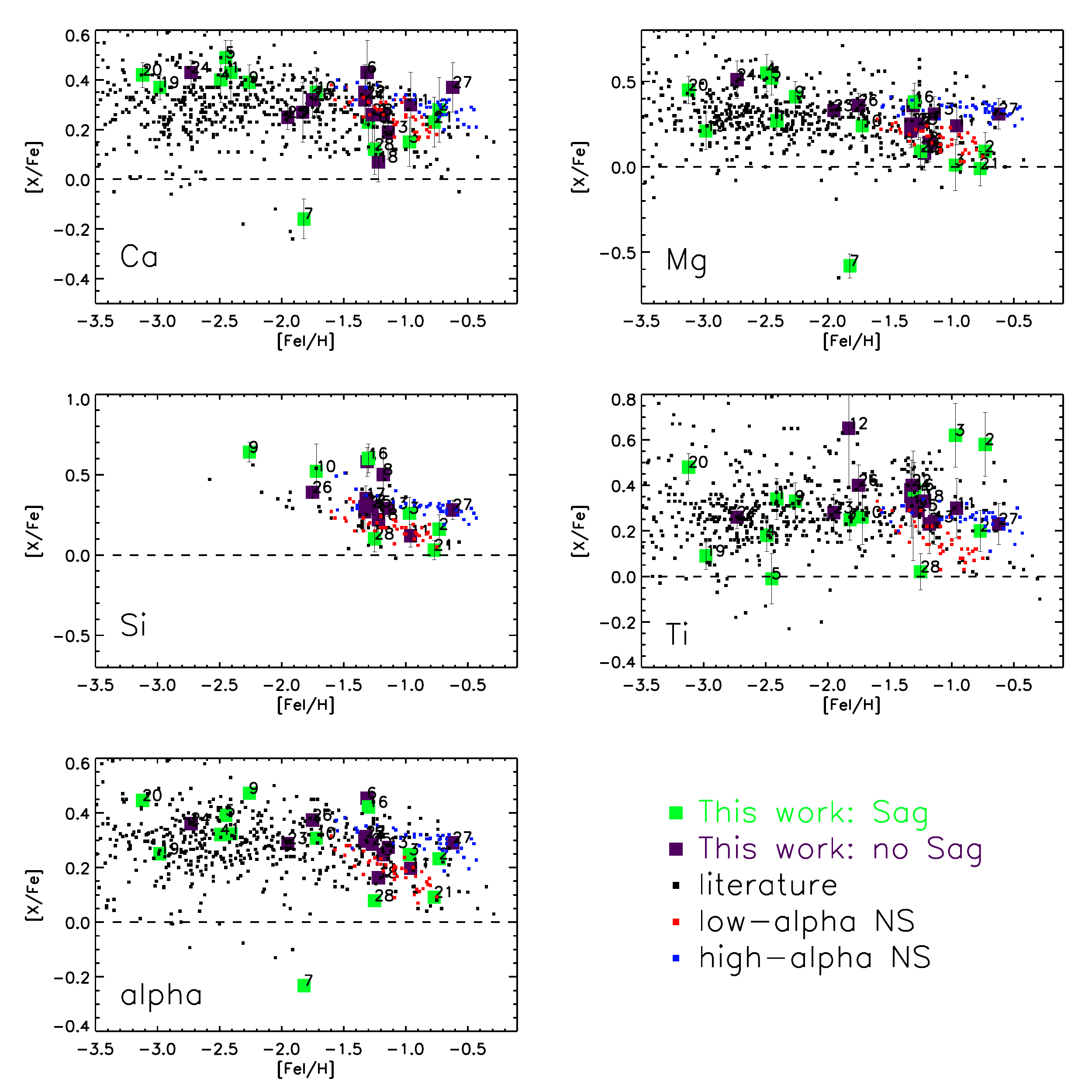}
\caption{Abundance of the $\alpha$ elements Ca, Mg, Si, \ion{Ti}{ii} and combined $\alpha$ relative
  to iron (\ion{Fe}{I}) as a function of [\ion{Fe}{i}/H]. The sample is split in stars
  that have spatial and kinematic properties compatible with membership to the Sag stream (green) and not
  compatible (purple). The small squares indicate the chemical abundances for literature samples of
  MW halo stars (black ones: Venn et al. (2004), Barklem et al. (2005), Ishigaki et al. (2012, 2013);
  red and blue: ``low-$alpha$'' and ``high-$alpha$'' populations
  as identified in Nissen \& Schuster 2010, 2011). We warn the reader that the global [$\alpha$/Fe] shown
  here and in Figs.~\ref{fig:alpha_dwarfs}, \ref{fig:baeu} has been calculated as the average of the
  abundance ratio of the individual $\alpha$- elements to Fe that were available in each of the catalogs.}
\label{fig:alpha_MW}
\end{figure*}

\begin{figure*}
\centering
\includegraphics[width=\hsize]{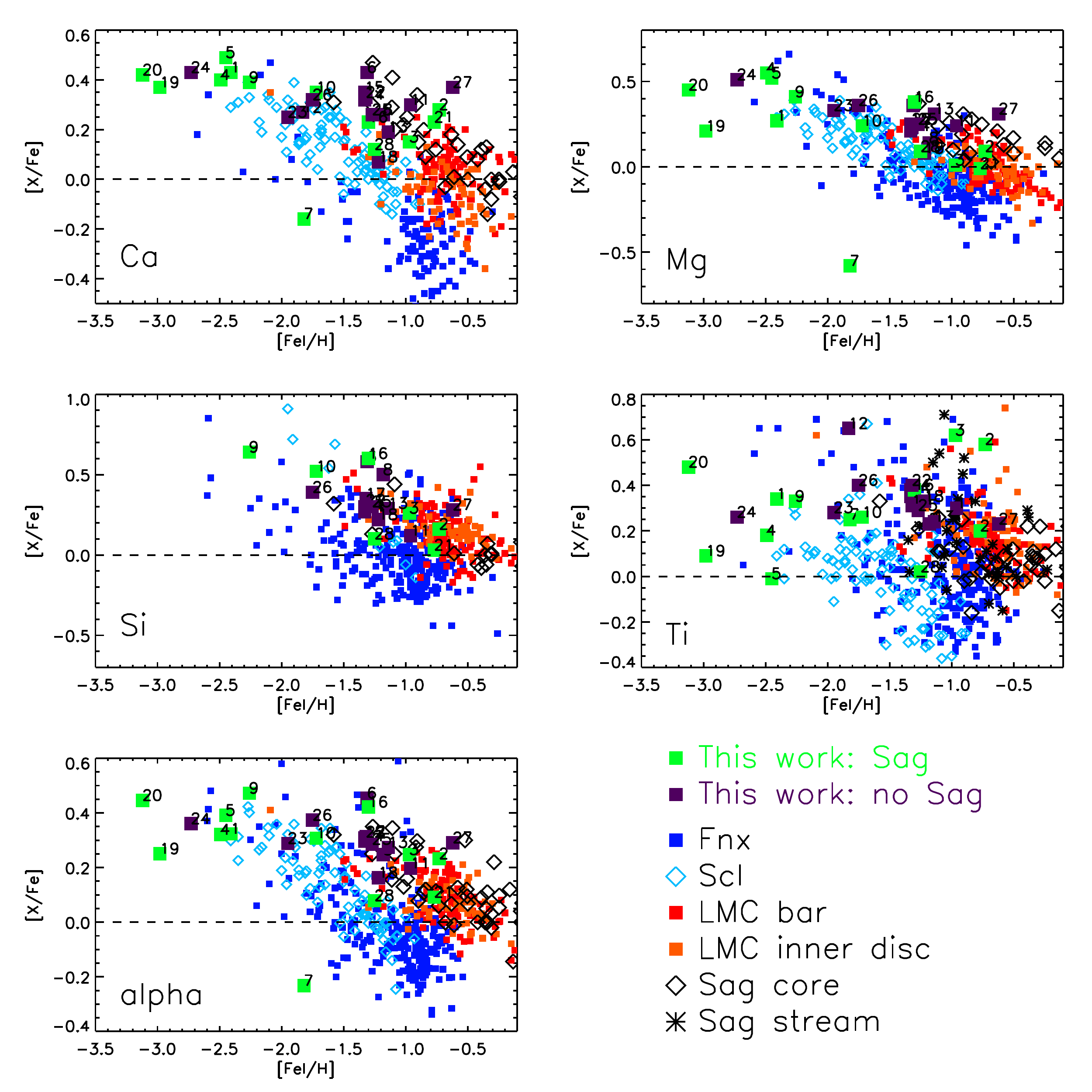}
\caption{Abundance of the $\alpha$ elements Ca, Mg, Si, \ion{Ti}{ii} and combined $\alpha$ relative
  to iron (\ion{Fe}{I}) as a function of [\ion{Fe}{i}/H]. The sample is split into stars that 
  have spatial and kinematic properties compatible with membership to the Sag stream (green) and
  those unlikely to be 
  compatible (purple). The other symbols indicate the chemical abundances for samples of
  stars in MW satellites, as described in the legend (Fnx = Fornax, Scl = Sculptor).}
\label{fig:alpha_dwarfs}
\end{figure*}

\begin{figure*}[!h]
\centering
\includegraphics[width=\hsize]{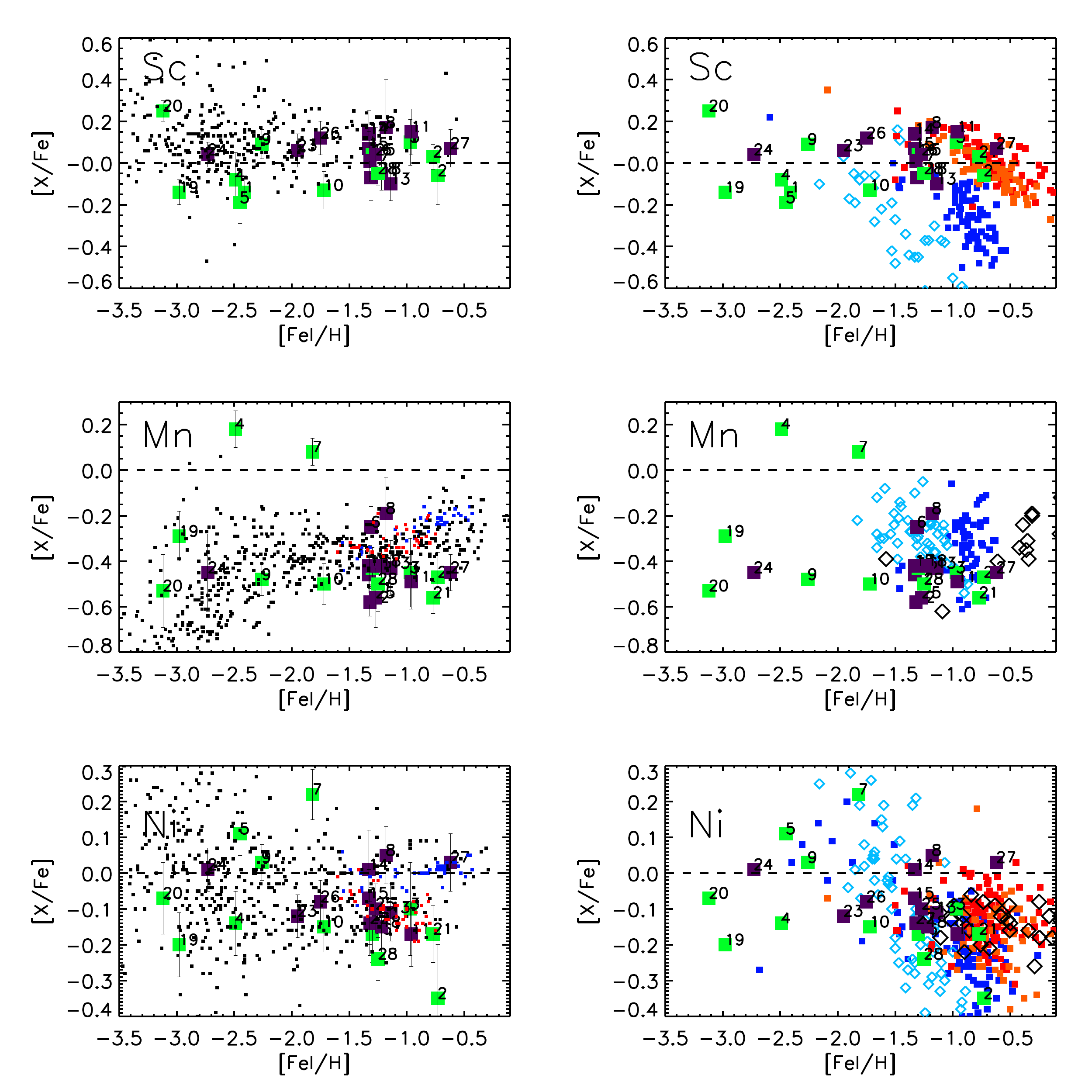}
\caption{As previous figure but for iron-peak elements Sc, Mn, Ni relative
  to iron (\ion{Fe}{I}) as a function of [\ion{Fe}{i}/H] (left: compared to MW halo samples; right: compared to
  samples of RGB stars in MW dwarf galaxies). The [Mn/Fe] abundances for MW dwarf galaxies are from \citet{NCJ12}.}
\label{fig:ironp}
\end{figure*}

\begin{figure*}
\centering
\includegraphics[width=\hsize]{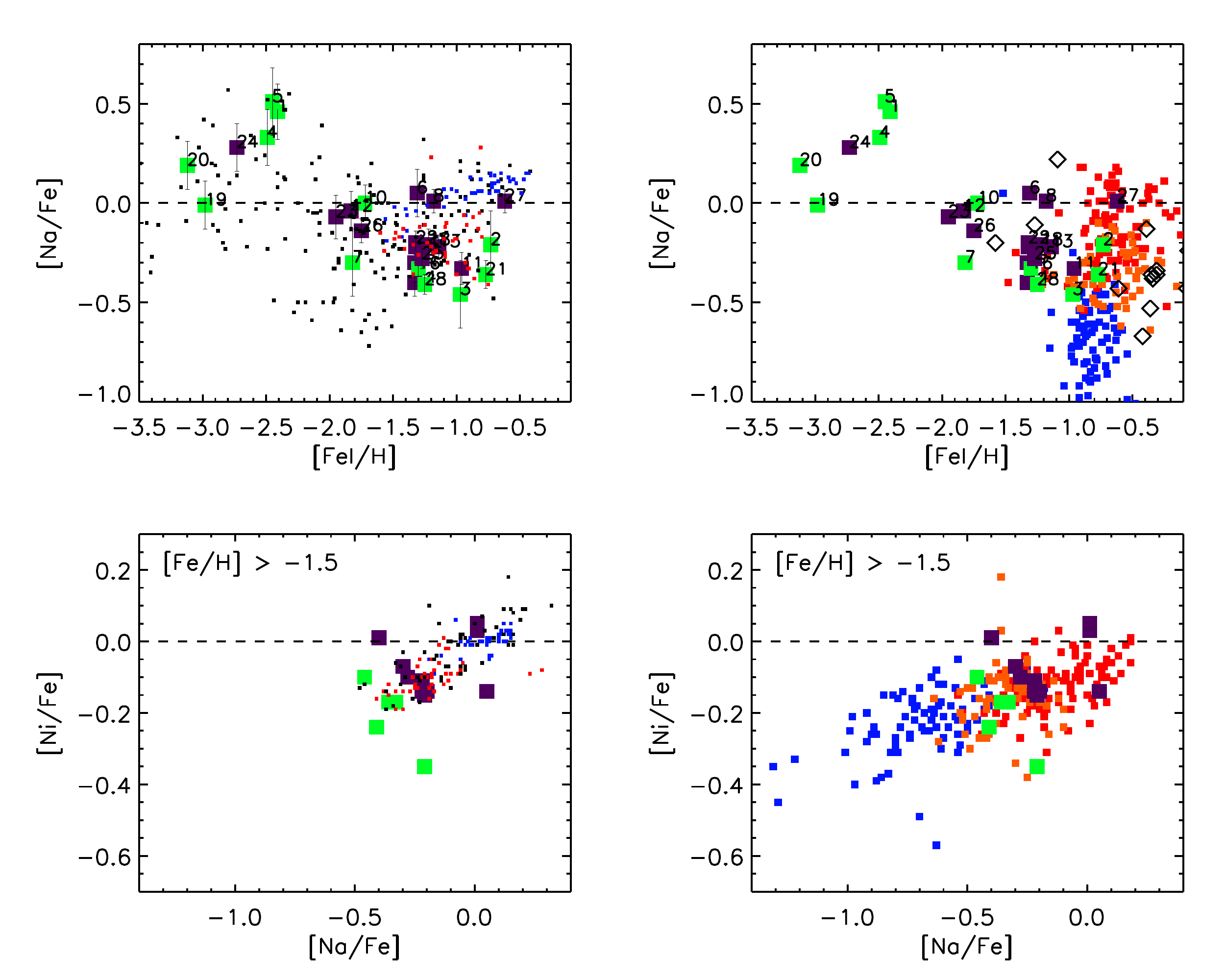}
\caption{Top: Abundance of Na relative
  to iron (\ion{Fe}{I}) as a function of [\ion{Fe}{i}/H]. Bottom: [Ni/Fe] vs [Na/Fe] for stars more metal-rich than [Fe/H]=-1.5.
Left: compared to MW halo samples; right: compared to
  samples of RGB stars in MW dwarf galaxies.}
\label{fig:nani}
\end{figure*}

\section{Chemical trends} \label{sec:trends}
In this section we compare the chemical abundances of our sample of 
distant MW halo stars to those of MW halo stars from solar neighborhood samples
\citep{venn04, barklem05, nissenschuster10, nissenschuster11, Ishigaki2012, ishigaki13} and a set of 
MW satellite galaxies, that is Sculptor \citep{Tolstoy2009},
Fornax \citep{letarte2010, hendricks14, lemasle14},
LMC (inner disk and bar: Van der Swaelmen et al. 2013), the
core of Sag \citep{mcwilliam05,carretta10} and a sample of Sag stream stars by \cite{chou10}. For the MW comparison
samples, given that our purpose is to compare
chemical trends, we specifically focus on studies dealing with halo samples (or studies that allow to select halo stars by providing a membership
probability to the various Galactic components) covering a large range in [Fe/H] and containing elemental abundances for a large
number of elements and stars. 

It should be kept in mind that solar neighborhood samples of (kinematically selected) halo stars
typically contain a mix of stars belonging to the inner- and outer-halo population,
in different proportions, whose orbits are such to place them at
present day in the vicinity of the Sun (for the outer-halo population this implies highly elongated orbits),
with the inner-halo population dominating at [Fe/H]$> -2$ \citep{Carollo2007, An2013, An2015}.
For example, \cite{schuster12} computed the 
orbital properties of the ``high'' and ``low''-$\alpha$ halo stars detected 
in \cite{nissenschuster10, nissenschuster11} and showed the former to be essentially confined 
within the inner-halo region (the orbits reach at most a maximum Galactocentric
distance of $\sim$15 kpc and a maximum height from the plane of 6-8 kpc),
while the latter reaches out to the outer-halo region (spanning a range of
maximum Galactocentric
distances from $\sim$5kpc to 30-40 kpc and a maximum height from the plane of $\lesssim$18 kpc). 
In the case of \citet{ishigaki13} the halo sample is estimated to have
maximum apocentric distances within 30-40kpc,
with the majority belonging to the inner-halo region. 

Although the large errors on the proper motion measurements 
of our sample of distant halo stars, which are typically
as large as the proper motions themselves in UCAC5,
prevents us from reconstructing
their orbital properties, the stars in our sample have large present-day Galactocentric distances
(from 12kpc to 73kpc, with a median value of 32kpc). Since the present-day distance is smaller than,
or can at most be equal to, the apocentric distance, this implies that
our sample probes on average much farther out in the 
MW outer-halo region with respect to our comparison MW halo samples. 

In this section, we display only the figures most relevant to highlight the main results,
while the abundance ratios for the full set of elements derived in this work are shown in the Appendix. We
will dedicate a separate subsection to discussing star~\#7, since its chemical properties appear distinct
from the bulk of the sample in several elements.

$\bullet$ {\bf Carbon} Figure~\ref{fig:cfe} shows that our measurements compare well with the
decreasing trend of [C/Fe] as a function of increasing luminosity observed by \citet{Kirby2015} in four MW dwarf spheroidals and in three
Galactic globular clusters. Clearly none of the stars in our sample is enhanced in [C/Fe], when comparing to the criterion of \citet{Aoki2007}
which takes into account evolutionary effects after the first dredge up. If we were to adopt the corrections by \citet{Placco2014}
as a function of stellar gravity and metallicity (see their Fig.~9), the [C/Fe] of our sample would increase to approximately solar,
overlapping the distribution of MW halo stars by \citet{Lee2013} for the same metallicity but landing on its lower envelope.
Probing out to R$\sim$20kpc and $\mid z \mid \, \sim$15kpc, 
\citet{Lee2017} used SDSS spectra to map the carbonicity of MW stars and found it to increase
when moving away from the plane of the Galaxy,
with the median [C/Fe] being $\gtrsim$+0.3-0.4 beyond $r_g \sim $20 kpc. The increase in carbonicity is accompanied by a decrease in the median
[Fe/H], with the Lee et al. (2017) sample being dominated by stars more metal-poor than [Fe/H]$=-1.7$ at those distances.
In this work we selected most of our stars from the Spaghetti survey, which targetted stars with relative featureless spectra in the
Washington C and M photometric bands. Since these bands include the G-band Carbon feature, we do not claim to have an unbiased sample in [C/Fe].
Additionally, since more than half of our
sample has [Fe/H]$> -1.7$ and the fraction of carbon-enhanced stars is known to increase with decreasing metallicity, these have a small chance of being carbon-rich anyway. When not distinguishing between CEMP-no and -s,-r/s and adopting a criterion of [C/Fe]$> + 0.7$,
the fraction of CEMP stars 
at [Fe/H]$<-2$ was found to be 13\% by \citet{Lee2013} and 30\% by \citet{Placco2014}.
In that regime our sample contains 7 stars; hence we cannot
exclude our non-detection of carbon-enhanced metal-poor stars being due to small number statistics. 

$\bullet$ {\bf $\alpha$-elements} The ratio of $\alpha$-elements (O, Mg, Si, S, Ca, Ti) over Fe is typically used to
trace the relative importance of the ISM chemical enrichment by SN~II and SN~Ia ejecta. At early times massive stars,
which conclude their life as SNe~II, are the main players in the chemical enrichment of the
ISM. SNe~II contribute mainly $\alpha$-elements and little Fe-peak elements,
on time-scales closely tracking star formation, due to the short life-times of SN~II progenitors.
On the other hand, SNe~Ia are the main producers of Fe and their progenitors can have long lifetimes
\citep[e.g.,][]{Tinsley1979}. Under the assumption of an homogeneously
chemically enriched ISM, a decline (``knee'') in
[$\alpha$/Fe] is then typically interpreted as a consequence of this time-delay \citep[e.g.,][]{Gilmore1989}, with stars
more metal-poor than the metallicity of the ``knee'' being born in an environment whose
chemical enrichment was largely driven by SN~II ejecta.

As visible in Fig.~\ref{fig:alpha_MW}, our distant halo stars display enhancements in the ratio of
$\alpha$-elements over Fe with respect to the Solar values and, given the errors on our measurements,
do not show significant differences from those of solar neighborhood halo (SoNH) stars
over the full range of metallicities explored. Although we cannot 
state robustly whether, for example, the [$\alpha$/Fe] are more compatible with the
``high''-$\alpha$ population (possibly formed in-situ) or the ``low''-$\alpha$ population (possibly accreted)
detected by Nissen \& Schuster (2010), in general we can say that these outer halo stars
do not appear to have formed from an ISM in which chemical enrichment from SN~Ia ejecta was dominant. 

As for SoNH stars, at [Fe/H]$\lesssim -1.5$ the abundance ratios
of our targets overlap those of red giant branch stars in
MW satellite galaxies (Fig.~\ref{fig:alpha_dwarfs}).  
At higher metallicities their chemical trends depart from those of systems like Sculptor and
Fornax, whose stars exhibit solar or even sub-solar ratios of $\alpha$-elements to iron; 
however, if we extend the comparison to massive dwarf galaxies such as
Sag (from samples in the core) or even the LMC, then we find a good agreement with the values
measured for our distant halo stars.
Recent work based on APOGEE data by \citet{Hasselquist2017}
  on larger samples of
  Sag stars suggests that for this dwarf galaxy the change from halo-like "high" $\alpha$- abundances
  to distinctly lower abundances is happening in the metallicity regime of our most metal-rich targets.
  Tentatively, we find indeed in this metallicity regime a ``low''-$\alpha$ population,
  in particular in Mg. As discussed in Sect.~\ref{sec:substructures}, a few of these might be genuine Sag stars from their position and velocity, others might be originating from a Sagittarius-like system.

We point out that we limit the comparison to systems as luminous as, or more luminous (massive) than, the Sculptor dSph,
because a) fainter systems have metallicity distribution functions (MDFs)
that barely, or do not, reach the largest [Fe/H] of our targets
\citep[see e.g.,][]{Kirby2013}; b) at a given [Fe/H],
the [$\alpha$/Fe] of fainter systems is even lower than in Sculptor,
which already exhibits clearly solar or sub-solar values
(depending on the $\alpha$-element) at [Fe/H]$\sim -1$. 

$\bullet$ {\bf Fe-peak elements}  The iron-peak elements (see Fig.~\ref{fig:ironp}
for Sc, Mn, Ni and Fig.~\ref{fig:ironp_app} in
the Appendix for V, Cr, Co, Cu, Zn)
are mainly formed in explosive nucleosynthesis
\citep[but see e.g.,][for production of small amounts of Cu and Zn in massive AGB stars]{Karakas2008, Karakas2010}.
Among them, Scandium is not synthesized in SNe~Ia \citep{Woosley2002}, and
indeed in MW dwarf galaxies [Sc/Fe] appears to show a similar behavior as the ratio of $\alpha$-elements to Fe,
with a ``knee''.
In our sample, [Sc/Fe] appears to be fairly constant, around the Solar value, at all metallicities,
exhibiting a range of values
again compatible with those of massive dwarf galaxies at [Fe/H]$\gtrsim -1.5$, as well as
with SoNH samples.

We find [Mn/Fe] to be sub-solar, around $-0.5$dex, for the
bulk of our sample, even at the largest metallicities we probe, while in SoNH samples [Mn/Fe] starts increasing at
[Fe/H]$\gtrsim -1$ (this is clearly seen in Nissen \& Schuster 2011, and to a lesser extent in
Sobeck et al. \citeyear{Sobeck2006}, using
stars from Fulbright \citeyear{Fulbright2000} and Simmerer et al. \citeyear{Simmerer2004}).
This increasing trend of [Mn/Fe] continues to higher metallicities for
thick and thin disk stars, see for example \citet{Battistini2015}. As these authors comment, below [Fe/H]$\lesssim -1$,
the low [Mn/Fe] ratios are mainly determined by the SN~II yields from massive stars \citep{Tsujimoto1998}, while the
increase seen with increasing metallicities is interpreted as a contribution from SNe~Ia \citep{Kobayashi2006},
with suggestions in the literature
that a dependence of Mn yields from SNe~Ia with metallicity may contribute to the increase in the [Mn/Fe] ratios
\citep{Cescutti2008}. 
Within this context, the constant sub-solar values shown by our sample, which are lower than in Sculptor and Fornax  
but at a similar level as in the Sag core (measurements are
not available for the LMC in the van der Swaelmen et al. \citeyear{VanderSwaelmen2013} work),
could again point to a lack of strong
contribution of SN~Ia ejecta to the enrichment of the ISM from which these stars were born. 
However, \citet{Battistini2015} also show that, when applying NLTE corrections to their sample of
thin and thick disk stars, the corresponding
Mn trend changes quite drastically, 
becoming essentially flat and pointing toward Mn sharing the same production site as Fe.
Since we are not applying NLTE corrections, we
then prefer not to over-interpret the results.

Even though star \#4 exhibits super-solar values,
it appears compatible with the large scatter in [Mn/Fe] found at low metallicities for SoNH stars.

$\bullet$ {\bf Nickel and sodium}
As for the other elements analyzed, 
also in [Ni/Fe] our outer halo stars cannot be distinguished
from SoNH stars of similar metallicities at [Fe/H]$\lesssim -1.5$; at larger metallicities,
while remaining compatible with the range of values exhibited by SoNH samples, our stars
preferentially occupy the sub-solar values end. The same behavior is seen in [Na/Fe].
When considering [Ni/Fe] vs [Na/Fe] (Fig.~\ref{fig:nani}), the ``low-$\alpha$'' and ``high-$\alpha$'' 
population of Nissen \& Schuster (2010) occupy distinct regions of the diagram, with 
our targets mostly sharing a similar location to the ``low-$\alpha$'' population.
However, this does not appear a particularly 
compelling diagnostic of whether a star is born in a dwarf galaxy: 
the RGB stars of similar metallicities in massive MW dwarf galaxies show a very broad range of values on this plane - 
smoothly transitioning
from negative [Ni/Fe] and [Na/Fe] to solar (or almost solar) values when moving from Fornax to the LMC inner disk and then bar - 
hence almost overlapping with the region occupied by the ``high-$\alpha$'' stars. 

$\bullet$ {\bf n-capture elements} We now move on to consider the light neutron-capture element 
Y- and the heavy n-capture elements
Ba- and Eu- (Figure~\ref{fig:ncapt}; see Figure~\ref{fig:ncapt_app} in the
Appendix for Sr, Zr, La and values in the Tables for Ce, Pe, Nd, Sm). Neutron-capture elements
are produced by adding neutrons to iron or iron-peak elements; depending on the rate of the
captures, the process is called {\it rapid, r} or {\it slow, s}.
While the contribution of
core collapse supernova from massive stars and compact objects to the {\it r-} process is still debated
\citep[e.g.,][]{Arnould2007}, the main {\it s-}process
is constrained to occur in thermally
pulsating AGB stars (1-4\msun) \citep[see e.g.,][]{Busso1999, Travaglio2004}. 
As discussed in Venn et al. (2004),
helium burning in massive stars (the {\it weak} s-process), only contributes elements lighter than or up to Zr.
Hence the {\it r-}process contributes to the chemical enrichment of the ISM with very little delay with respect to the
formation time of the sites of production, while the {\it s-}process contributes
with a delay of a few 100s Myrs with respect to when the stars were born, therefore tracing chemical enrichment 
on a slightly longer timescales. Importantly,
while Y and Ba can be produced both via the {\it r-} and {\it s-} process,
Eu is considered as an almost
purely {\it r-} process element \citep[e.g.,][]{Truran1981, Travaglio1999}.

At [Fe/H]$\lesssim -1.5$ the [Y, Ba, Eu / Fe] ratios of our sample overlaps completely with the
range of values exhibited by SoNH samples, with [Y/Fe] however occupying the lower end of the SoNH stars 
[Y/Fe] distribution.
On the other hand, significant differences are seen at higher metallicities: most of the stars
with [Fe/H]$\sim -1.3$ group at similar values of [Y/Fe], [Ba/Fe] and [Eu/Fe], with the [Y/Fe]
and [Eu/Fe] concentrating on the low and high end, respectively, of the range of values
exhibited by SoNH samples at similar metallicity, while the [Ba/Fe] is
clearly above the approximately solar values of SoNH samples
($+0.4 \lesssim$ [Ba/Fe] $\lesssim +1.0$). The departure of
distant halo stars from the chemical abundances of SoNH samples becomes even more evident at
[Fe/H]$\gtrsim -1$ in these 3 abundance ratios.

When we compare to RGB stars in dwarf galaxies, [Y/Fe] does not provide
much information, due to the large scatter shown by measurements in dwarf galaxies.
On the other hand, 
an increase of [Ba/Fe] to super-solar values similar to those observed in these distant halo
stars ($+0.4 \lesssim$ [Ba/Fe] $\lesssim +1.0$) is observed for massive systems such as
Fornax and the LMC but not in a relatively small galaxy such as the 
Sculptor dSph ($L_V \sim 2.3 \times 10^6$\lsun, as compared to $L_V \sim 2 \times 10^7$\lsun of Fornax
and $\sim 1.5 \times 10^9$\lsun of the LMC, see compilation by McConnachie \citeyear{McConnachie2012}).
In Fornax and the LMC the increase is seen
at about 0.3dex higher metallicity than in our sample stars, and at these metallicities
RGB stars in these galaxies also show similar enhancements of [Eu/Fe]. This difference in the
metallicity at which the high [Ba/Fe] and [Eu/Fe] values kick in are likely a consequence of
a difference in the initial star formation rate of the various systems. We remind the reader that, while one
might be tempted to interpret the trends observed for the outer halo stars in terms of chemical enrichment
within one galactic environment, there is no evidence that these stars belong all to one,
disrupted massive
system (see next section); hence what we are seeing here is likely stars formed in environments
each following a separate evolutionary path; this should caution against providing an interpretation
within one single chemical enrichment history. 

In Fig.~\ref{fig:baeu} we consider the ratios of [Y/Eu] and [Ba/Eu] in order to gain some insight into the
relative contribution of the {\it s-} and {\it r-} process. [Ba/Eu] for a pure $r-$process is predicted
to be at [Ba/Eu]$_r = -0.69$ \citep{Arlandini1999} or $\sim -0.8$dex \citep{Sneden2008, Bisterzo2014}, while the pure
{\it s-}process [Ba/Eu]$_s = +1$ \citep{Arlandini1999} or $+1.15$ \citep{Bisterzo2014};
our stars are within these values, 
therefore it is likely that the ISM from which they were born was enriched through
both the $r-$ and $s-$ process and that we are
likely detecting a component of enrichment from AGB stars. No particular differences are
seen with respect to the behavior of SoNH stars or MW dwarf galaxies stars at similar metallicities, which are also
thought to have been polluted by AGBs material at [Fe/H]$\gtrsim -2.0$ as witnessed by the rise in [Ba/Eu].

As discussed in Venn et al. (2004), a possible interpretation for a low [Y/Eu] is contributions
from metal-poor AGB stars:  Y belongs to the first peak
that builds through rapid captures around neutron 
number N=50, while Ba (and La) belong to the
second peak that builds around N=82, and at low-metallicity first-peak elements would be bypassed in favor of second-peak elements,
because there are fewer nuclei to absorb the available neutrons. They also argue that a high [Ba/Y] would be compatible with the
yields of low-metallicity AGB stars. Interestingly, \citet{fenner2006} show the [Ba/Y] yields of AGB stars of different
metallicities as a function of the stellar mass and lifetime (see references therein): at metallicity [M/H] $\sim -1.5$,
[Ba/Y] $\sim$ +0.5 would be contributed by AGB stars with lifetimes between 150-300Myr, while at the upper end ([Ba/Y]$\sim$+0.8) 
would be contributed by AGB stars with lifetimes from 800 Myr to several Gyrs.
It is beyond the scope of this paper to trace the nucleosynthetic site of the
various elements: what we point out though is that, once again, at [Fe/H]$\gtrsim -1.5$ the chemical signature of
our outer halo stars departs from the one of SoNH stars, showing a [Ba/Y]$\sim +0.5$, while the latter are scattered around
Solar values; also in this case, our sample resembles the pattern exhibited by massive MW dwarf galaxies, thought to show
enrichment by low-metallicity AGB stars. It is then possible that our distant targets formed in an environment where 
pollution of the ISM by AGB stars played a more dominant role with respect to the environment where
inner halo stars 
formed. Due to the delayed contribution of AGB stars, this might imply that our distant halo stars formed in an environment
with a slower chemical enrichment than those in the SoNH samples. We note that even the
\cite{nissenschuster11} ``low-$\alpha$'' stars at [Fe/H]$>-1.5$, which the authors argue to have originated in
accreted systems, have a much lower [Ba/Y] than we measure at similar metallicity \citep[see also][]{Fishlock2017}: this would suggest a different
chemical enrichment path of the systems that deposited their tidal debris in the outskirts of the MW halo. 

Finally, the bottom right panel of Fig.~\ref{fig:baeu}
shows the location of our targets on the [Ba/Fe] vs [$\alpha$/Fe] plane to summarize some of
the main similarities and
differences with respect to MW satellites and SoNH stars at [Fe/H]$> -1.5$.
As previously mentioned, at these metallicities, only systems more
massive than Sculptor show enhancements in [Ba/Fe] as large as those we detect, which can tentatively 
point to a lower limit in the mass of the accreted satellite galaxies in which our [Fe/H]$> -1.5$
outer halo stars formed. Since the high [Ba/Fe] (and high [Ba/Y]) can be explained by enrichment of low-metallicity AGB stars
with lifetimes from 150Myr to several billion years,
these massive accreted systems must have experienced a slower chemical enrichment with respect to MW halo stars
found in the more central regions, which have solar [Ba/Fe] and [Ba/Y] at the same metallicity.
This might translate into a slightly later accretion time of
these massive systems whose shredded stellar component deposited debris at the Galactocentric distances we are probing 
(at [Fe/H]$>-1.5$: $12 \le r_g \: \mathrm{[kpc]} \le 60$, with a median of 25.5kpc)
with respect to the putative accreted population of Nissen \& Schuster (2010) whose
maximum apocenter reaches to 30-40kpc. Looking at [$\alpha$/Fe], the bulk of our outer halo
stars with [Fe/H]$\sim -1.5$ does not appear to have formed from an ISM dominated by pollution from SNe~Ia, hence
not long after the ``knee'', with the possible exception of \#21 and \#28, which show the lowest [$\alpha$/Fe]
and among the highest [Ba/Fe]; even though this does not provide a stringent 
upper limit on the accretion time, it would exclude accretion after several Gyrs from the start of star formation,
because in that case we would have expected to detect lower values of [$\alpha$/Fe].
Nonetheless it is interesting to notice that the range of combined [$\alpha$/Fe] and
[Ba/Fe] of RGB stars in massive MW satellites does overlap those of stars in our samples, while this is not
the case for the majority of SoNH stars; not only does this highlight again a significant difference
in the chemical abundance of halo stars when moving to the outskirts of the halo, but also tells us that
chemical abundances as those we are detecting are not ``unheard of'' among the MW satellites when
comparing to systems such as the LMC.

Further comparison to the chemical evolution of the massive MW satellites would need to take into account 
the age distribution of the stars in the various spectroscopic samples, which not only varies according to the
specific star formation history of the dwarf galaxy, but also depends on the spatial location where the spectroscopic samples
were gathered, due to the age gradients generally present in dwarf galaxies. However, our findings appear to
point to the fact that the MW outer halo could be at least partially made by accretion of systems experiencing a similar
chemical enrichment to those of massive MW satellites. This is in agreement with the conclusions reached by \citet{Zinn2014, Fiorentino2015}
on the basis of the period distribution of RRLyrae stars in the MW halo and MW satellite galaxies.

\subsection{Star 07}
The elemental abundance ratios of star \#07 stand out with respect to the rest of the sample,
exhibiting a much lower 
[$\alpha$/Fe] and [Ba/Fe], higher [Mn/Fe]$\sim +0.1$ and marginally higher 
[Ni/Fe] ([Mg/Fe]$\sim -0.6$, [Ca/Fe]$\sim -0.15$), [Ba/Fe]$\sim -1.6$, [Mn/Fe]$\sim +0.1$). The differences in the abundances of star \#07 are large enough to be seen visually in the spectrum.   Figure~\ref{fig:spectra} exhibits two regions of the HET spectrum of star \#07 and a star with similar stellar parameters, star \#26. We note that the Fe and Ni features have roughly the same strength while the Ba and Mg features are much weaker in Star \#07.

A rise in [Mn/Fe] is seen in dSph galaxies at lower metallicities than in the MW \citep{NCJ12}, which can
be explained by a metallicity dependent yield from SNae~Ia. This would point to the low [$\alpha$/Fe] and
low [Ba/Fe] being due to a high Fe abundance, perhaps from having originated in/close to a SN~Ia pocket.

\begin{figure*}
\centering
\includegraphics[width=\hsize]{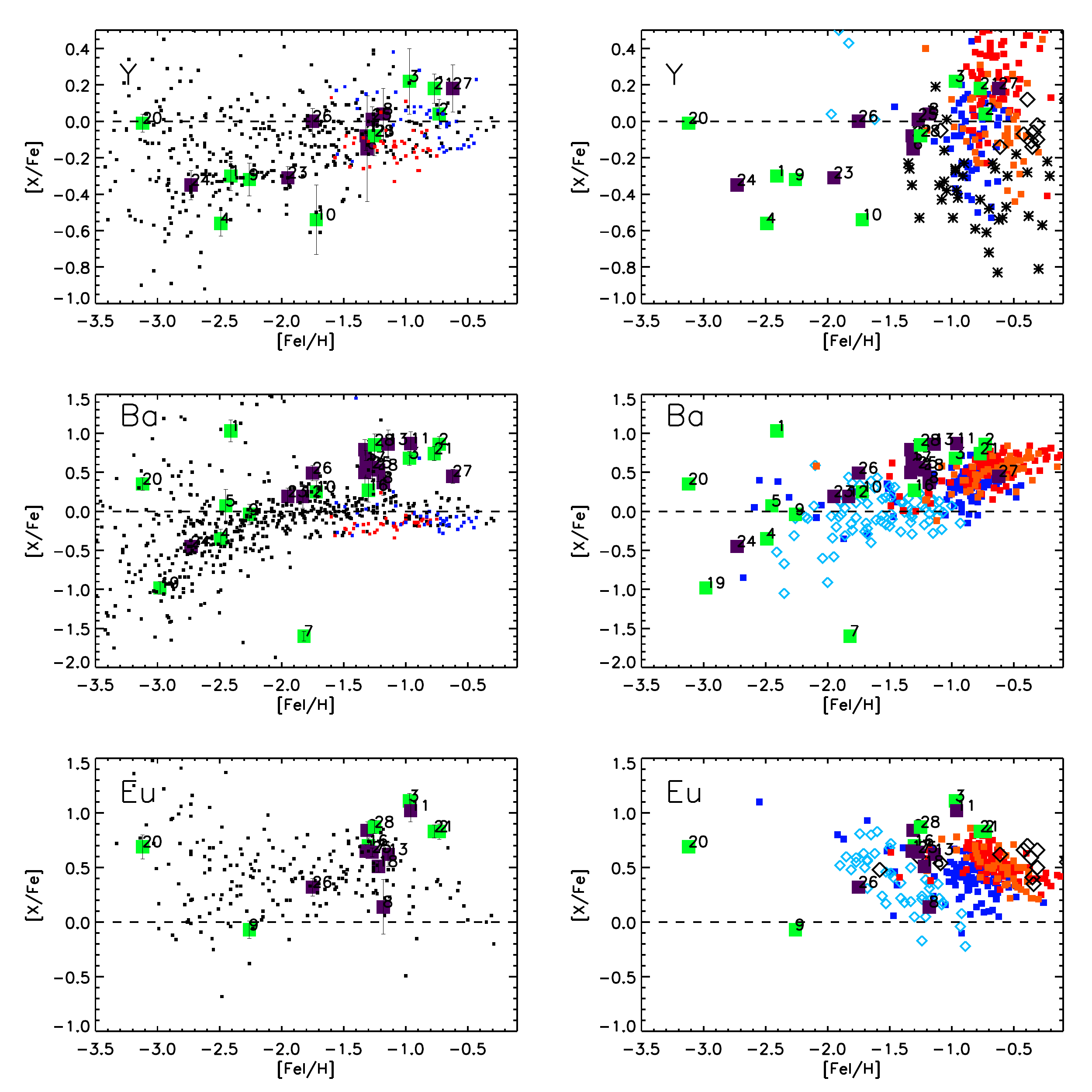}
\caption{As previous figure but for n-capture elements Y, Ba, Eu (left: compared to MW halo samples; right: compared to
  samples of RGB stars in MW dwarf galaxies).}
\label{fig:ncapt}
\end{figure*}

\begin{figure*}
\centering
\includegraphics[width=\hsize]{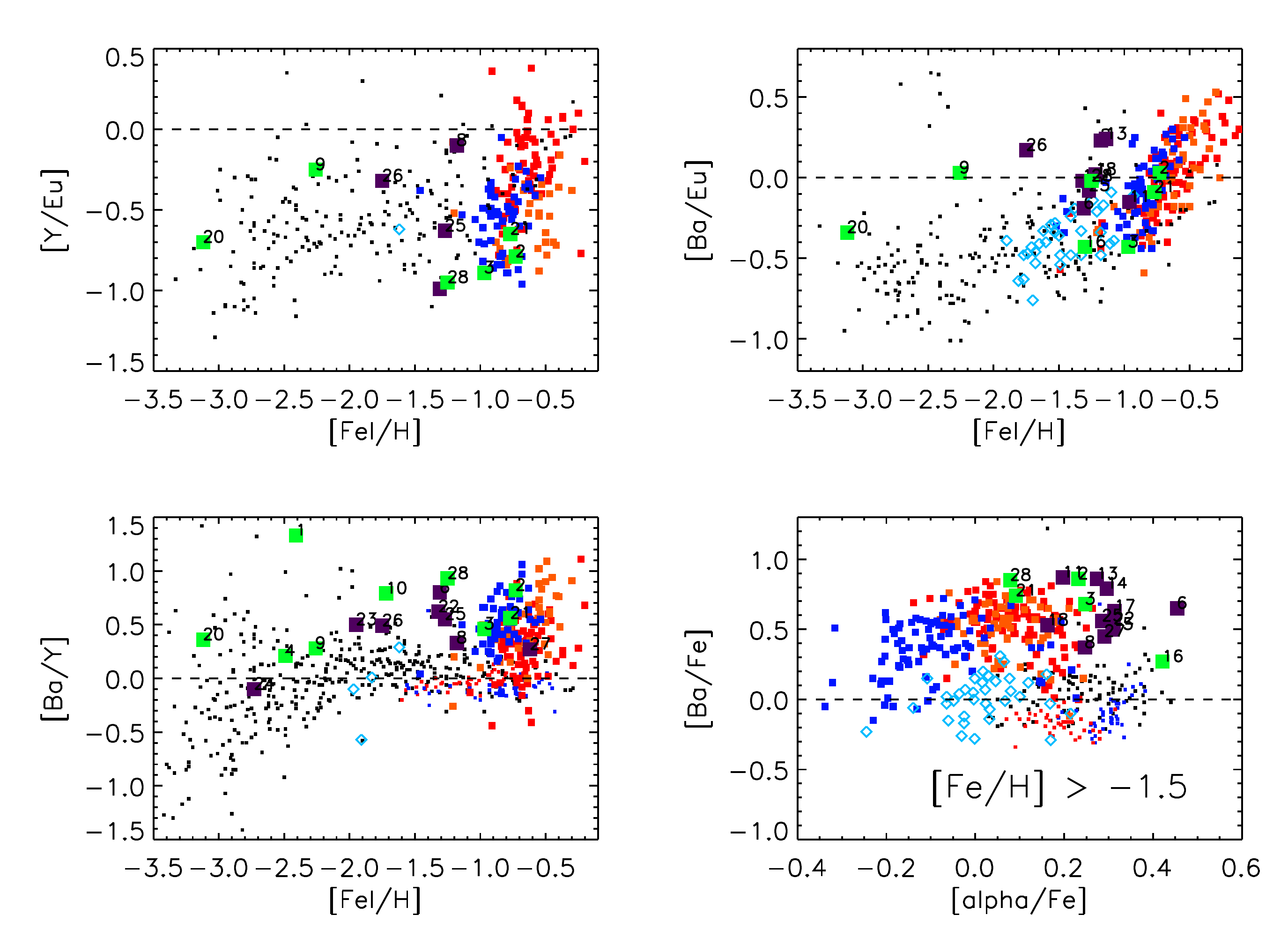}
\caption{[Y/Eu], [Ba/Eu], [Ba/Y] vs [Fe/H] at all metallicities,
  and [Ba/Fe] vs [$\alpha$/Fe] of stars more metal-rich than [Fe/H]$=-1.5$ in our sample compared to literature samples
  of MW halo stars and of stars in MW dwarf galaxies (for the symbols see legend in Fig.~\ref{fig:alpha_MW} and \ref{fig:alpha_dwarfs}).}
\label{fig:baeu}
\end{figure*}

\begin{figure*}
\centering
\includegraphics[width=0.7\hsize]{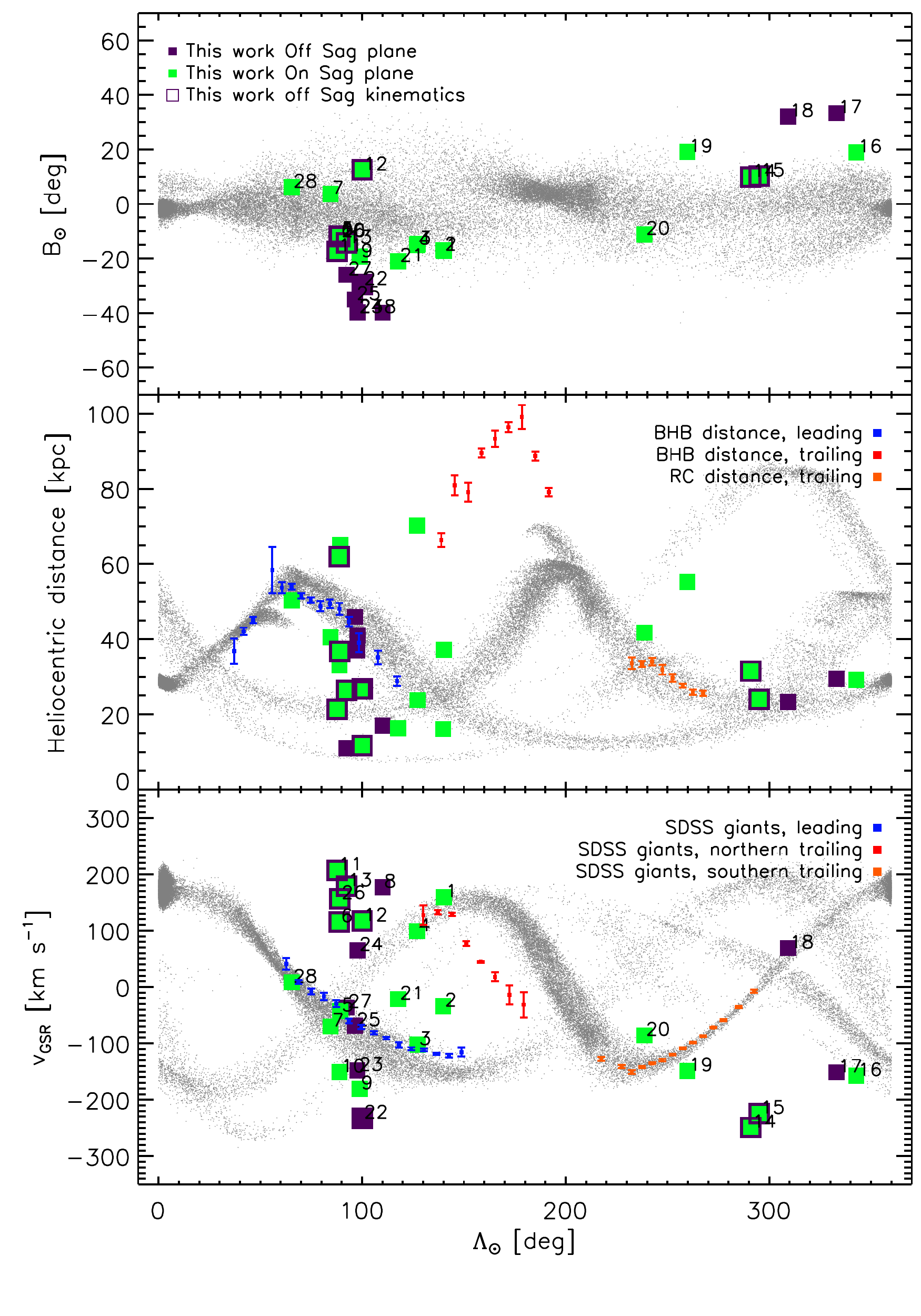}
\caption{
  Location and kinematics of our sample stars (green and purple squared) compared to expectations for the Sag stream
  from the \citet{LM10} model (gray points; only particles either still bound or stripped in the most recent five
  pericentric passages are plotted) and from various sources of observations (see legend and main text).
  The top panel shows the distribution as a function of longitude $\Lambda_{\odot}$ and latitude $B_{\odot}$ in a reference system where
  the equator is aligned with the Sag stream trailing tail \citep[see][]{Majewski2003} but with
  modifications proposed by \citet{Belokurov2014}, that is $\Lambda$ increasing in the direction of the Sag motion
  and latitude axis pointing to the North Galactic Pole. Middle: Heliocentric distance versus $\Lambda_{\odot}$. Bottom:
  Galactic standard of Rest velocities versus $\Lambda_{\odot}$.   Our targets are shown with large filled squares
  (green and purple squared having latitude $|B_{\odot}|$ smaller 
  and larger than 30deg, respectively; green squares with purple border show the sample stars with v$_{\rm GSR}$
  incompatible with the expectations/observations of the Sag stream), while the trend derived in the literature
  using the tracers as in the legend are shown with small (red, blue, and orange) squares with errorbars
  (see main text for the references to the original articles). We note that our v$_{\rm GSR}$ and those
    for the Sag particles in the LM10 model have been calculated assuming
    $R_{\odot}$=8kpc, a Local Standard of Rest (LSR) velocity of 220 \kms and the Solar motion from \citet{Dehnen1998}; in this figure,
    a factor has been applied to both our velocities and those from the particles in the LM10 model to
    correct for the different solar motion and v$_{\rm LSR}$ used in \citet{Belokurov2014}.}
\label{fig:sag}
\end{figure*}

\section{Possible relation to known substructures} \label{sec:substructures}

Although our targets are not associated to any known MW satellite galaxy or globular cluster, 
Fig.~\ref{fig:location} shows that several of them project onto known MW halo substructures, such as
the bright and faint Sag
streams, the Orphan stream, the Virgo Overdensity and approximately in the Tri/And region.
This raises the question of whether the chemical abundance trends that we have traced, which are
distinct from those of halo stars in solar neighborhood samples at [Fe/H]$> -1.5$, are due to stars belonging to
known MW halo substructures, or are to be ascribed to other features.

\subsection{Sagittarius stream} 
The tidal shredding of the Sag dwarf galaxy \citep{Ibata1994} has produced the most impressive substructure visible in the halo
of our Galaxy. While the inner regions (``core'') of the Sag dwarf galaxy
are still gravitationally bound and form a spatially and kinematically confined structure,
its stellar tidal debris are spread across a very large area on the sky and strongly overlap with halo stars both in distance,
radial velocity and metallicity. To date the Sag stream has been traced both in the northern and southern hemisphere, with the
various wraps encompassing heliocentric distances from
$\sim$25 to 100 kpc, and Galactic Standard of Rest (GSR) velocities\footnote{These are
  line-of-sight heliocentric velocities corrected for the Sun motion and Local Standard of Rest motion,
 where we use the values from Dehnen \& Binney (1998) and v$_{\rm LSR}= 220$\kms, respectively.} about $\pm$150 \kms
(e.g., Koposov et al. \citeyear{Koposov2012},  Belokurov et al. \citeyear{Belokurov2014} and references therein;
see Fig.~\ref{fig:sag}). Recently, \citet{Hasselquist2017} have compared Sag core and MW stars in the metallicity range $-1.2 < $ [Fe/H] $ < 0$, showing that at [Fe/H] $\lesssim -0.8$ Sag core stars exhibit similar chemical
patterns than MW stars at high latitude (mostly consistent of halo stars); this makes a distinction based on chemistry alone difficult in the metallicity regime of our targets.

Given the complexity of the Sag system, N-body simulations 
modeling its orbital evolution in a MW-like gravitational potential offer a
useful aid for a first order identification of which stars are most unlikely to be part of the
stream. It should be kept in mind though that associating stars with substructures via comparison with models in various phase space parameters inherits the models' limitations, for example with respect to modeling older wraps of streams and parts of the stream that are far from the main body. The shape of the dark matter halo assumed will influence the modeling of the Sag stream, and in reality such a shape might be complex, time-variable and show variation at different radii
\citep[e.g.,][]{Vera-Ciro2011}. Additionally, the influence from large perturbers, like the LMC, can not always be ignored
\citep{Vera-Ciro2013}.

Here we compare the location on the sky and radial velocities of our targets with the
predictions from the \citet[hereafter LM10]{LM10} model (Fig.~\ref{fig:sag}), following their
recommendation of using only particles stripped in the most recent five pericentric passages. 
We transform the equatorial coordinates of the stars in our sample to a heliocentric coordinate 
system aligned with the Sag stream; in particular, in this reference frame, the 
equator is defined by the mid-plane of the Sag trailing tail debris as proposed by 
\citet{Majewski2003}. We follow the convention of \citet{Belokurov2014} where 
the Sag longitude $\Lambda_{\odot}$ increases in the direction of Sag' motion and 
the Sag latitude axis $B_{\odot}$ points to the North Galactic pole.

Due to the known mismatches between the predictions of Sag stream N-body models and
some sets of observables (see Fig.~\ref{fig:sag}), we let our comparison be guided also by the observed signatures of
portions of the stream, as traced in distance by BHB and red clump (RC) stars and in kinematics by
SDSS spectroscopy of giants (using the estimates given in Koposov et al. 2012, Belokurov et al. 2014).

According to the LM10 model, Sag debris can be found at different heights above the
mid-plan of the trailing tail, with some dependence on the longitude $\Lambda_{\odot}$ range under consideration, with 99.4\% of the particles having $-23\degr < B_{\odot} < +21\degr$, approximately. We then adopt
a very conservative cut of $|B_{\odot}| > $23deg to tag which stars in our sample have a location on the
sky that makes them unlikely to be associated with the Sag stream (see Fig.~\ref{fig:sag}).
Eight of the stars with $|B_{\odot}| \le 23$deg have GSR velocities well beyond the range of values
expected for the stream  at the corresponding $\Lambda_{\odot}$, both in terms of predictions from the  
LM10 model and of the observations, and therefore we also consider them as unlikely to belong to the
Sag stream (see Fig.~\ref{fig:sag}): this leaves us with 13 stars being possibly associated and 15 unlikely to belong to
the stream.

Of the stars in our sample, none was classified as belonging to the Sag stream (nor other groups) by Janesh et al. (2016); on the other hand, Starkenburg et al. (2010) tag \#28 as part of Sag stream, \#20 possibly of early stripped Sag tidal debris
(if the MW dark matter halo is prolate), while
\#10 and \#26 are most likely associated to the Virgo overdensity, although it cannot be excluded that they
belong to the Sag northern leading arm, in particular if the MW DM halo is oblate.

Both samples of unlikely Sag stream members (purple) and possible Sag stream members (green) contain stars
in the metallicity regime where
we detect differences with SoNH stars (see e.g., Figs.~\ref{fig:alpha_MW}-\ref{fig:baeu}). Nonetheless,
it is evident that the group of stars
which show the most distinct chemical patterns with respect to SoNH stars, in particular those with [Fe/H]$\sim$-1.3
that ``clump'' in [Mg, Si, Ti, Mn, Ni, Na, Ba, Eu/Fe] and mildly in [Co/Fe]
are not due to possible Sag stream members.

We emphasize that with our selection we are likely to have overestimated the number of stars possibly belonging to the Sag stream, 
due to our 
very conservative Sag latitude selection. For example, the 5th and 95th percentiles of $B_{\odot}$ for the particles
lost in the most recent five pericentric passages in the LM10 model are $-11$deg and 9deg; if we were to adopt this selection,
only stars \#7 and \#28 would have position and kinematics compatible with membership to the stream. Hence,
assuming that the LM10 model is not missing any important Sag feature in the regime of sky locations, distances
and velocities we are exploring, it seems unlikely that the distinct chemical trends we are detecting 
at [Fe/H] $> -1.5$ between our distant outer halo stars and halo samples in the solar neighborhood are due to 
``contamination'' by Sag stream stars. 
Therefore, it appears we might be probing the signature of other massive systems accreted by the MW
in the outer parts of the stellar halo. In the next
section, we will explore possible membership to other known MW halo substructures.

It is noteworthy that the only clearly chemically peculiar star in our sample (\#7) would survive as
Sag stream member also to the most stringent
selection cut in $B_{\odot}$; however, its chemical properties do not resemble those of any 
Sag core stars studied at high resolution to date.

\subsection{Other known substructures}
Substructures in the MW halo have been detected in a wealth of works in the literature, 
surveying different portions of the visible sky and using a variety of stellar tracers
(M-giants, horizontal branch stars, RRLyrae, main-sequence-turn-off etc.). This sometimes has led to detections
of supposedly different structures, which only
later have been linked back to the same features. Reviewing all these studies is clearly beyond the scope of the present work, hence
we have taken the review article by \citet{Grillmair2016} as reference for the location and general properties of
streams and clouds that are known to-date.
For those stars in our sample whose location on the sky broadly coincides with any of the substructures listed in that
article, we have explored
further their possible physical association considering the velocity, and (to a less extent, due to the large uncertainties)
the distance, information, going back to the original sources to check more in detail the expected trends in
distance and line-of-sight (l.o.s.) velocity as a function of, for instance, galactic coordinates. 

The region that encompasses the range 150 $<$ RA [deg] $<$ 220, $-20 <$ DEC [deg] $< +20$ , where the majority of
  our targets falls, is a complex one, being home to the 
  features generally known as the ``Virgo Overdensity'' (VOD), including the Virgo Stellar Stream (VSS) 
and partially projecting also onto the
Sag stream. The VOD is a poorly understood overdensity with (tentatively) associated to it
many different components \cite[e.g.,][and references therein]{Duffau2014}. The heliocentric velocities
typically
associated to substructures in this area span the range (200, 360) \kms for the VOD and $\sim$130 \kms for the VSS, while
  distances of stars associated to these substructures are expected to be approximately less than 20\,kpc.

  Another part of the sky rich in substructures is in the Triangulum-Andromeda region, 
where several features have been detected: the Segue~2 ultra-faint object \citep{Belokurov2009}, 
the Triangulum Stream \citep{Bonaca2012}, and a number of other features detected as overdensities
of main sequence (MS) stars, main sequence turn-off (MSTO) stars,
or K- and M-giants, dubbed Tri-And~1 and Tri-And~2  
\citep{Majewski2004, Rocha-Pinto2004, Martin2007}, where some of them may be related 
to each other and have their origin in the disruption of the same small galactic system  
\citep[see e.g.,][]{Deason2014, Sheffield2014}. 

Only a handful of our targets appears to possibly belong to any of the 25 streams and clouds listed
in Grillmair \& Carlin (2016).
Star~\#18 has position, distance, line-of-sight (l.o.s.) velocity and metallicity that make it compatible with membership
  to the Hercules-Aquila Cloud. Stars~\#8,11,13 position, distance, l.o.s. velocity and metallicity fall well
  within the range of values listed for the VOD, hence they might be members of this structure;
  on the other hand, stars \#1,24,26 have positions, l.o.s. velocities and literature distance estimates compatible
  with those of VOD stars,
  but our revised distance values place them beyond 35kpc, making membership to the VOD unlikely.
  The metallicity does not  provide a very tight constrain, as stars in the VOD have been measured to have a 
broad range of values, 
  within $-2 \lesssim$[Fe/H]$\lesssim -1$; the metallicity of \#1 \& 24 could be considered too low ([Fe/H]$= -2.4$ and $-2.7$,
  respectively), but one cannot exclude that the MDF of VOD stars extends to lower values, in particular if the VOD is all or in part the
  result of the accretion of a dwarf galaxy.

  In any case, of the stars with [Fe/H]$\sim -1.3$ clumping in [Mg, Si, Ti, Y, Ba, Eu, Na and Ni/Fe] and [Co/Fe], only \#13 and \#18 would
  appear to belong to known MW halo substructures other than Sag (VOD and Hercules-Aquila, respectively).
  Hence, overall, the aforementioned chemical trend does not appear to be due to stars belonging to known
  MW substructures.  

\section{Summary and conclusions} \label{sec:summary}

We have obtained VLT/UVES, Magellan/MIKE and HET/HRS high resolution optical spectroscopy for a sample of 28 halo stars with heliocentric distances
12 $\le d_h \: \mathrm{[kpc]} \le $73 (median $d_h=$ 32kpc), for which we have derived chemical abundances
for 27 elements.

The large present-day distances of the stars in our sample place them in 
the outer-halo region of the MW and allow us to explore the chemical properties of MW halo stars over a
considerably larger volume than literature studies based on high resolution spectroscopy of
larger samples of halo stars currently found in the solar neighborhood. The sample size is even competitive with respect to
what is currently provided by high resolution spectroscopic
surveys such as APOGEE at similar distances and
provides abundances for 11 elements in common and 16 not available from APOGEE.

At [Fe/H] $\lesssim -1.5$ the chemical properties of our sample stars mostly overlap those exhibited by SoNH stars.
However, at [Fe/H] $\gtrsim -1.5$ differences are seen in particular for [Mn, Ni, Na, Ba, Eu/Fe], [Ba/Y] and [Ba/Eu], either
as the chemical properties of our sample departing from those of SoNH stars or ``clumping'' in a restricted region of values with respect to
the broader distribution shown by SoNH stars of similar metallicities (this can also be appreciated in [Mg, Si, Ti/Fe]).
For the majority of stars at high metallicity the detected trends do not appear to be due to stars belonging to known MW substructures, including the Sag stream.  

We analyze this behavior in the light of the chemical abundances measured for RGB stars in MW dwarf galaxies, including in
the comparison also massive systems such as Sag and the LMC. Super-solar values of [Ba/Fe] as those measured at
[Fe/H] $\gtrsim -1.5$ in our sample appear only in the most luminous (massive) satellites (Fornax, Sag and the LMC),
but not in the Sculptor dSph ($L_V \sim 2.3 \times 10^6$ \lsun), for which [Ba/Fe] remains constant to a solar-value in the high metallicity regime.
On the other hand, the ratio of $\alpha$-elements to iron do not resemble
those observed for RGB stars of similar metallicity in the Fornax dSph, while compare well to Sag and LMC stars. 
We find that the elemental abundances of stars
in the most luminous dwarf galaxies show similar trends to those seen in our data for the elements for which we see differences
with respect to SoNH stars at [Fe/H] $\gtrsim -1.5$. We conclude that, if MW outer halo stars have originated in the shredded stellar
component of accreted dwarf galaxies, then the MW outer halo is partially formed by massive accreted systems.  
The large super-solar values we measure for [Ba/Fe] and [Ba/Y], which can be interpreted as the result of pollution by metal-poor AGB stars
over time-scales of few 100s Myr to almost 1Gyr, compared to the solar values of SoNH samples at similar metallicities,
might indicate that star formation in the accreted systems was able to proceed on longer time-scales than the accreted systems
that possibly contributed to the build-up of the more internal halo regions, possibly because of later accretion times.

Due to our target selection (see Sect.~\ref{sec:data}), half of our targets have [Fe/H]$ > -1.6$, which is the regime
where we detect the most interesting differences with SoNH samples. 
Data from SDSS low resolution optical spectroscopy \citep{Fernandez-Alvar2015}
suggest that at large Galactocentric distances, in particular at $r_g \gtrsim 30$kpc,
stars with $-1.5\lesssim$ [Fe/H]$\lesssim -0.8$ , are not very common and
essentially occupy the high metallicity tail of the metallicity distribution of MW stars. If stars in the MW outer halo originated
in accreted dwarf galaxies, this would imply that
this metallicity regime is sampling the end point of their chemical evolution. We note that
the metallicity distribution function of MW satellite stars are very wide, encompassing a few dex in [Fe/H] even for the
systems that have stopped star formation at $z>2$; hence if we are probing the high metallicity tail of accreted systems,
these will have carried with them a significant, or even larger,
portion of more metal-poor stars that are now spread out in the outer halo.
It will be interesting to see if just a few massive systems might have formed a considerable fraction of the MW outer halo. 

\begin{acknowledgements}
  The authors are indebted to the International Space Science Institute (ISSI), Bern, Switzerland,
  for supporting and funding the international team "First stars in dwarf galaxies", where the idea for
  this project was born.  The authors acknowledge the referee, T. Beers, for useful comments and
  H.~Morrison for providing clarification about possible
  selection effects in the Spaghetti survey with respect to the star's Carbon abundances. 
  GB gratefully acknowledges financial support by the Spanish Ministry of
  Economy and Competitiveness (MINECO) under the Ramon y Cajal Programme (RYC-2012-11537) and the grant AYA2014-56795-P.
  ES gratefully acknowledges funding by the Emmy Noether program from the Deutsche Forschungsgemeinschaft (DFG).
  The Hobby-Eberly Telescope (HET) is a joint project of the University of Texas at Austin,
the Pennsylvania State University, Stanford University, Ludwig-Maximilians-Universit\"at
M\"unchen, and Georg-August-Universit\"at G\"ottingen. The HET is named in honor of its
principal benefactors, William P. Hobby and Robert E. Eberly.
This work has made use of data from the European Space Agency (ESA)
mission {\it Gaia} (\url{https://www.cosmos.esa.int/gaia}), processed by
the {\it Gaia} Data Processing and Analysis Consortium (DPAC,
\url{https://www.cosmos.esa.int/web/gaia/dpac/consortium}). Funding
for the DPAC has been provided by national institutions, in particular
the institutions participating in the {\it Gaia} Multilateral Agreement.

\end{acknowledgements}

\bibliographystyle{aa}
\bibliography{halo}

\clearpage

\section*{Appendix: Additional figures on chemical abundance trends}

\begin{figure*}[!ht]
\centering
\includegraphics[width=0.7\hsize]{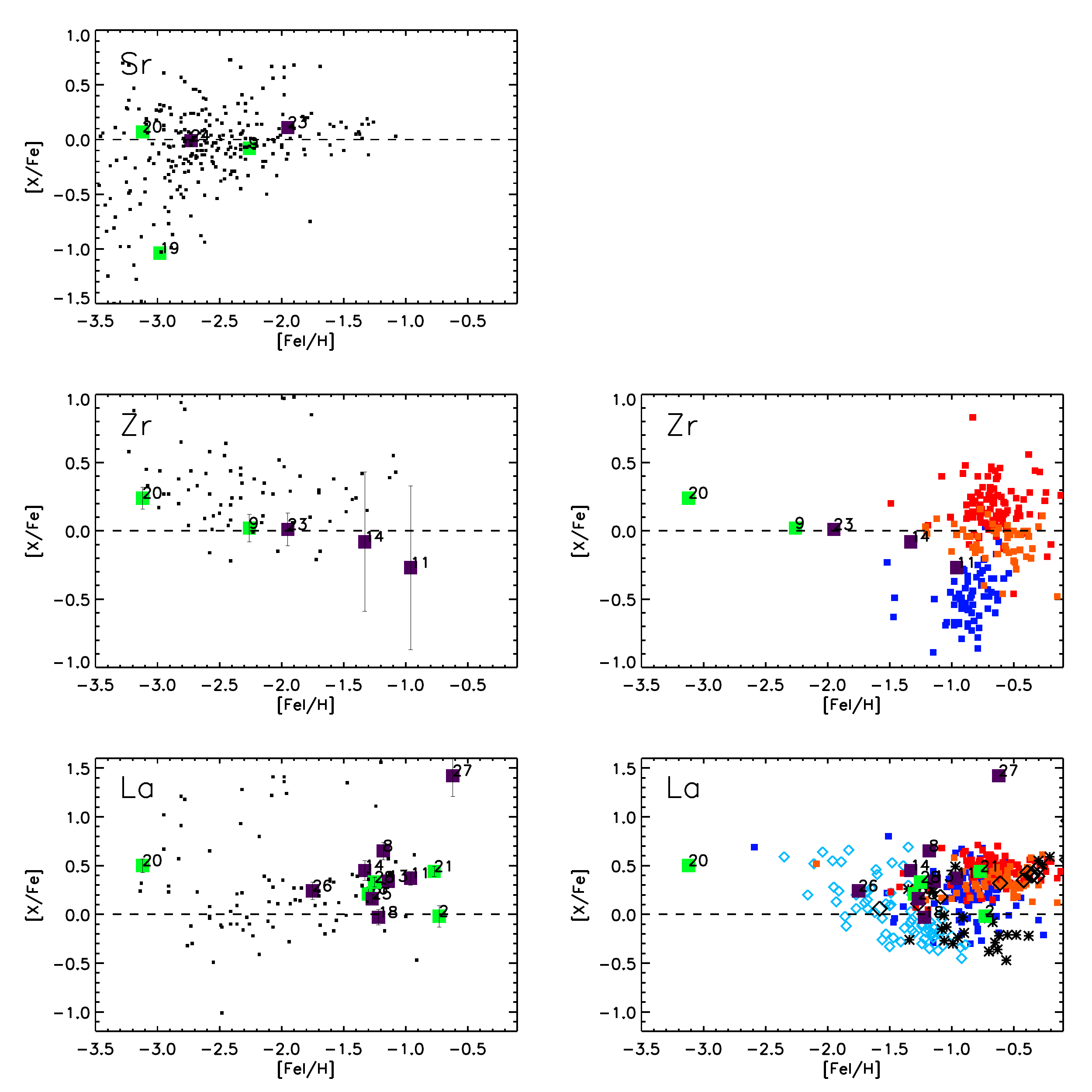}
\caption{Abundance of the n-capture elements Sr, Zr, La relative
  to iron (\ion{Fe}{I}) as a function of [\ion{Fe}{i}/H].}
\label{fig:ncapt_app}
\end{figure*}

\begin{figure*}
\centering
\includegraphics[width=0.8\hsize]{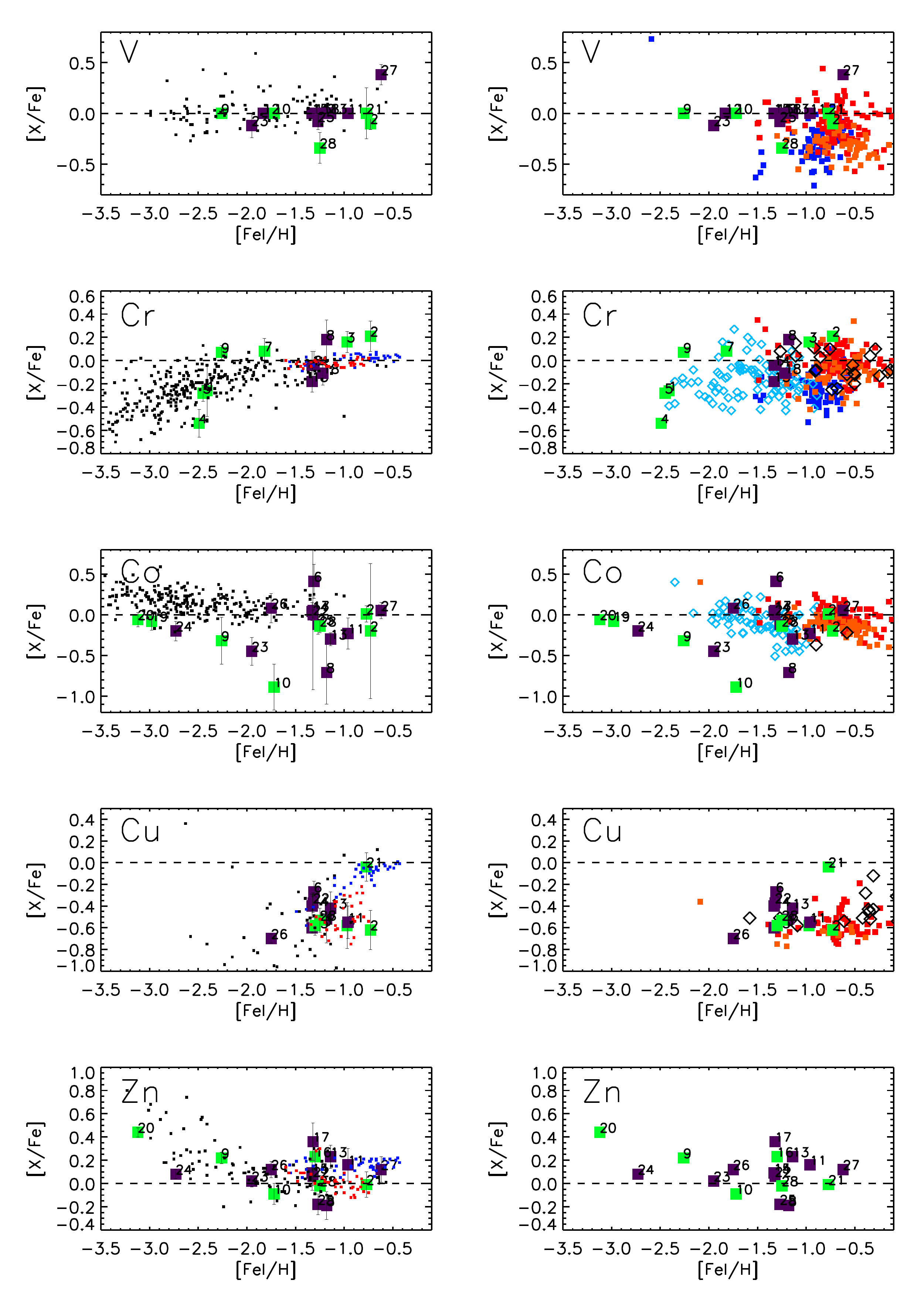}
\caption{Abundance of the Fe-peak elements \ion{V}{I}, \ion{Cr}{I}, Co, Cu, Zn relative
  to iron (\ion{Fe}{I}) as a function of [\ion{Fe}{i}/H].}
\label{fig:ironp_app}
\end{figure*}

\begin{landscape}
\begin{table}[h]
\caption{Atmospheric stellar parameters and details of observations for the sample stars. The table
    format is the following: (1) ID; (2), (3), (4) \& (5): 
spectroscopic $T_\mathrm{eff}$ and $\log g$, [Fe/H] of the atmosphere model, microturbulence velocity; (6) 
photometric $\log g$; (7) Julian day for the middle of the exposure (in case of single
exposures, otherwise we give the days of the first and last exposures); (8) total exposure time; (9) 
radial velocity and error; (10) signal-to-noise ratios at $\sim4000$~\AA,
$\sim4900$~\AA, and $\sim6700$~\AA; (11) instrument. For the HET and UVES stars, the first line gives
the average radial velocity of all exposures, while the following lines give the Julian day and
heliocentric radial velocity of each exposure.
The photometric $\log g$ of most HRS stars had asymmetric error bars; the error quoted here is 
the larger one of the upper and lower errors.
Sources: All stars observed with HRS and UVES, and stars \#8, 9, 10 \& 12 with MIKE, were selected
from the Spaghetti survey, as listed in \citep{Starkenburg2011}; star \#11 comes from the APOGEE survey;
stars from \#13-18 were drawn from the Xue et al.(2014) catalog. 
}
\label{table:atmo}
\tiny
\centering
\begin{tabular}{lrrrrrrrrrrrl} \hline \hline
\multicolumn{1}{c}{ID}
&\multicolumn{1}{c}{$T_\mathrm{eff}^\mathrm{sp}$}
&\multicolumn{1}{c}{$\log g$}
&\multicolumn{1}{c}{[Fe/H]}
&\multicolumn{1}{c}{v$_\mathrm{turb}^\mathrm{sp}$}
&\multicolumn{1}{c}{$\log g_\mathrm{pho}$}
&\multicolumn{1}{c}{JD$-$}
&\multicolumn{1}{c}{$t_\mathrm{exp}$}
&\multicolumn{1}{c}{V$_\mathrm{rad}$}
&\multicolumn{3}{c}{S/N}
&\multicolumn{1}{c}{Instrument,} \\
&\multicolumn{1}{c}{[K]}
&\multicolumn{1}{c}{[cgs]}
&\multicolumn{1}{c}{model}
&\multicolumn{1}{c}{[km\,s$^{-1}$]}
&\multicolumn{1}{c}{[cgs]}
&\multicolumn{1}{c}{$2456000$}
&\multicolumn{1}{c}{[s]}
&\multicolumn{1}{c}{[km\,s$^{-1}$]}
&\multicolumn{1}{c}{$4000$}
&\multicolumn{1}{c}{$4900$}
&\multicolumn{1}{c}{$6700$}
&\multicolumn{1}{c}{Remark} \\ \hline
           &	  &	&	&      &	   &	      &       & 	       &    &	 &     &\\
01     &$4750$&$1.5$&$-2.40$&$1.10$&$1.7\pm0.2$&$ 358-414$&$ 7200$&$ 284.70\pm0.17$&$ -$&$36$&$ 59$&HRS\\
           &	  &	&	&      &	   &$ 357.650$&       &$ 284.39\pm0.77$&    &	 &     &\\
           &	  &	&	&      &	   &$ 358.665$&       &$ 285.17\pm0.77$&    &	 &     &\\
           &	  &	&	&      &	   &$ 369.617$&       &$ 284.60\pm1.30$&    &	 &     &\\
           &	  &	&	&      &	   &$ 413.671$&       &$ 284.62\pm0.72$&    &	 &     &\\
02     &$4654$&$1.5$&$-0.75$&$1.30$&$1.5\pm0.5$&$ 339-398$&$ 7200$&$  90.48\pm0.15$&$ -$&$27$&$ 44$&HRS\\
           &	  &	&	&      &	   &$ 338.716$&       &$  90.18\pm0.32$&    &	 &     &\\
           &	  &	&	&      &	   &$ 388.754$&       &$  90.52\pm0.35$&    &	 &     &\\
           &	  &	&	&      &	   &$ 397.722$&       &$  90.69\pm0.26$&    &	 &     &\\
03     &$4800$&$2.2$&$-0.95$&$1.10$&$1.7\pm0.3$&$ 365-424$&$ 9000$&$  13.95\pm7.59$&$ -$&$27$&$ 42$&HRS, spectroscopic binary\\
           &	  &	&	&      &	   &$ 364.671$&       &$   2.57\pm0.26$&    &	 &     &\\
           &	  &	&	&      &	   &$ 397.757$&       &$  10.69\pm0.29$&    &	 &     &\\
           &	  &	&	&      &	   &$ 423.676$&       &$  28.30\pm0.26$&    &	 &     &\\
04     &$4486$&$0.7$&$-2.50$&$1.70$&$0.9\pm0.1$&$ 341-386$&$ 9600$&$ 216.43\pm0.42$&$ -$&$32$&$ 50$&HRS\\
           &	  &	&	&      &	   &$ 340.910$&       &$ 214.94\pm1.80$&    &	 &     &\\
           &	  &	&	&      &	   &$ 363.681$&       &$ 216.45\pm1.16$&    &	 &     &\\
           &	  &	&	&      &	   &$ 383.629$&       &$ 216.79\pm0.77$&    &	 &     &\\
           &	  &	&	&      &	   &$ 385.617$&       &$ 216.53\pm0.29$&    &	 &     &\\
     &$4600$&$1.6$&$-1.00$&$1.10$&$1.9\pm0.4$&$ 366-412$&$ 7200$&$ 117.00\pm0.18$&$ -$&$27$&$ 52$&HRS, \#21 UVES\\
           &	  &	&	&      &	   &$ 365.719$&       &$ 117.04\pm0.27$&    &	 &     &\\
           &	  &	&	&      &	   &$ 387.660$&       &$ 116.63\pm0.51$&    &	 &     &\\
           &	  &	&	&      &	   &$ 411.709$&       &$ 117.22\pm0.38$&    &	 &     &\\
     &$4700$&$1.8$&$-1.70$&$1.10$&$1.8\pm0.5$&$ 357-445$&$12000$&$ 247.88\pm0.19$&$ -$&$54$&$ 75$&HRS, \#26 UVES\\ 
           &	  &	&	&      &	   &$ 356.851$&       &$ 248.31\pm0.34$&    &	 &     &\\
           &	  &	&	&      &	   &$ 364.824$&       &$ 248.39\pm0.74$&    &	 &     &\\
           &	  &	&	&      &	   &$ 365.827$&       &$ 248.01\pm0.64$&    &	 &     &\\
           &	  &	&	&      &	   &$ 420.747$&       &$ 247.43\pm0.31$&    &	 &     &\\
           &	  &	&	&      &	   &$ 444.672$&       &$ 247.63\pm0.32$&    &	 &     &\\
05     &$4650$&$0.7$&$-2.45$&$1.90$&$1.1\pm0.5$&$ 391-444$&$12000$&$  45.16\pm0.39$&$ -$&$34$&$ 63$&HRS\\
           &	  &	&	&      &	   &$ 390.841$&       &$  45.64\pm1.06$&    &	 &     &\\
           &	  &	&	&      &	   &$ 427.730$&       &$  44.36\pm1.18$&    &	 &     &\\
           &	  &	&	&      &	   &$ 428.724$&       &$  45.28\pm0.96$&    &	 &     &\\
           &	  &	&	&      &	   &$ 442.685$&       &$  44.19\pm0.74$&    &	 &     &\\
           &	  &	&	&      &	   &$ 443.682$&       &$  46.23\pm0.76$&    &	 &     &\\
06     &$4350$&$0.8$&$-1.30$&$1.50$&$1.1\pm0.6$&$ 395-417$&$ 9000$&$ 204.69\pm0.15$&$ -$&$18$&$ 41$&HRS\\
           &	  &	&	&      &	   &$ 394.739$&       &$ 204.93\pm0.32$&    &	 &     &\\
           &	  &	&	&      &	   &$ 413.762$&       &$ 204.40\pm0.37$&    &	 &     &\\
           &	  &	&	&      &	   &$ 416.682$&       &$ 204.70\pm0.67$&    &	 &     &\\
07     &$4700$&$1.3$&$-1.80$&$1.40$&$1.7\pm0.2$&$ 387-445$&$12000$&$ -47.41\pm0.15$&$ -$&$54$&$ 96$&HRS\\
           &	  &	&	&      &	   &$ 386.744$&       &$ -46.95\pm0.52$&    &	 &     &\\
           &	  &	&	&      &	   &$ 412.844$&       &$ -47.85\pm0.27$&    &	 &     &\\
           &	  &	&	&      &	   &$ 427.807$&       &$ -47.45\pm0.34$&    &	 &     &\\
           &	  &	&	&      &	   &$ 429.790$&       &$ -47.27\pm0.26$&    &	 &     &\\
           &	  &	&	&      &	   &$ 444.751$&       &$ -47.23\pm0.44$&    &	 &     &\\
     &$4529$&$1.1$&$-0.85$&$1.40$&$0.8\pm0.2$&$ 392-398$&$ 6000$&$   4.75\pm0.36$&$ -$&$12$&$ 21$&HRS, \#28 UVES\\
           &	  &	&	&      &	   &$ 391.805$&       &$   4.46\pm0.29$&    &	 &     &\\
           &	  &	&	&      &	   &$ 397.798$&       &$   5.18\pm0.44$&    &	 &     &\\
\hline
\end{tabular}
\end{table}
\end{landscape}

\setcounter{table}{0}
\begin{landscape}
\begin{table}[h]
  \caption{(continued)}
\tiny
\centering
\begin{tabular}{lrrrrrrrrrrrl} \hline \hline
\multicolumn{1}{c}{ID}
&\multicolumn{1}{c}{$T_\mathrm{eff}^\mathrm{sp}$}
&\multicolumn{1}{c}{$\log g$}
&\multicolumn{1}{c}{[Fe/H]}
&\multicolumn{1}{c}{v$_\mathrm{turb}^\mathrm{sp}$}
&\multicolumn{1}{c}{$\log g_\mathrm{pho}$}
&\multicolumn{1}{c}{JD$-$}
&\multicolumn{1}{c}{$t_\mathrm{exp}$}
&\multicolumn{1}{c}{V$_\mathrm{rad}$}
&\multicolumn{3}{c}{S/N}
&\multicolumn{1}{c}{Instrument,} \\
&\multicolumn{1}{c}{[K]}
&\multicolumn{1}{c}{[cgs]}
&\multicolumn{1}{c}{model}
&\multicolumn{1}{c}{[km\,s$^{-1}$]}
&\multicolumn{1}{c}{[cgs]}
&\multicolumn{1}{c}{$2456000$}
&\multicolumn{1}{c}{[s]}
&\multicolumn{1}{c}{[km\,s$^{-1}$]}
&\multicolumn{1}{c}{$4000$}
&\multicolumn{1}{c}{$4900$}
&\multicolumn{1}{c}{$6700$} 
&\multicolumn{1}{c}{Remark} \\ \hline
&	  &	&	&      &	   &	      &       & 	       &    &	 &     &\\
08 &$5000$&$2.3$&$-1.20$&$1.25$&$2.2\pm0.4$&$ 725.573$&$10800$&$ 371.81\pm0.11$&$20$&$30$&$ 65$&MIKE\\
09 &$4748$&$1.8$&$-2.30$&$1.50$&$1.9\pm0.3$&$ 725.712$&$10800$&$ -56.04\pm0.51$&$45$&$60$&$150$&MIKE\\
10 &$4842$&$1.8$&$-1.70$&$1.50$&$1.8\pm0.4$&$ 725.851$&$10800$&$ -61.13\pm0.36$&$30$&$50$&$ 80$&MIKE\\
11 &$4444$&$1.4$&$-0.90$&$1.10$&$1.5\pm0.4$&$ 828.537$&$ 4400$&$ 317.78\pm0.03$&$20$&$50$&$110$&MIKE\\
12 &$5264$&$3.1$&$-1.80$&$0.70$&$2.5\pm0.4$&$ 828.597$&$ 5400$&$ 111.95\pm0.15$&$ -$&$12$&$ 45$&MIKE\\
 &	  &	&	&      &	   &	      &       & 	&    &    &	&\\
13 &$4243$&$0.8$&$-1.15$&$1.50$&$1.0\pm0.3$&$ 828.476$&$ 5400$&$ 281.00\pm0.04$&$30$&$60$&$ 90$&MIKE\\
14 &$4513$&$1.3$&$-1.30$&$1.14$&$1.6\pm0.3$&$ 828.898$&$ 4800$&$-342.89\pm0.09$&$25$&$45$&$ 70$&MIKE\\
15 &$4570$&$1.4$&$-1.30$&$1.20$&$1.5\pm0.3$&$ 828.847$&$ 3600$&$-320.57\pm0.08$&$25$&$50$&$ 75$&MIKE\\
16 &$4253$&$0.8$&$-1.25$&$1.40$&$0.9\pm0.2$&$ 828.677$&$ 5400$&$-282.22\pm0.18$&$20$&$40$&$ 80$&MIKE\\
17 &$4550$&$1.3$&$-1.30$&$1.15$&$1.3\pm0.2$&$ 828.750$&$ 5400$&$-328.43\pm0.13$&$25$&$35$&$ 80$&MIKE\\
18 &$4350$&$1.0$&$-1.20$&$1.35$&$0.9\pm0.3$&$ 828.804$&$ 3600$&$-107.42\pm0.14$&$25$&$45$&$ 95$&MIKE\\
 \hline
           &	  &	&	&      &	   &	      &       & 	       &    &	 &     &\\
19 &$4603$&$1.2$&$-2.95$&$1.70$&$1.4\pm0.2$&$ 873-889$&$15000$&$-259.91\pm0.22$&$35$&$75$&$120$&UVES\\
           &	  &	&	&      &	   &$ 872.888$&       &$-259.65\pm0.06$&    &	 &     &\\
           &	  &	&	&      &	   &$ 877.884$&       &$-260.30\pm0.12$&    &	 &     &\\
           &	  &	&	&      &	   &$ 888.798$&       &$-259.85\pm0.39$&    &	 &     &\\
           &	  &	&	&      &	   &$ 888.835$&       &$-260.14\pm0.25$&    &	 &     &\\
           &	  &	&	&      &	   &$ 888.872$&       &$-260.38\pm0.36$&    &	 &     &\\
20 &$4500$&$1.3$&$-3.10$&$1.75$&$1.1\pm0.2$&$ 875-896$&$ 7200$&$ -54.67\pm0.22$&$32$&$60$&$ 80$&UVES\\
           &	  &	&	&      &	   &$ 874.900$&       &$ -54.23\pm0.22$&    &	 &     &\\
           &	  &	&	&      &	   &$ 893.890$&       &$ -54.98\pm0.15$&    &	 &     &\\
           &	  &	&	&      &	   &$ 895.878$&       &$ -54.88\pm0.09$&    &	 &     &\\
21 &$4800$&$2.3$&$-0.75$&$0.90$&$2.0\pm0.4$&$747-1071$&$15000$&$ 116.40\pm0.13$&$35$&$50$&$ 85$&UVES, also observed with HRS\\
           &	  &	&	&      &	   &$ 746.604$&       &$ 117.12\pm0.26$&    &    &     &\\
           &	  &	&	&      &	   &$ 746.640$&       &$ 117.26\pm0.27$&    &    &     &\\
           &	  &	&	&      &	   &$ 826.505$&       &$ 116.57\pm0.21$&    &    &     &\\
           &	  &	&	&      &	   &$1070.793$&       &$ 115.81\pm0.10$&    &    &     &\\
           &	  &	&	&      &	   &$1070.829$&       &$ 115.07\pm0.07$&    &    &     &\\
22 &$4700$&$1.8$&$-1.30$&$1.15$&$1.6\pm0.3$&$837-1070$&$15000$&$ -71.88\pm0.10$&$35$&$55$&$ 80$&UVES\\
           &	  &	&	&      &	   &$ 837.500$&       &$ -72.36\pm0.04$&    &    &     &\\
           &	  &	&	&      &	   &$1042.843$&       &$ -72.08\pm0.05$&    &    &     &\\
           &	  &	&	&      &	   &$1069.679$&       &$ -71.71\pm0.14$&    &    &     &\\
           &	  &	&	&      &	   &$1069.715$&       &$ -71.17\pm0.12$&    &    &     &\\
           &	  &	&	&      &	   &$1069.751$&       &$ -72.29\pm0.13$&    &    &     &\\
23 &$4730$&$1.7$&$-1.90$&$1.40$&$1.7\pm0.3$&$1128-1158$&$15000$&$  42.22\pm0.13$&$38$&$65$&$ 90$&UVES\\
           &	  &	&	&      &	   &$1127.544$&       &$  41.40\pm0.28$&    &    &     &\\
           &	  &	&	&      &	   &$1127.580$&       &$  40.51\pm0.12$&    &    &     &\\
           &	  &	&	&      &	   &$1128.586$&       &$  42.07\pm0.11$&    &    &     &\\
           &	  &	&	&      &	   &$1157.561$&       &$  42.93\pm0.11$&    &    &     &\\
           &	  &	&	&      &	   &$1157.601$&       &$  43.77\pm0.16$&    &    &     &\\
24 &$4600$&$1.2$&$-2.70$&$1.55$&$1.3\pm0.3$&$780-1129$&$ 7200$&$ 255.05\pm0.18$&$32$&$55$&$ 75$&UVES\\
           &	  &	&	&      &	   &$ 779.611$&       &$ 254.29\pm0.12$&    &	 &     &\\
           &	  &	&	&      &	   &$1070.866$&       &$ 255.65\pm0.16$&    &	 &     &\\
           &	  &	&	&      &	   &$1128.549$&       &$ 255.18\pm0.19$&    &	 &     &\\
\hline
\end{tabular}
\end{table}
\end{landscape}

\setcounter{table}{0}
\begin{landscape}
\begin{table}[h]
  \caption{(continued)}
\tiny
\centering
\begin{tabular}{lrrrrrrrrrrrl} \hline \hline
\multicolumn{1}{c}{ID}
&\multicolumn{1}{c}{$T_\mathrm{eff}^\mathrm{sp}$}
&\multicolumn{1}{c}{$\log g$}
&\multicolumn{1}{c}{[Fe/H]}
&\multicolumn{1}{c}{v$_\mathrm{turb}^\mathrm{sp}$}
&\multicolumn{1}{c}{$\log g_\mathrm{pho}$}
&\multicolumn{1}{c}{JD$-$}
&\multicolumn{1}{c}{$t_\mathrm{exp}$}
&\multicolumn{1}{c}{V$_\mathrm{rad}$}
&\multicolumn{3}{c}{S/N}
&\multicolumn{1}{c}{Instrument,} \\
&\multicolumn{1}{c}{[K]}
&\multicolumn{1}{c}{[cgs]}
&\multicolumn{1}{c}{model}
&\multicolumn{1}{c}{[km\,s$^{-1}$]}
&\multicolumn{1}{c}{[cgs]}
&\multicolumn{1}{c}{$2456000$}
&\multicolumn{1}{c}{[s]}
&\multicolumn{1}{c}{[km\,s$^{-1}$]}
&\multicolumn{1}{c}{$4000$}
&\multicolumn{1}{c}{$4900$}
&\multicolumn{1}{c}{$6700$}
&\multicolumn{1}{c}{Remark} \\ \hline
&	  &	&	&      &	   &	      &       & 	       &    &	 &     &\\
25 &$4300$&$1.0$&$-1.25$&$1.40$&$1.2\pm0.2$&$ 742-785$&$15000$&$ 107.25\pm0.13$&$30$&$60$&$ 80$&UVES\\
           &	  &	&	&      &	   &$ 741.580$&       &$ 106.80\pm0.22$&    &    &     &\\
           &	  &	&	&      &	   &$ 774.661$&       &$ 108.04\pm0.08$&    &    &     &\\
           &	  &	&	&      &	   &$ 779.646$&       &$ 106.66\pm0.03$&    &    &     &\\
           &	  &	&	&      &	   &$ 784.539$&       &$ 107.38\pm0.09$&    &    &     &\\
           &	  &	&	&      &	   &$ 784.575$&       &$ 107.37\pm0.10$&    &    &     &\\
26 &$4720$&$1.9$&$-1.70$&$1.15$&$2.0\pm0.3$&$858-1126$&$15000$&$ 247.26\pm0.14$&$40$&$60$&$ 80$&UVES, also observed with HRS\\
           &	  &	&	&      &	   &$ 858.503$&       &$ 247.05\pm0.27$&    &    &     &\\
           &	  &	&	&      &	   &$ 859.505$&       &$ 246.91\pm0.14$&    &    &     &\\
           &	  &	&	&      &	   &$1069.839$&       &$ 247.29\pm0.13$&    &    &     &\\
           &	  &	&	&      &	   &$1125.606$&       &$ 247.91\pm0.20$&    &    &     &\\
           &	  &	&	&      &	   &$1125.642$&       &$ 247.01\pm0.16$&    &    &     &\\
27 &$4800$&$2.7$&$-0.65$&$0.75$&$2.1\pm0.3$&$784-1128$&$15000$&$ 108.42\pm0.10$&$30$&$45$&$ 65$&UVES\\
           &	  &	&	&      &	   &$ 784.499$&       &$ 108.44\pm0.13$&    &    &     &\\
           &	  &	&	&      &	   &$ 838.509$&       &$ 108.49\pm0.15$&    &    &     &\\
           &	  &	&	&      &	   &$1127.621$&       &$ 109.19\pm0.13$&    &    &     &\\
           &	  &	&	&      &	   &$1127.657$&       &$ 107.39\pm0.05$&    &    &     &\\
           &	  &	&	&      &	   &$1127.693$&       &$ 108.43\pm0.12$&    &    &     &\\
28 &$4450$&$0.9$&$-1.25$&$1.50$&$0.7\pm0.2$&$ 742-841$&$15000$&$   4.60\pm0.11$&$40$&$45$&$ 80$&UVES, also observed with HRS\\ 
           &	  &	&	&      &	   &$ 741.851$&       &$   4.23\pm0.17$&    &    &     &\\
           &	  &	&	&      &	   &$ 775.797$&       &$   5.06\pm0.04$&    &    &     &\\
           &	  &	&	&      &	   &$ 830.598$&       &$   4.03\pm0.17$&    &    &     &\\
           &	  &	&	&      &	   &$ 830.637$&       &$   4.22\pm0.14$&    &    &     &\\
           &	  &	&	&      &	   &$ 840.574$&       &$   5.29\pm0.20$&    &    &     &\\
\hline
\end{tabular}
\end{table}
\end{landscape}

\begin{landscape}
\begin{table}[h]
  \caption{Coordinates, magnitude, distance and velocity of the stars in the sample.
    Column (1): ID; (2)-(3)-(4)-(5)-(6): first release Gaia coordinates and G magnitude;
    (7)-(8)-(9): heliocentric distance, Galactocentric distance and Galactic Standard of Rest velocity;
    (10)-(11) longitude and latitude in a system aligned with the Sag stream; (12) name of the
    known MW substructure to which the star is possibly associated. 
    The coordinates are given in the International Celestial Reference System (ICRS) and correspond
approximately to the equatorial J2000.0 coordinates, while the reference year is 2015. The errors
on the equatorial coordinates vary between $0.1$ and $30$ milliarcseconds (mas),
depending on the number of measurements.}
\label{table:Gaia}
\tiny
\centering
\begin{tabular}{rrrrrrrrrrrrll} \hline \hline
\multicolumn{1}{c}{ID}
&\multicolumn{1}{c}{Gaia ID}
&\multicolumn{1}{c}{$\alpha$ (J2000)}
&\multicolumn{1}{c}{$\delta$ (J2000)}
&\multicolumn{1}{c}{$\ell$}
&\multicolumn{1}{c}{$b$}
&\multicolumn{1}{c}{m$_\mathrm{G}$}
&\multicolumn{1}{c}{d$_h$}
&\multicolumn{1}{c}{$r_{\rm gal}$}
&\multicolumn{1}{c}{$v_{\rm GSR}$}
&\multicolumn{1}{c}{$\Lambda_{\odot}$}
&\multicolumn{1}{c}{$B_{\odot}$}
&\multicolumn{1}{c}{Substructure}
&\multicolumn{1}{c}{Comment}\\
&
&\multicolumn{1}{c}{[decimal \degr]}
&\multicolumn{1}{c}{[decimal \degr]}
&\multicolumn{1}{c}{[decimal \degr]}
&\multicolumn{1}{c}{[decimal \degr]}
&\multicolumn{1}{c}{[mag]}
 &\multicolumn{1}{c}{[kpc]}
&\multicolumn{1}{c}{[kpc]}
&\multicolumn{1}{c}{\kms}
&\multicolumn{1}{c}{[decimal \degr]}
&\multicolumn{1}{c}{[decimal \degr]}
&
&\\
\hline
             &                    &                 &                  &                 &                 &      &       &         &          &            &      &  & \\
HRS-01 & 612887125855203968 &$ 145.722736176$ &$  11.064412509$  &$ 223.454513677$ &$  42.974341222$ &$ 16.282$ & 37.2  &    42.0 &    170.9 &   140.2182 &   -17.2112 & Sag & \\ 
      02 & 612943888143412096 &$ 146.141816105$ &$  11.279705430$  &$ 223.446107779$ &$  43.438397170$ &$ 16.393$ & 16.1  &    21.4 &    -22.3 &   139.8422 &   -16.9177 & Sag  & \\ 
      03 &3870240041683456000 &$ 158.659559972$ &$  10.347341776$  &$ 233.831860702$ &$  53.709170985$ &$ 17.099$ & 23.8  &    27.6 &    -91.4 &   127.2325 &   -14.7064 & Sag  & \\ 
      04 &3870165687209304192 &$ 158.695828606$ &$   9.973295653$  &$ 234.395945105$ &$  53.538110317$ &$ 16.558$ & 70.2  &    73.4 &    109.9 &   127.0807 &   -15.0523 & Sag  & \\
\#21 UVES&3805383905209904768 &$ 165.104636554$ &$   1.395482629$  &$ 252.281919882$ &$  52.983728567$ &$ 16.664$ &       &         &          &            &            &      &  \\ 
\#26 UVES&3682458195986672640 &$ 193.727319152$ &$  -2.341630089$  &$ 304.694033858$ &$  60.518242433$ &$ 16.637$ &       &         &          &            &            &       &   \\ 
      05 &3688559970125163904 &$ 193.960593490$ &$  -1.613468534$  &$ 305.219928945$ &$  61.238713105$ &$ 16.987$ & 65.0 &    63.2 &    -34.3 &    89.2721 &   -11.2485  & Sag  & \\ 
      06 &3688452286707840640 &$ 194.119215967$ &$  -2.199181707$  &$ 305.500614133$ &$  60.647434102$ &$ 17.181$ & 62.0 &    60.2 &    123.9 &    88.8395 &   -11.6827  &     & clump \\ 
      07 &3724956004027595392 &$ 205.560857699$ &$   9.019925634$  &$ 338.848956766$ &$  68.273464814$ &$ 16.446$ & 40.5 &    38.5 &    -67.4 &    84.4498 &     3.7067  & Sag  & \\
\#28 UVES&3652114389480042880 &$ 223.211492264$ &$   1.496296732$  &$ 356.701963111$ &$  51.228426070$ &$ 16.883$ &       &         &          &            &            &       &  \\ 
 \hline
         &                    &                 &                  &                 &                 &          &       &          &            &         &            &       &\\
 MIKE-08 &3555759031576592000 &$ 163.824823166$ &$ -18.508049407$  &$ 268.066167485$ &$  36.380672547$ &$ 16.921$ & 17.0 &    19.0 &    194.5 &   110.0752 &   -39.8714  & VOD & [Fe/H] of clump\\
          &                    &                 &                  &                 &                 &          &       &          &            &         &             &       &   but Eu,Mn,Ni off \\ 
      09 &3597524083837476864 &$ 182.062459753$ &$  -4.416150382$  &$ 283.001779175$ &$  56.775105125$ &$ 15.667$ & 26.2 &    26.4 &   -169.1 &    98.7544 &   -19.3433  & Sag  & \\ 
      10 &3688450087681660416 &$ 194.035314734$ &$  -2.273345489$  &$ 305.324070376$ &$  60.576556987$ &$ 16.659$ & 33.1 &    31.8 &   -142.3 &    88.8772 &   -11.7883  & Sag  & \\ 
      11 &3675598136782397824 &$ 192.270660520$ &$  -7.729364732$  &$ 301.911128674$ &$  55.137716988$ &$ 15.590$ & 21.4 &    20.5 &    217.4 &    87.6893 &   -17.4041  & VOD  & \\ 
      12 &3957130734774308736 &$ 196.207779183$ &$  24.320979562$  &$ 350.727065501$ &$  85.879785221$ &$ 16.970$ & 11.7 &    13.7 &    117.2 &    99.9986 &    12.4984  &   & \\ 
      13 &3682924973032975744 &$ 189.870793093$ &$  -2.916219697$  &$ 296.987047134$ &$  59.817468183$ &$ 15.135$ & 26.4 &    25.8 &    188.6 &    92.2963 &   -14.3636  & VOD & clump \\ 
      14 &2435815144861784832 &$ 355.468325584$ &$  -9.188827808$  &$  77.182067445$ &$ -65.674537158$ &$ 16.213$ & 31.5 &    31.8 &   -258.0 &   290.9595 &     9.8968  &  & clump\\ 
      15 &2436884729157603840 &$ 351.627842024$ &$ -10.473751039$  &$  68.661339153$ &$ -63.980925733$ &$ 15.787$ & 24.0 &    24.1 &   -233.4 &   295.0057 &    10.2066  &  & clump \\ 
      16 &6880970854429743232 &$ 303.805525046$ &$ -11.519145229$  &$  31.805016791$ &$ -23.753969365$ &$ 15.847$ & 29.2 &    23.5 &   -168.7 &   342.7682 &    18.9589  & Sag & \\ 
      17 &1733990241423159936 &$ 311.867125660$ &$   3.229578126$  &$  50.168829709$ &$ -23.903942909$ &$ 16.132$ & 29.5 &    25.6 &   -167.3 &   333.0929 &    33.3503  &  & clump\\ 
      18 &2720307593796059136 &$ 331.814984733$ &$   5.764033021$  &$  66.378274064$ &$ -38.465250833$ &$ 15.196$ & 23.3 &    22.1 &     52.8 &   309.2599 &    32.1026  & Her-Aqu & clump\\ 
 \hline
       &                    &                 &                    &                 &                 &          &      &          &            &         &    \\
UVES-19&2583930459319327872 &$  18.041523460$  &$  12.998214278$ &$ 130.729521882$ &$ -49.559449053$  &$ 16.365$  & 55.2 &    59.0 &   -158.9 &   259.7272 &    19.0744  & Sag  & \\ 
     20&3261503881461950208 &$  51.549431319$  &$  -2.494533042$ &$ 186.031431352$ &$ -45.549794858$  &$ 15.556$  & 41.7 &    47.6 &    -83.3 &   238.6173 &   -11.1438  & Sag & \\ 
     21&3805383905209904768 &$ 165.104636554$  &$   1.395482629$ &$ 252.281919882$ &$  52.983728567$  &$ 16.664$  & 16.3 &    19.4 &     -8.9 &   117.7666 &   -21.0405  & Sag & \\ 
     22&3585463609511300864 &$ 176.168096220$  &$ -12.897982785$ &$ 278.781672081$ &$  46.819132340$  &$ 16.595$  & 26.8 &    27.2 &   -217.9 &   100.1554 &   -29.5518  &    & clump\\ 
     23&3540980289631278208 &$ 172.801776544$  &$ -22.909035199$ &$ 279.841303528$ &$  36.339371983$  &$ 16.413$  & 37.0 &    36.8 &   -130.9 &    97.7173 &   -39.8698  &  & \\ 
     24&3541034852893925632 &$ 172.784730895$  &$ -22.767590125$ &$ 279.756572860$ &$  36.464282297$  &$ 15.999$  & 41.0 &    40.7 &     82.1 &    97.8268 &   -39.7552  &  &  \\ 
     25&3543472882489394304 &$ 176.277546542$  &$ -19.199845600$ &$ 282.021491710$ &$  40.963207478$  &$ 16.546$  & 45.9 &    45.3 &    -52.8 &    96.4867 &   -35.0405  &  & clump \\ 
     26&3682458195986672640 &$ 193.727319152$  &$  -2.341630089$ &$ 304.694033858$ &$  60.518242433$  &$ 16.637$  & 36.7 &    35.3 &    165.2 &    89.1178 &   -11.9979  &  & \\ 
     27&3576989398518840576 &$ 184.185698087$  &$ -13.077444623$ &$ 290.013749161$ &$  48.922114525$  &$ 16.926$  & 11.0 &    12.1 &    -23.0 &    92.2993 &   -25.9764  &  & \\ 
     28&3652114389480042880 &$ 223.211492264$  &$   1.496296732$ &$ 356.701963111$ &$  51.228426070$  &$ 16.883$  & 50.3 &    45.7 &      8.3 &    65.4251 &     6.1280  & Sag & \\ 
 \hline
\end{tabular}
\end{table}
\end{landscape}

\begin{table}
\caption{Carbon abundance obtained from the CH G band, for stars observed with MIKE
(except for star \#12, see text) and UVES. The error given is the formal
  one provided by the fit and should be considered as an indication of its quality.
  A more realistic error would amount to $\sim 0.2 dex$, taking into
account the uncertainties on the continuum, oxygen abundance, and stellar parameters
(especially T$_\mathrm{eff}$). The adopted C- solar abundance is 8.55 as in Jablonka et al. (2015).} 
\label{tab:carbon_ab}
\centering
\begin{tabular}{cccccc}
        \hline \hline
Star \#& [C/H] & [C/Fe] & Star \# & [C/H]& [C/Fe]  \\ 
        \hline\\
08 & $6.92 \pm 0.04$ & $-0.43$ & 19 & $5.01\pm 0.01$ & $-0.56$ \\
09 & $5.79 \pm 0.01$ & $-0.50$ & 20 & $4.48\pm 0.02$ & $-0.95$ \\
10 & $6.52 \pm 0.03$ & $-0.31$ & 21 & $7.21\pm 0.02$ & $-0.57$ \\
11 & $6.97 \pm 0.04$ & $-0.62$ & 22 & $6.77\pm 0.02$ & $-0.46$ \\
13 & $6.69 \pm 0.05$ & $-0.72$ & 23 & $6.12\pm 0.01$ & $-0.49$ \\
14 & $6.80 \pm 0.03$ & $-0.42$ & 24 & $5.30\pm 0.01$ & $-0.52$ \\
15 & $6.72 \pm 0.02$ & $-0.50$ & 25 & $6.64\pm 0.03$ & $-0.64$ \\
16 & $6.57 \pm 0.04$ & $-0.68$ & 26 & $6.27\pm 0.01$ & $-0.53$ \\
17 & $6.79 \pm 0.02$ & $-0.44$ & 27 & $7.75\pm 0.03$ & $-0.18$ \\
18 & $6.68 \pm 0.03$ & $-0.65$ & 28 & $6.41\pm 0.02$ & $-0.89$ \\
        \hline
\end{tabular}
\end{table}

\begin{landscape}
\begin{table}[h]
\begin{center}
  \caption{Average abundance ratios. Column (1) gives the star identification (after the instrument name at the head of each block); 
    columns (2) and (3) give the neutral and ionized iron abundance relative to solar, respectively; 
    the remaining columns give the abundance ratios for each of the specie listed in the Table header relative to neutral iron and normalized to solar
    ([X/FeI]). For each star, the abundance ratios are given in the first line and the corresponding errors in the second line.
    Below each of the element, we list the corresponding solar abundance according to \citet{GS98},
    on the usual scale where the H abundance is 12. For reasons of space only TiII, VI, and CrI are given in this table.}
\tiny
\tabcolsep 4pt
\label{tab:abundances}
\begin{tabular}{l|rr|rrrrrrrrrrrrrrrrrrrrrrrrr}
\hline
ID&$\mathrm{\left[\frac{FeI}{H}\right]}$
&$\mathrm{\left[\frac{FeII}{H}\right]}$
                           & O    &  Na   & Mg    & Al    & Si    & Ca    & Sc    & TiII  &   VI  &  CrI  &  Mn   & Co    & Ni    & Cu     & Zn    & Sr    & Y     & Zr    & Ba    & La    & Ce    & Pr    & Nd    & Sm    & Eu     \\
  &       &       &       &       &       &       &       &       &       &       &       &       &       &       &       &       &       &       &       &       &       &       &       &     &     &        \\ 
\hline
   & $7.50$& $7.50$&$8.83$&$6.33$ & $7.58$& $6.47$& $7.55$& $6.36$& $3.17$& $5.02$& $4.00$& $5.67$& $5.39$& $4.92$& $6.25$& $4.21$ & $4.60$& $2.97$& $2.24$& $2.60$& $2.13$& $1.17$& $1.58$& $0.71$& $1.50$& $1.01$& $0.51$ \\              
\hline
     HRS:&       &       &       &       &       &       &       &       &       &       &       &       &       &       &       &       &       &       &       &       &       &       &       &     &     &        \\ 
         01&$-2.41$&$-2.25$&$  -  $&$ 0.46$&$ 0.27$&$  -  $&$  -  $&$ 0.43$&$-0.14$&$ 0.34$&$  -  $&$-0.26$&$  -  $&$  -  $&$  -  $&$  -  $&$  -  $&$  -  $&$-0.30$&$  -  $&$ 1.03$&$  -  $&$  -  $&$  -  $&$ 0.64$&$  -  $&$  -  $ \\
           &$ 0.09$&$ 0.16$&$  -  $&$ 0.14$&$ 0.08$&$  -  $&$  -  $&$ 0.13$&$ 0.08$&$ 0.09$&$  -  $&$ 0.22$&$  -  $&$  -  $&$  -  $&$  -  $&$  -  $&$  -  $&$ 0.08$&$  -  $&$ 0.14$&$  -  $&$  -  $&$  -  $&$ 0.06$&$  -  $&$  -  $ \\
         02&$-0.73$&$-0.72$&$ 0.42$&$-0.21$&$ 0.09$&$  -  $&$ 0.16$&$ 0.28$&$-0.07$&$ 0.58$&$-0.10$&$ 0.21$&$-0.47$&$  -  $&$-0.36$&$-0.63$&$  -  $&$  -  $&$ 0.04$&$  -  $&$ 0.86$&$-0.02$&$  -  $&$  -  $&$ 0.33$&$  -  $&$ 0.83$ \\
           &$ 0.11$&$ 0.09$&$ 0.13$&$ 0.17$&$ 0.11$&$  -  $&$ 0.09$&$ 0.13$&$ 0.14$&$ 0.14$&$ 0.06$&$ 0.13$&$ 0.11$&$  -  $&$ 0.15$&$ 0.18$&$  -  $&$  -  $&$ 0.08$&$  -  $&$ 0.10$&$ 0.11$&$  -  $&$  -  $&$ 0.08$&$  -  $&$ 0.07$ \\
\end{tabular}
\end{center}
\end{table}
\end{landscape}

\setcounter{table}{3}
\begin{landscape}
\begin{table}[h]
\begin{center}
\caption{(continued)}
\tiny
\tabcolsep 4pt
\label{tab:abundances_cont}
\begin{tabular}{l|rr|rrrrrrrrrrrrrrrrrrrrrrrrr}
\hline
ID&$\mathrm{\left[\frac{FeI}{H}\right]}$
&$\mathrm{\left[\frac{FeII}{H}\right]}$
                           & O     & Na    & Mg    & Al    & Si    & Ca    & Sc    & TiII  &   VI  &  CrI  &  Mn   & Co    & Ni    & Cu    & Zn    & Sr    & Y     & Zr    & Ba    & La    & Ce    & Pr    & Nd    & Sm    & Eu \\
\hline
& $7.50$& $7.50$& $8.83$&$6.33$ & $7.58$& $6.47$& $7.55$& $6.36$& $3.17$& $5.02$& $4.00$& $5.67$& $5.39$& $4.92$& $6.25$& $4.21$& $4.60$& $2.97$& $2.24$& $2.60$& $2.13$& $1.17$& $1.58$& $0.71$& $1.50$& $1.01$& $0.51$ \\              
      \hline
     UVES: &       &       &       &       &       &       &       &       &       &       &       &       &       &       &       &       &       &       &       &       &       &       &       &       &     &     &        \\ 
         19&$-2.98$&$-2.84$&$  -  $&$-0.01$&$ 0.21$&$-0.58$&$  -  $&$ 0.37$&$-0.14$&$ 0.09$&$  -  $&$-0.26$&$-0.29$&$-0.08$&$-0.20$&$  -  $&$  -  $&$-1.04$&$  -  $&$  -  $&$-0.98$&$  -  $&$  -  $&$  -  $&$  -  $&$  -  $&$  -  $ \\ 
           &$ 0.05$&$ 0.05$&$  -  $&$ 0.12$&$ 0.11$&$ 0.15$&$  -  $&$ 0.05$&$ 0.06$&$ 0.06$&$  -  $&$ 0.07$&$ 0.11$&$ 0.11$&$ 0.09$&$  -  $&$  -  $&$ 0.17$&$  -  $&$  -  $&$ 0.08$&$  -  $&$  -  $&$  -  $&$  -  $&$  -  $&$  -  $ \\ 
         20&$-3.12$&$-2.92$&$  -  $&$ 0.19$&$ 0.45$&$  -  $&$  -  $&$ 0.42$&$ 0.25$&$ 0.48$&$  -  $&$-0.32$&$-0.53$&$-0.06$&$-0.07$&$  -  $&$ 0.44$&$ 0.07$&$-0.01$&$ 0.23$&$ 0.35$&$ 0.50$&$ 0.64$&$  -  $&$  -  $&$  -  $&$ 0.69$ \\ 
           &$ 0.05$&$ 0.08$&$  -  $&$ 0.12$&$ 0.08$&$  -  $&$  -  $&$ 0.05$&$ 0.05$&$ 0.06$&$  -  $&$ 0.06$&$ 0.16$&$ 0.09$&$ 0.10$&$  -  $&$ 0.04$&$ 0.12$&$ 0.05$&$ 0.06$&$ 0.08$&$ 0.07$&$ 0.04$&$  -  $&$  -  $&$  -  $&$ 0.11$ \\ 
         21&$-0.77$&$-0.74$&$  -  $&$-0.36$&$-0.01$&$  -  $&$ 0.03$&$ 0.23$&$ 0.03$&$ 0.20$&$ 0.00$&$-0.11$&$-0.57$&$ 0.01$&$-0.17$&$-0.04$&$-0.01$&$  -  $&$ 0.18$&$  -  $&$ 0.74$&$ 0.44$&$ 0.24$&$  -  $&$ 0.70$&$  -  $&$ 0.83$ \\
\end{tabular}
\end{center}
\end{table}
\end{landscape}

\onecolumn
\tiny
\begin{longtable}{llrr|rr|rr|rr|rr|rr}
\caption{\label{table:lines_a}Line parameters, observed equivalent widths and corresponding abundances
of stars 01, 02, 03, 04, and \#21 UVES observed with HRS.}\\
%\hline\hline
El.&\multicolumn{1}{c}{$\lambda$}&$\chi_\mathrm{ex}$&$\log(gf)$&\multicolumn{10}{c}{EW$\pm\sigma$(EW)~~$\log\varepsilon(X)+12$~[dex]}\\
   & & & &\multicolumn{2}{c}{01}&\multicolumn{2}{c}{02}&\multicolumn{2}{c}{03}&\multicolumn{2}{c}{04}&\multicolumn{2}{c}{\#21 UVES}\\
\endfirsthead
\caption{continued.}\\
%%\hline\hline
El.&\multicolumn{1}{c}{$\lambda$}&$\chi_\mathrm{ex}$&$\log(gf)$&\multicolumn{10}{c}{EW$\pm\sigma$(EW)~~$\log\varepsilon(X)+12$~[dex]}\\
   & & & &\multicolumn{2}{c}{01}&\multicolumn{2}{c}{02}&\multicolumn{2}{c}{03}&\multicolumn{2}{c}{04}&\multicolumn{2}{c}{\#21 UVES}\\
\hline
\endhead
\hline
\endfoot
\hline\hline
 O 1&$ 6363.776$&$ 0.02$&$ -10.260$&$        -     $&$    -  $&$  25.7\pm  5.9$&$    8.53$&$	    -	  $&$	  -  $&$	-     $&$     -  $&$	    -	  $&$	  -  $ \\
Na 1&$ 5682.633$&$ 2.10$&$  -0.706$&$        -     $&$    -  $&$	-     $&$     -  $&$  28.8\pm  3.3$&$	 4.76$&$	-     $&$     -  $&$	    -	  $&$	  -  $ \\
Na 1&$ 5688.205$&$ 2.10$&$  -0.452$&$        -     $&$    -  $&$	-     $&$     -  $&$	    -	  $&$	  -  $&$  21.7\pm  3.4$&$    4.23$&$	    -	  $&$	  -  $ \\
Na 1&$ 5889.951$&$ 0.00$&$   0.108$&$        -     $&$    -  $&$ 344.3\pm 24.7$&$    5.08$&$	    -	  $&$	  -  $&$	-     $&$     -  $&$	    -	  $&$	  -  $ \\
Na 1&$ 5895.924$&$ 0.00$&$  -0.194$&$ 174.8\pm 13.3$&$   4.38$&$ 306.0\pm 22.5$&$    5.22$&$ 286.5\pm 18.4$&$	 5.11$&$ 189.2\pm 13.3$&$    3.88$&$	    -	  $&$	  -  $ \\
Na 1&$ 6154.226$&$ 2.10$&$  -1.547$&$        -     $&$    -  $&$  35.5\pm  3.5$&$    5.61$&$	    -	  $&$	  -  $&$	-     $&$     -  $&$	    -	  $&$	  -  $ \\
Mg 1&$ 5172.684$&$ 2.71$&$  -0.450$&$ 210.1\pm 12.8$&$   5.44$&$	-     $&$     -  $&$	    -	  $&$	  -  $&$ 242.2\pm 13.7$&$    5.41$&$	    -	  $&$	  -  $ \\
Mg 1&$ 5183.604$&$ 2.72$&$  -0.239$&$        -     $&$    -  $&$	-     $&$     -  $&$	    -	  $&$	  -  $&$ 269.9\pm 16.3$&$    5.37$&$	    -	  $&$	  -  $ \\
\end{longtable}
%\onecolumn

%\tiny
\begin{longtable}{llrr|rr|rr|rr|rr|rr}
\caption{\label{table:lines_b}Line parameters, observed equivalent widths and corresponding abundances
of stars \#26 UVES, 05, 06, 07, and \#28 UVES observed with HRS.}\\
%\hline\hline
El.&\multicolumn{1}{c}{$\lambda$}&$\chi_\mathrm{ex}$&$\log(gf)$&\multicolumn{10}{c}{EW$\pm\sigma$(EW)~~$\log\varepsilon(X)+12$~[dex]}\\
   & & & &\multicolumn{2}{c}{\#26 UVES}&\multicolumn{2}{c}{05}&\multicolumn{2}{c}{06}&\multicolumn{2}{c}{07}&\multicolumn{2}{c}{\#28 UVES}\\
\endfirsthead
\caption{continued.}\\
%%\hline\hline
El.&\multicolumn{1}{c}{$\lambda$}&$\chi_\mathrm{ex}$&$\log(gf)$&\multicolumn{10}{c}{EW$\pm\sigma$(EW)~~$\log\varepsilon(X)+12$~[dex]}\\
   & & & &\multicolumn{2}{c}{\#26 UVES}&\multicolumn{2}{c}{05}&\multicolumn{2}{c}{06}&\multicolumn{2}{c}{07}&\multicolumn{2}{c}{\#28 UVES}\\
\hline
\endhead
\hline
\endfoot
\hline\hline
Na 1&$ 5688.205$ &$ 2.10$&$  -0.452$&$  27.0\pm  2.6$&$   4.43$&$	 -     $&$    -  $&$  84.7\pm  7.3$&$	5.08$&$        -     $&$    -  $&$	  -	$&$    -  $ \\
Na 1&$ 5889.951$ &$ 0.00$&$   0.108$&$ 239.9\pm 18.0$&$   4.40$&$ 221.5\pm 15.8$&$   4.24$&$	    -	  $&$	 -  $&$ 210.7\pm 19.7$&$   4.25$&$	  -	$&$    -  $ \\
Na 1&$ 5895.924$ &$ 0.00$&$  -0.194$&$ 229.4\pm 15.0$&$   4.64$&$ 218.6\pm 14.7$&$   4.51$&$	    -	  $&$	 -  $&$ 178.8\pm 11.6$&$   4.19$&$	  -	$&$    -  $ \\
Mg 1&$ 5172.684$ &$ 2.71$&$  -0.450$&$        -     $&$    -  $&$ 233.5\pm 13.4$&$   5.57$&$	    -	  $&$	 -  $&$ 197.9\pm 10.8$&$   5.29$&$	  -	$&$    -  $ \\
Mg 1&$ 5183.604$ &$ 2.72$&$  -0.239$&$        -     $&$    -  $&$ 261.9\pm 14.7$&$   5.56$&$	    -	  $&$	 -  $&$ 213.4\pm 11.2$&$   5.21$&$	  -	$&$    -  $ \\
Mg 1&$ 5528.405$ &$ 4.35$&$  -0.498$&$ 132.7\pm  7.0$&$   6.22$&$ 109.0\pm  6.1$&$   5.82$&$ 197.2\pm 10.2$&$	6.63$&$  63.5\pm  3.2$&$   5.13$&$ 207.6\pm 11.5$&$   6.88$ \\
Mg 1&$ 5711.088$ &$ 4.35$&$  -1.724$&$  52.8\pm  3.1$&$   6.18$&$	 -     $&$    -  $&$ 104.1\pm  6.7$&$	6.64$&$        -     $&$    -  $&$  69.3\pm  6.0$&$   6.25$ \\
Si 1&$ 5645.613$ &$ 4.93$&$  -2.140$&$        -     $&$    -  $&$	 -     $&$    -  $&$  43.0\pm  3.3$&$	6.97$&$        -     $&$    -  $&$	  -	$&$    -  $ \\
\end{longtable}

\newpage

\begin{longtable}{llrr|rr|rr|rr|rr|rr}
\caption{\label{table:lines_c}Line parameters, observed equivalent widths and corresponding abundances
of stars 08, 09, 10, 11, and 12 observed with MIKE.}\\
El.&\multicolumn{1}{c}{$\lambda$}&$\chi_\mathrm{ex}$&$\log(gf)$&\multicolumn{10}{c}{EW$\pm\sigma$(EW)~~$\log\varepsilon(X)+12$~[dex]}\\
   & & & &\multicolumn{2}{c}{08}&\multicolumn{2}{c}{09}&\multicolumn{2}{c}{10}&\multicolumn{2}{c}{11}&\multicolumn{2}{c}{12}\\
\endfirsthead
\caption{continued.}\\
El.&\multicolumn{1}{c}{$\lambda$}&$\chi_\mathrm{ex}$&$\log(gf)$&\multicolumn{10}{c}{EW$\pm\sigma$(EW)~~$\log\varepsilon(X)+12$~[dex]}\\
   & & & &\multicolumn{2}{c}{08}&\multicolumn{2}{c}{09}&\multicolumn{2}{c}{10}&\multicolumn{2}{c}{11}&\multicolumn{2}{c}{12}\\
\hline
\endhead
\hline
\endfoot
\hline\hline
Li 1&$ 6707.764$&$    0$&$  -0.036$&$        -     $&$    -  $&$  19.3\pm  2.0$&$   1.01$&$       -      $&$    -  $&$        -     $&$    -  $&$        -     $&$    -  $ \\
 O 1&$ 6300.304$&$    0$&$  -9.750$&$        -     $&$    -  $&$        -     $&$    -  $&$       -      $&$    -  $&$  37.9\pm  2.7$&$   8.13$&$        -     $&$    -  $ \\
 O 1&$ 6363.776$&$ 0.02$&$ -10.260$&$        -     $&$    -  $&$        -     $&$    -  $&$   9.7\pm  3.0$&$  <7.90$&$        -     $&$    -  $&$        -     $&$    -  $ \\
Na 1&$ 5682.633$&$ 2.10$&$  -0.706$&$  43.3\pm  3.8$&$   5.16$&$        -     $&$    -  $&$  20.5\pm  3.4$&$   4.61$&$  63.5\pm  5.6$&$   5.11$&$        -     $&$    -  $ \\
Na 1&$ 5895.924$&$    0$&$  -0.194$&$        -     $&$    -  $&$        -     $&$    -  $&$       -      $&$    -  $&$        -     $&$    -  $&$ 159.9\pm 11.0$&$   4.46$ \\
Na 1&$ 6160.747$&$ 2.10$&$  -1.246$&$        -     $&$    -  $&$        -     $&$    -  $&$       -      $&$    -  $&$  27.5\pm  3.6$&$   5.00$&$        -     $&$    -  $ \\
Mg 1&$ 4571.096$&$ 0.00$&$  -5.623$&$ 112.7\pm 10.1$&$   6.62$&$  94.1\pm  6.3$&$   5.77$&$ 108.0\pm  8.9$&$   6.15$&$        -     $&$    -  $&$        -     $&$    -  $ \\
Mg 1&$ 4702.991$&$ 4.35$&$  -0.440$&$ 134.8\pm  9.3$&$   6.31$&$ 101.0\pm  6.3$&$   5.68$&$ 116.8\pm  7.7$&$   6.00$&$        -     $&$    -  $&$        -     $&$    -  $ \\
\end{longtable}
\begin{longtable}{llrr|rr|rr|rr|rr|rr}
\caption{\label{table:lines_d}Line parameters, observed equivalent widths and corresponding abundances
of stars 13, 14, 15, 16, and 17 observed with MIKE.}\\
El.&\multicolumn{1}{c}{$\lambda$}&$\chi_\mathrm{ex}$&$\log(gf)$&\multicolumn{10}{c}{EW$\pm\sigma$(EW)~~$\log\varepsilon(X)+12$~[dex]}\\
   & & & &\multicolumn{2}{c}{13}&\multicolumn{2}{c}{14}&\multicolumn{2}{c}{15}&\multicolumn{2}{c}{16}&\multicolumn{2}{c}{17}\\
\endfirsthead
\caption{continued.}\\
El.&\multicolumn{1}{c}{$\lambda$}&$\chi_\mathrm{ex}$&$\log(gf)$&\multicolumn{10}{c}{EW$\pm\sigma$(EW)~~$\log\varepsilon(X)+12$~[dex]}\\
   & & & &\multicolumn{2}{c}{13}&\multicolumn{2}{c}{14}&\multicolumn{2}{c}{15}&\multicolumn{2}{c}{16}&\multicolumn{2}{c}{17}\\
\hline
\endhead
\hline
\endfoot
\hline\hline
 O 1&$ 6300.304$&$ 0.00$&$  -9.750$&$  65.5\pm  4.9$&$   8.11$&$        -     $&$    -  $&$  51.1\pm  5.2$&$   8.19$&$  40.6\pm  2.8$&$   7.74$&$  44.4\pm  6.0$&$   8.04$ \\
 O 1&$ 6363.776$&$ 0.02$&$ -10.260$&$  24.9\pm  1.7$&$   8.03$&$  21.4\pm  1.7$&$   8.15$&$        -     $&$    -  $&$        -     $&$    -  $&$        -     $&$    -  $ \\
Na 1&$ 5682.633$&$ 2.10$&$  -0.706$&$  69.5\pm  4.6$&$   4.98$&$  30.4\pm  3.5$&$   4.60$&$  30.3\pm  3.1$&$   4.64$&$        -     $&$    -  $&$  38.8\pm  4.3$&$   4.79$ \\
Na 1&$ 5688.205$&$ 2.10$&$  -0.452$&$        -     $&$    -  $&$        -     $&$    -  $&$  56.6\pm  6.0$&$   4.85$&$        -     $&$    -  $&$        -     $&$    -  $ \\
Na 1&$ 6160.747$&$ 2.10$&$  -1.246$&$  35.3\pm  3.3$&$   4.96$&$        -     $&$    -  $&$        -     $&$    -  $&$  21.6\pm  3.5$&$   4.70$&$        -     $&$    -  $ \\
Mg 1&$ 4167.271$&$ 4.35$&$  -0.745$&$        -     $&$    -  $&$        -     $&$    -  $&$ 143.7\pm 10.0$&$   6.64$&$        -     $&$    -  $&$        -     $&$    -  $ \\
Mg 1&$ 4571.096$&$    0$&$  -5.623$&$        -     $&$    -  $&$        -     $&$    -  $&$ 142.0\pm  9.4$&$   6.56$&$        -     $&$    -  $&$ 138.7\pm  9.1$&$   6.47$ \\
Mg 1&$ 4702.991$&$ 4.35$&$  -0.440$&$        -     $&$    -  $&$        -     $&$    -  $&$ 152.2\pm 10.2$&$   6.40$&$        -     $&$    -  $&$ 153.7\pm  9.0$&$   6.44$ \\
\end{longtable}

\begin{longtable}{llrr|rr}
\caption{\label{table:lines_e}Line parameters, observed equivalent widths and corresponding abundances
of star 18 observed with MIKE.}\\
El.&\multicolumn{1}{c}{$\lambda$}&$\chi_\mathrm{ex}$&$\log(gf)$&\multicolumn{2}{c}{EW$\pm\sigma$(EW)~~$\log\varepsilon(X)+12$~[dex]}\\
   & & & &\multicolumn{2}{c}{18}\\
\endfirsthead
\caption{continued.}\\
El.&\multicolumn{1}{c}{$\lambda$}&$\chi_\mathrm{ex}$&$\log(gf)$&\multicolumn{2}{c}{EW$\pm\sigma$(EW)~~$\log\varepsilon(X)+12$~[dex]}\\
   & & & &\multicolumn{2}{c}{18}\\
\hline
\endhead
\hline
\endfoot
\hline\hline
 O 1&$ 6300.304$ &$ 0.00$ &$  -9.750$ &$  42.5\pm  2.3$ &$   7.89$ \\
 O 1&$ 6363.776$ &$ 0.02$ &$ -10.260$ &$  19.9\pm  1.7$ &$   8.00$ \\
Na 1&$ 6160.747$ &$ 2.10$ &$  -1.246$ &$  26.7\pm  3.8$ &$   4.90$ \\
Mg 1&$ 5711.088$ &$ 4.35$ &$  -1.724$ &$  88.5\pm  5.9$ &$   6.44$ \\
Si 1&$ 5701.104$ &$ 4.93$ &$  -2.050$ &$  26.5\pm  2.7$ &$   6.60$ \\
Si 1&$ 5772.146$ &$ 5.08$ &$  -1.750$ &$  28.0\pm  2.7$ &$   6.52$ \\
Si 1&$ 5948.541$ &$ 5.08$ &$  -1.230$ &$  72.2\pm  5.3$ &$   6.84$ \\
Si 1&$ 6244.465$ &$ 5.62$ &$  -1.091$ &$  21.6\pm  2.6$ &$   6.35$ \\
\end{longtable}

\begin{longtable}{llrr|rr|rr|rr|rr|rr}
\caption{\label{table:lines_f}Line parameters, observed equivalent widths and corresponding abundances
of stars 19, 20, 21, 22, and 23 observed with UVES.}\\
El.&\multicolumn{1}{c}{$\lambda$}&$\chi_\mathrm{ex}$&$\log(gf)$&\multicolumn{10}{c}{EW$\pm\sigma$(EW)~~$\log\varepsilon(X)+12$~[dex]}\\
   & & & &\multicolumn{2}{c}{19}&\multicolumn{2}{c}{20}&\multicolumn{2}{c}{21}&\multicolumn{2}{c}{22}&\multicolumn{2}{c}{23}\\
\endfirsthead
\caption{continued.}\\
El.&\multicolumn{1}{c}{$\lambda$}&$\chi_\mathrm{ex}$&$\log(gf)$&\multicolumn{10}{c}{EW$\pm\sigma$(EW)~~$\log\varepsilon(X)+12$~[dex]}\\
   & & & &\multicolumn{2}{c}{19}&\multicolumn{2}{c}{20}&\multicolumn{2}{c}{21}&\multicolumn{2}{c}{22}&\multicolumn{2}{c}{23}\\
\hline
\endhead
\hline
\endfoot
\hline\hline
 O 1&$ 6300.304$&$    0$&$  -9.750$&$        -     $&$    -  $&$        -     $&$    -  $&$        -     $&$    -  $&$  31.7\pm  2.2$&$   8.07$&$        -     $&$    -  $ \\
Na 1&$ 5682.633$&$ 2.10$&$  -0.706$&$        -     $&$    -  $&$        -     $&$    -  $&$  49.0\pm  3.0$&$   5.15$&$  35.0\pm  2.8$&$   4.82$&$        -     $&$    -  $ \\
Na 1&$ 5688.205$&$ 2.10$&$  -0.452$&$        -     $&$    -  $&$        -     $&$    -  $&$  71.6\pm  3.8$&$   5.29$&$  49.6\pm  2.9$&$   4.83$&$        -     $&$    -  $ \\
Na 1&$ 5889.951$&$ 0.00$&$   0.108$&$ 148.0\pm  7.9$&$   3.34$&$ 167.2\pm  9.7$&$   3.40$&$        -     $&$    -  $&$        -     $&$    -  $&$ 213.1\pm 13.6$&$   4.21$ \\
Na 1&$ 5895.924$&$ 0.00$&$  -0.194$&$ 128.9\pm  7.0$&$   3.32$&$ 145.5\pm  7.6$&$   3.40$&$        -     $&$    -  $&$ 237.4\pm 14.4$&$   4.73$&$ 198.8\pm 10.8$&$   4.39$ \\
Mg 1&$ 3829.355$&$ 2.71$&$  -0.227$&$        -     $&$    -  $&$ 166.1\pm 12.8$&$   4.85$&$        -     $&$    -  $&$        -     $&$    -  $&$        -     $&$    -  $ \\
Mg 1&$ 3832.304$&$ 2.71$&$   0.125$&$ 176.0\pm 10.0$&$   4.85$&$        -     $&$    -  $&$        -     $&$    -  $&$        -     $&$    -  $&$        -     $&$    -  $ \\
Mg 1&$ 5172.684$&$ 2.71$&$  -0.450$&$ 162.3\pm  9.5$&$   4.76$&$ 190.0\pm 11.5$&$   4.82$&$        -     $&$    -  $&$        -     $&$    -  $&$        -     $&$    -  $ \\
\end{longtable}

\begin{longtable}{llrr|rr|rr|rr|rr|rr}
\caption{\label{table:lines_g}Line parameters, observed equivalent widths and corresponding abundances
of stars 24, 25, 26, 27, and 28 observed with UVES.}\\
El.&\multicolumn{1}{c}{$\lambda$}&$\chi_\mathrm{ex}$&$\log(gf)$&\multicolumn{10}{c}{EW$\pm\sigma$(EW)~~$\log\varepsilon(X)+12$~[dex]}\\
   & & & &\multicolumn{2}{c}{24}&\multicolumn{2}{c}{25}&\multicolumn{2}{c}{26}&\multicolumn{2}{c}{27}&\multicolumn{2}{c}{28}\\
\endfirsthead
\caption{continued.}\\
El.&\multicolumn{1}{c}{$\lambda$}&$\chi_\mathrm{ex}$&$\log(gf)$&\multicolumn{10}{c}{EW$\pm\sigma$(EW)~~$\log\varepsilon(X)+12$~[dex]}\\
   & & & &\multicolumn{2}{c}{24}&\multicolumn{2}{c}{25}&\multicolumn{2}{c}{26}&\multicolumn{2}{c}{27}&\multicolumn{2}{c}{28}\\
\hline
\endhead
\hline
\endfoot
\hline\hline
 O 1&$ 6300.304$&$    0$&$  -9.750$&$        -     $&$    -  $&$        -     $&$    -  $&$        -     $&$    -  $&$  30.8\pm  2.2$&$   8.70$&$        -     $&$    -  $ \\
 O 1&$ 6363.776$&$ 0.02$&$ -10.260$&$        -     $&$    -  $&$  21.2\pm  1.3$&$   8.00$&$        -     $&$    -  $&$        -     $&$    -  $&$        -     $&$    -  $ \\
Na 1&$ 5682.633$&$ 2.10$&$  -0.706$&$        -     $&$    -  $&$  54.0\pm  3.0$&$   4.81$&$        -     $&$    -  $&$        -     $&$    -  $&$        -     $&$    -  $ \\
Na 1&$ 5688.205$&$ 2.10$&$  -0.452$&$        -     $&$    -  $&$  75.2\pm  4.0$&$   4.89$&$  28.4\pm  2.1$&$   4.47$&$ 110.9\pm  5.7$&$   5.93$&$  53.1\pm  2.9$&$   4.67$ \\
Na 1&$ 5889.951$&$ 0.00$&$   0.108$&$ 183.1\pm  9.7$&$   3.84$&$        -     $&$    -  $&$ 229.6\pm 13.8$&$   4.32$&$        -     $&$    -  $&$        -     $&$    -  $ \\
Na 1&$ 5895.924$&$ 0.00$&$  -0.194$&$ 165.2\pm  9.2$&$   3.92$&$        -     $&$    -  $&$ 209.3\pm 11.4$&$   4.48$&$        -     $&$    -  $&$        -     $&$    -  $ \\
Na 1&$ 6154.226$&$ 2.10$&$  -1.547$&$        -     $&$    -  $&$        -     $&$    -  $&$        -     $&$    -  $&$  32.7\pm  2.2$&$   5.67$&$        -     $&$    -  $ \\
Na 1&$ 6160.747$&$ 2.10$&$  -1.246$&$        -     $&$    -  $&$  20.9\pm  1.4$&$   4.72$&$        -     $&$    -  $&$  50.6\pm  2.9$&$   5.71$&$        -     $&$    -  $ \\
\end{longtable}

\end{document}